\documentclass[fleqn,usenatbib]{mnras}

\usepackage{newtxtext,newtxmath}

\usepackage[T1]{fontenc}

\DeclareRobustCommand{\VAN}[3]{#2}
\let\VANthebibliography\thebibliography
\def\thebibliography{\DeclareRobustCommand{\VAN}[3]{##3}\VANthebibliography}

\usepackage{graphicx}	
\usepackage{amsmath}	
\usepackage{fix-cm}
\usepackage{subfig}
\usepackage{enumitem}
\usepackage{hyperref}
\usepackage{pdflscape}
\usepackage{longtable}
\usepackage[most]{tcolorbox}

\LTcapwidth=\textwidth
\newcommand{\hi}{\mbox{H\,{\sc i}}} 

\title[ML Classification of \hi\ 21-cm Absorption Spectra]{\hi\ 21-cm Absorption Spectra Classification using Machine Learning}

\author[D. Mondal, A. S. Nemmani and A. Banerjee]{Debasish Mondal,$^{1}$\thanks{E-mail: debasishmondal@labs.iisertirupati.ac.in}
Anirudh S. Nemmani$^{1,2}$\thanks{E-mail: anemmani@camk.edu.pl}
and Arunima Banerjee$^{1}$\thanks{E-mail: arunima@iisertirupati.ac.in}
\\
$^{1}$Department of Physics, Indian Institute of Science Education and Research (IISER) Tirupati, Yerpedu, Tirupati 517619, Andhra Pradesh, India\\
$^{2}$Nicolaus Copernicus Astronomical Center, Polish Academy of Sciences, Bartycka 18 00-716 Warsaw, Poland
}

\date{Accepted XXX. Received YYY; in original form ZZZ}

\pubyear{\the\year{}}

\begin{document}
\label{firstpage}
\pagerange{\pageref{firstpage}--\pageref{lastpage}}
\maketitle

\begin{abstract}
\hi\ 21-cm absorption, an extremely useful tool to study the cold atomic hydrogen gas, can arise either from the intervening galaxies along the line-of-sight towards the background radio source or from the radio source itself. Determining whether \hi\ 21-cm absorption lines detected as part of large, blind surveys are `intervening' or `associated' using optical spectroscopy would be unfeasible. We therefore investigate a more efficient, machine learning (ML)-based method to classify \hi\ 21-cm absorption lines. Using a sample of 118 known \hi\ 21-cm absorption lines from the literature, we train six ML models (Gaussian naive Bayes, logistic regression, decision tree, random forest, SVM and XGBoost) on the spectral parameters obtained by fitting the Busy function to the absorption spectra. We found that a random forest model trained on these spectral parameters gives the most reliable classification results, with an accuracy of 89\%, a $F_1$-score of 0.9 and an AUC score of 0.94. We note that the linewidth parameter $w_{20}$ is the most significant spectral parameter that regulates the classification performance of this model. Retraining this random forest model only with this linewidth and the integrated optical depth parameters yields an accuracy of 88\%, a $F_1$-score of 0.88 and an AUC score of 0.91. We have applied this retrained random forest model to predict the type of 30 new \hi\ 21-cm absorption lines detected in recent blind surveys, viz. FLASH, illustrating the potential of the techniques developed in this work for future large \hi\ surveys with the Square Kilometre Array.
\end{abstract}




\section{Introduction} 
\label{sec:1}

Atomic hydrogen (\hi) gas, being the major constituent of the interstellar medium and the reservoir for the formation of molecules and stars in galaxies, plays a crucial role in the baryon cycle and galaxy evolution \citep[e.g.][]{Peroux2020,McClure-Griffiths2023}. The \hi\ 21-cm spectral line, which occurs due to the hyperfine transition in the ground state of the hydrogen atom, is an extremely powerful tool to probe the atomic hydrogen gas in galaxies \citep[see][for a review]{Dutta2022}. However, due to the faintness of this line, it can be detected in emission only from nearby galaxies ($z \lesssim 0.2$) in reasonable integration times with current radio telescopes \citep[e.g.][]{Fernandez2016}. On the other hand, \hi\ 21-cm absorption can be used to trace the cold ($T\sim100-1000$\,K) atomic gas in galaxies independent of redshift provided there is a radio-loud background source such as a quasar or a radio-loud galaxy.

\hi\ 21-cm absorption has been used extensively to study the cold gas in normal galaxies that lie between the background source and the observer \citep[see][for a review]{Dutta2019}, as well as in active radio galaxies that are the background source themselves \citep[see][for a review]{Morganti2018}. In the former case, the absorption is termed as `intervening', and in the latter, it is termed as `associated'. While intervening \hi\ 21-cm absorption observations have shed light on the distribution and physical properties of cold atomic gas in and around galaxies \citep[e.g.][]{Kanekar2009b,Dutta2017a,Dutta2017b}, associated \hi\ 21-cm absorption has been used to probe AGN feeding and feedback processes and the AGN-galaxy co-evolution \citep[e.g.][]{Gereb2015,Allison2016,Murthy2021}. However, the number of detections of \hi\ 21-cm absorption has been limited for various reasons, including bias against dust-obscured systems due to optical pre-selection, relatively narrow frequency bandwidths, and radio frequency interference that restricts searchable frequency ranges \citep[e.g.][]{Curran2006,Curran2008}.

With the recent technological advancements made with the Square Kilometre Array \citep[SKA;][]{Weltman2020} precursor telescopes such as Australian SKA Pathfinder \citep[ASKAP;][]{Johnston2008} and MeerKAT \citep{Jonas2009}, it is now possible to conduct blind searches for \hi\ 21-cm absorption over large sky areas and continuous frequency coverage in radio-quiet sites. Thus, ongoing and upcoming surveys with these telescopes expect to detect hundreds to thousands of new \hi\ 21-cm absorbers \citep[e.g.][]{Gupta2021,Allison2022}. To infer the nature of the absorbing gas and to conduct statistical studies with large samples of \hi\ 21-cm absorbers, it is imperative to determine whether these lines are due to associated or intervening absorption. In other words, the redshift of the radio source is required to determine whether the \hi\ 21-cm absorption line is arising from the vicinity of the radio source or from a foreground galaxy. 

However, obtaining the redshifts of the radio sources towards which \hi\ 21-cm absorption is detected would require follow-up deep optical/infrared spectroscopy \citep[e.g.][]{Allison2015}. Moreover, spectroscopic determination of redshifts will not be possible for those that exhibit weak or no spectral lines such as blazars \citep[e.g.][]{Yan2012,Yan2016}. Instead of spectroscopic redshifts, it may be possible to predict the photometric redshifts of the radio sources by training machine learning (ML) models such as neural networks on multi-band photometric data \citep[e.g.][]{Beck2021,Henghes2022}. However, for accurate predictions, such models typically require multi-band measurements from the near-infrared to far-ultraviolet for training, which may not be available for all the sources. Therefore, obtaining the absorber type from the photometric redshifts predicted by training ML models for a large number of radio sources using spectroscopy data or on extensive multi-band photometry data would be time-consuming and impractical. Alternatively, the \hi\ 21-cm absorption line properties themselves could be used for training ML models to predict the absorber type.

The objective of this manuscript is to develop ML classification models to categorize large samples of \hi\ 21-cm absorption lines that are expected to be detected in SKA surveys into intervening or associated. Previously, \citet{Curran2016} had trained five ML classification models (Bayesian network, sequential minimal optimization, classification via regression, logistic model tree and random forest) using a sample of 98 \hi\ 21-cm absorbers. Subsequently, \citet{Curran2021} used a similar approach to train four ML classification models with a sample of 136 \hi\ 21-cm absorbers. Both the studies used Gaussian profiles to fit the absorption spectra (obtained through digitization) and extract properties such as linewidth, optical depth and number of Gaussian components, which were used to train the ML models. In both cases, test accuracy of $\approx 80\%$ was obtained. Gaussian profiles are symmetric and can not accurately describe the characteristics of the broad, asymmetric double-horn profiles of spiral galaxies. The flanks of Gaussian profiles are not steep enough to reflect the sharp rise generally seen, particularly in the spectra of large spiral galaxies. Also, the central trough does not resemble the broad, flat troughs seen in many disc galaxies \citep{Stewart2014}. Multiple Gaussians may bear resemblance to the double-horn galaxy profiles, but still differ sometimes from the actual shape of most galaxy spectra due to these limitations \citep{Koch2021}. The Busy function is versatile when it comes to fitting spectral profiles of different shapes due to its two constituent error functions and one polynomial function \citep[see][]{Westmeier2014}. Thus, we aim to investigate whether using the Busy function to fit a sample of 118 \hi\ 21-cm absorption spectra, and training the ML models on a different and larger set of extracted spectral parameters, leads to any improvement in the classification results over Gaussian fitting. We have trained six different ML Classification models -- Gaussian naive Bayes, logistic regression, decision tree, random forest, support vector machine (SVM) and extreme gradient boosting (XGBoost) -- to categorize the \hi\ 21-cm absorption spectra into associated or intervening using the spectral parameters extracted via Busy function fitting.

The rest of this manuscript is structured as follows. Section \ref{sec:2} provides a description of our \hi\ 21-cm absorber sample and the fitting of spectra using the Busy function. In Section \ref{sec:3}, the ML classification algorithms used in the work for model training are described briefly. Section \ref{sec:4} presents the results from the different ML classification models and discusses their implications. Finally, the conclusions from this work are outlined in Section \ref{sec:5}.

\section{Data} 
\label{sec:2}

\subsection{Absorber sample}
\label{sec:2.1}
Our data sample comprises 118 \hi\ 21-cm absorption line spectra, of which 74 are associated and 44 are intervening. The associated absorbers are defined as those whose redshifts fall within $\pm$ 3000 km $\text{s}^{-1}$ of the systemic redshift of the radio AGN, and therefore are likely to be physically associated with the radio AGN \citep{Ellison2002,Prochaska2008}; the remaining absorbers are defined as intervening. All the spectra of our data sample are collected through a thorough literature survey (see Table \ref{tab:b1} for references). Only a small fraction ($<10$\%) of the spectra (marked with $\ast$ in Table \ref{tab:b1}) are obtained from digitized versions \citep[e.g.][]{Curran2021} using ADS's \texttt{Dexter Data Extraction Applet} \citep{Demleitner2001}. Remaining are obtained from the corresponding authors in \texttt{ASCII} format. We collected our data samples independently; however, we have 52 spectra (20 associated and 32 intervening) in common with \citep{Curran2021}. In our data sample, we include only the spectra from the literature with confirmed \hi\ 21-cm detections, for which we could obtain reliable Busy function fits (see Section \ref{sec:2.2}). Having access to high-quality spectra in \texttt{ASCII} format mitigated errors in the fitting process and led to more accurate fit parameters. 

\subsection{Spectral fitting using Busy function}
\label{sec:2.2}

\hi\ 21-cm absorption lines can provide us with information on various physical properties of the atomic gas such as the spin temperature and kinematics \citep{Field1959,Kulkarni1988}. Thus, robust spectral fitting methods are required to extract meaningful information from them. The Gaussian function has been used extensively in the literature to fit \hi\ 21-cm absorption lines \citep{Heiles2003,Roy2013}, but this method is subjective to the number of components used. \citet{Gereb2015} used a more robust function known as the Busy function \citep{Westmeier2014} to characterise \hi\ 21-cm absorption lines.

The Busy function is a continuously differentiable analytic function used to model spectral lines. A continuously differentiable function means the function is differentiable within its domain, and its derivative is a continuous function. Such functions are required to evaluate the partial derivatives with respect to the function’s free parameters for the purpose of least-squares fitting on a given spectral dataset. The Busy function is formed by multiplying a polynomial by two error functions. A simplified functional form of the generalised Busy function is as follows: 
\begin{equation}
\label{eq:1}
\begin{split}
B(x) = \frac{a}{4} & \times (\text{erf}[b_1\{w + x - x_e\}] + 1)\\
                   & \times (\text{erf}[b_2\{w - x + x_e\}] + 1) \times (c|x - x_p|^n + 1),
\end{split}
\end{equation}
where the form of error function is, $\text{erf}(x) = \frac{2}{\sqrt{\pi}} \int_{0}^{x} e^{-t^2} dt$ and $x$ represents the spectral axis. It has eight free parameters, namely, $a$ - total amplitude scaling factor; $b_1$ and $b_2$ - slopes of the two error functions, $x_{\text{e}}$ - offset of the two error functions, $c$ - amplitude of the central trough of the fitted polynomial, $x_{\text{p}}$ - offset of the fitted polynomial, $n$ - degree of the fitted polynomial, $w$ - half-width of the fitted profile. Apart from these eight parameters, five more parameters are extracted from the fitted spectral profile, namely, $x_0$ - centroid of the spectrum, $\tau_{\text{peak}}$ - peak optical depth, $\tau_{\text{int}}$ - integrated optical depth, $w_{50}$ and $w_{20}$ - the spectral linewidths at 20\% and 50\% of $\tau_{\text{peak}}$, respectively.   

\begin{figure*}
\centering
\subfloat[Busy function fit\label{fig:1a}]{\includegraphics[height=0.3\textwidth,width=0.48\textwidth]{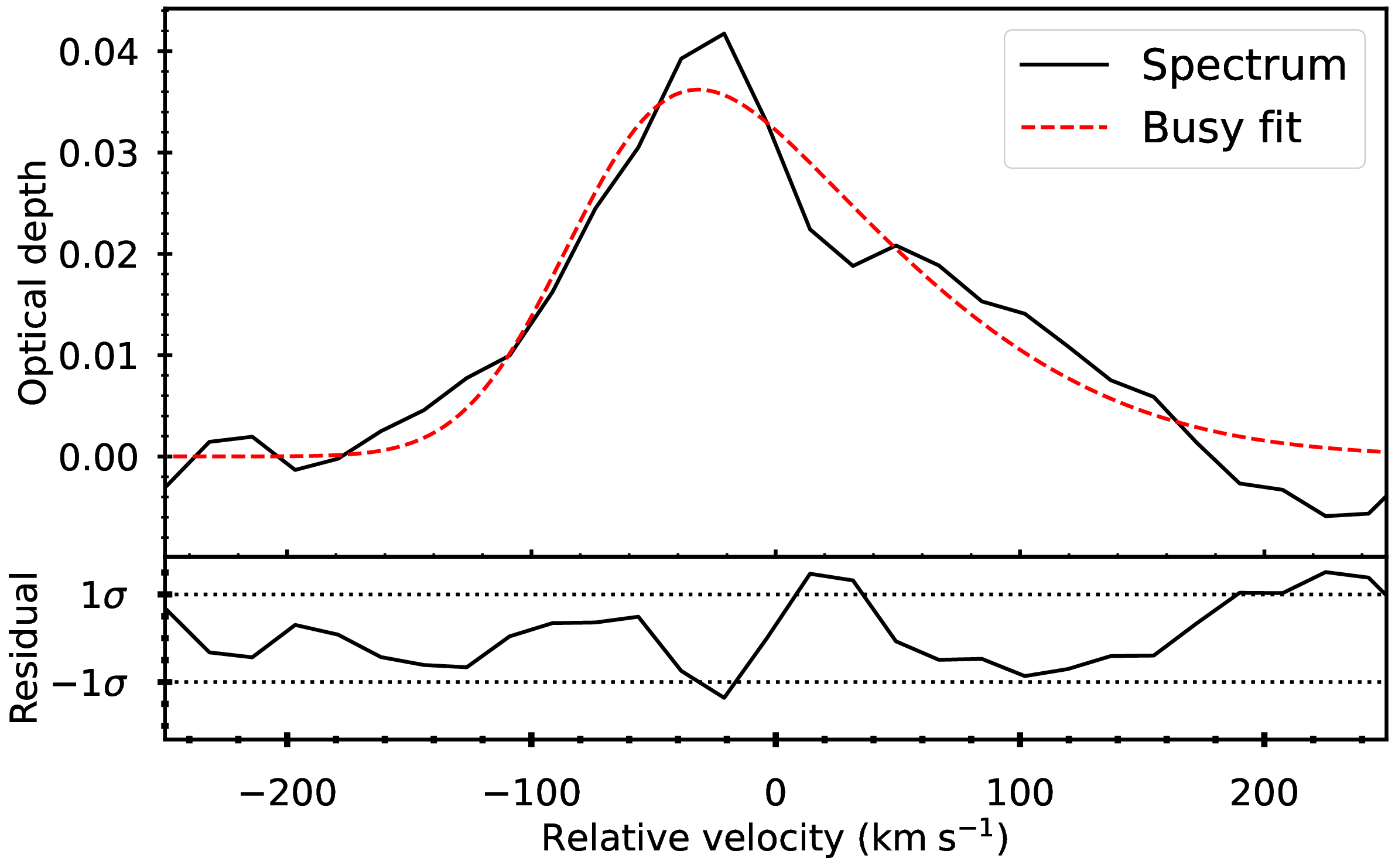}}
\hfill
\subfloat[Multi-Gaussian function fit\label{fig:1b}]{\includegraphics[height=0.3\textwidth,width=0.48\textwidth]{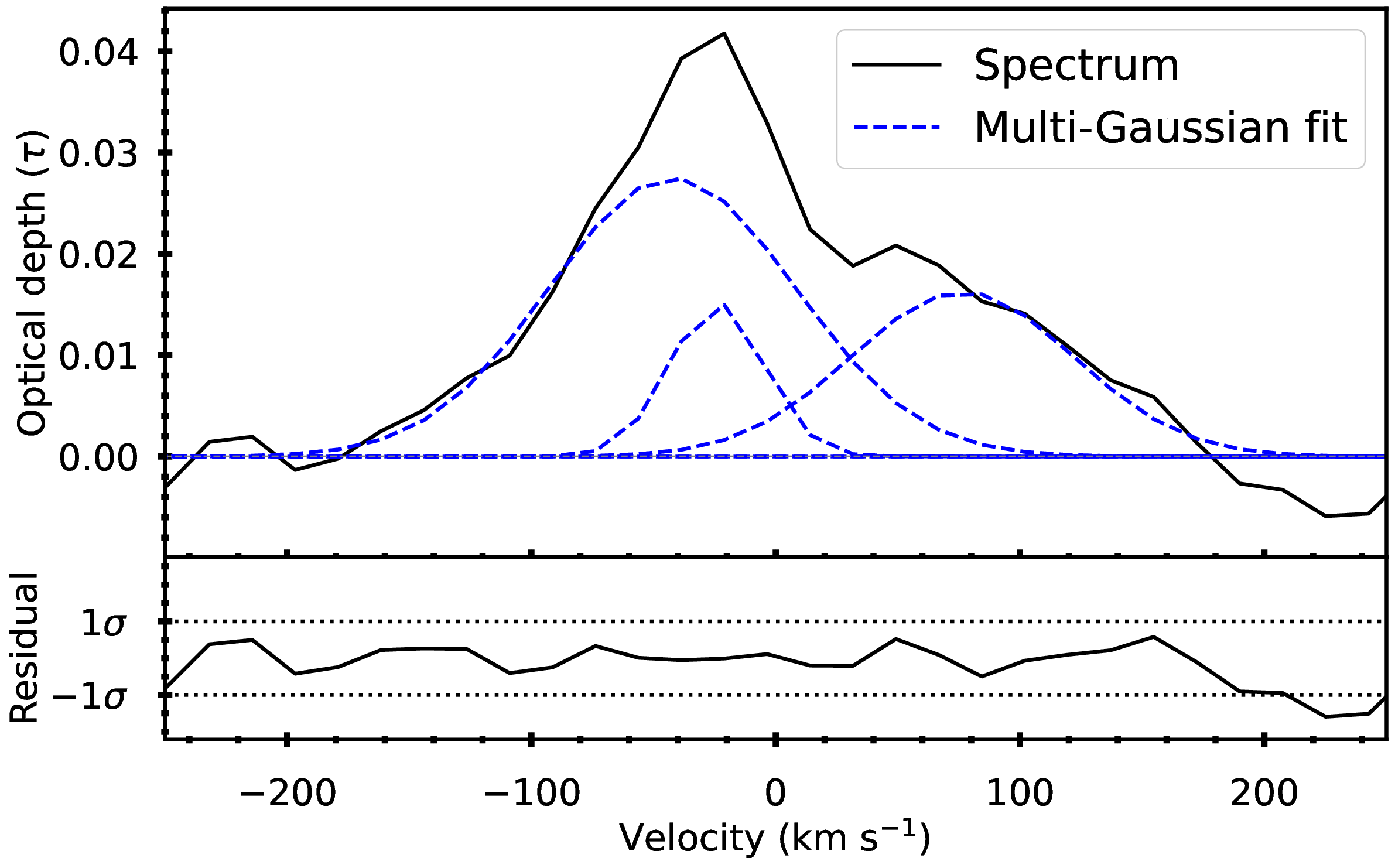}}
\caption{Busy function vs. multi-Gaussian function fit to a galaxy spectrum in our sample -- SDSS J075756.71+395936.1. The data shown in red and blue dotted lines are for the Busy function fit and the multi-Gaussian function fit, respectively. The residuals of each fit are plotted at the bottom of the plot along with the $\pm1\sigma$ lines (black dotted lines).}
\label{fig:1}
\end{figure*}

Compared to Gaussian, which can only model symmetric spectral profiles, one can use the Busy function to model spectral profiles of different shapes. The Busy function can model the steep flanks often seen in the double-horned \hi\ spectra while also recovering the characteristic trough and sharp, narrow peaks of the spectrum with the help of its two constituent error functions and one polynomial function (see Eq. \ref{eq:1}). Each parameter of the Busy function has a unique physical meaning regarding the shape of all types of spectral profiles. By carefully choosing appropriate values of spectral parameters $b_1$, $b_2$, $w$, $c$ and $n$, nearly any shape of double-horned galaxy spectral profile can be reproduced using the Busy function. Here, the two error functions representing the flanks of the spectrum are mainly regulated by the parameters $b_1$, $b_2$ and $w$, which are used to model a steep rise in the spectrum. The polynomial part of the Busy function is mainly regulated by the parameters $c$ and $n$, which are used to model the central trough of the spectrum \citep[for reference see Figs. 2 and 3 of][]{Westmeier2014}. Thus, the Busy function provides an efficient and uniform way of modelling \hi\ 21-cm double-horned absorption lines with a wide range in shape and width parameters, including asymmetric line shapes in a more accurate manner. Fig. \ref{fig:1} provides the Busy function and the Gaussian function fits to a spectrum from our sample for illustration.

We successfully fitted each of the 118 \hi\ 21-cm absorption spectra in our sample using the generalised Busy function (Eq. \ref{eq:1}) via the BusyFit software\footnote{https://gitlab.com/SoFiA-Admin/BusyFit}\citep{Westmeier2014}, and obtained the 13 spectral parameters mentioned above for each absorber. Also, for each, the signal-to-noise (SNR) ratio is calculated using the \texttt{snr\_derived} module of the \texttt{specutils Python} library. The best-fit parameters and the SNR of each absorber are listed in Table \ref{tab:b1}.

\begin{figure*}
\centering
\subfloat[\label{fig:2a}]{\includegraphics[width=0.24\textwidth]{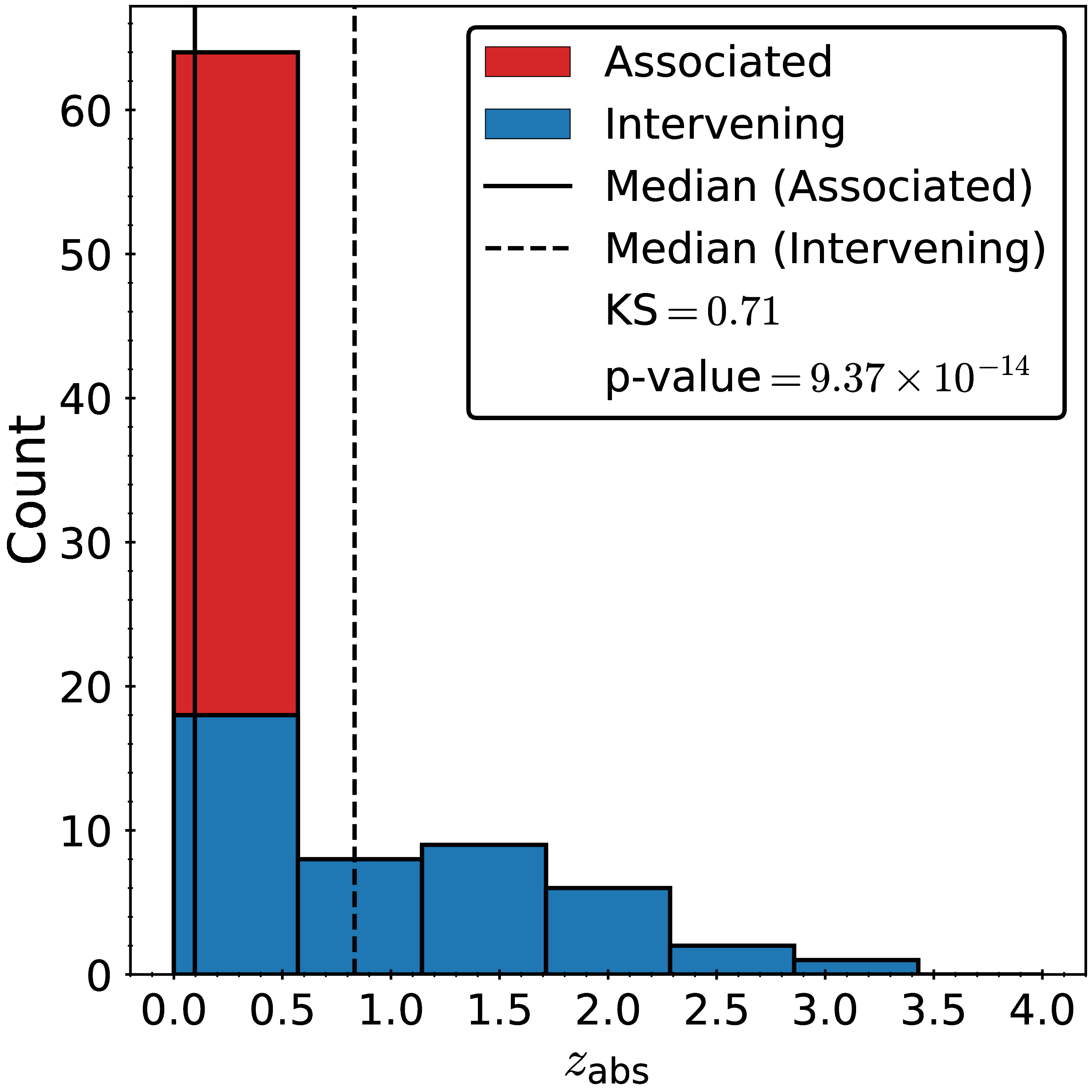}}
\hfill
\subfloat[\label{fig:2b}]{\includegraphics[width=0.24\textwidth]{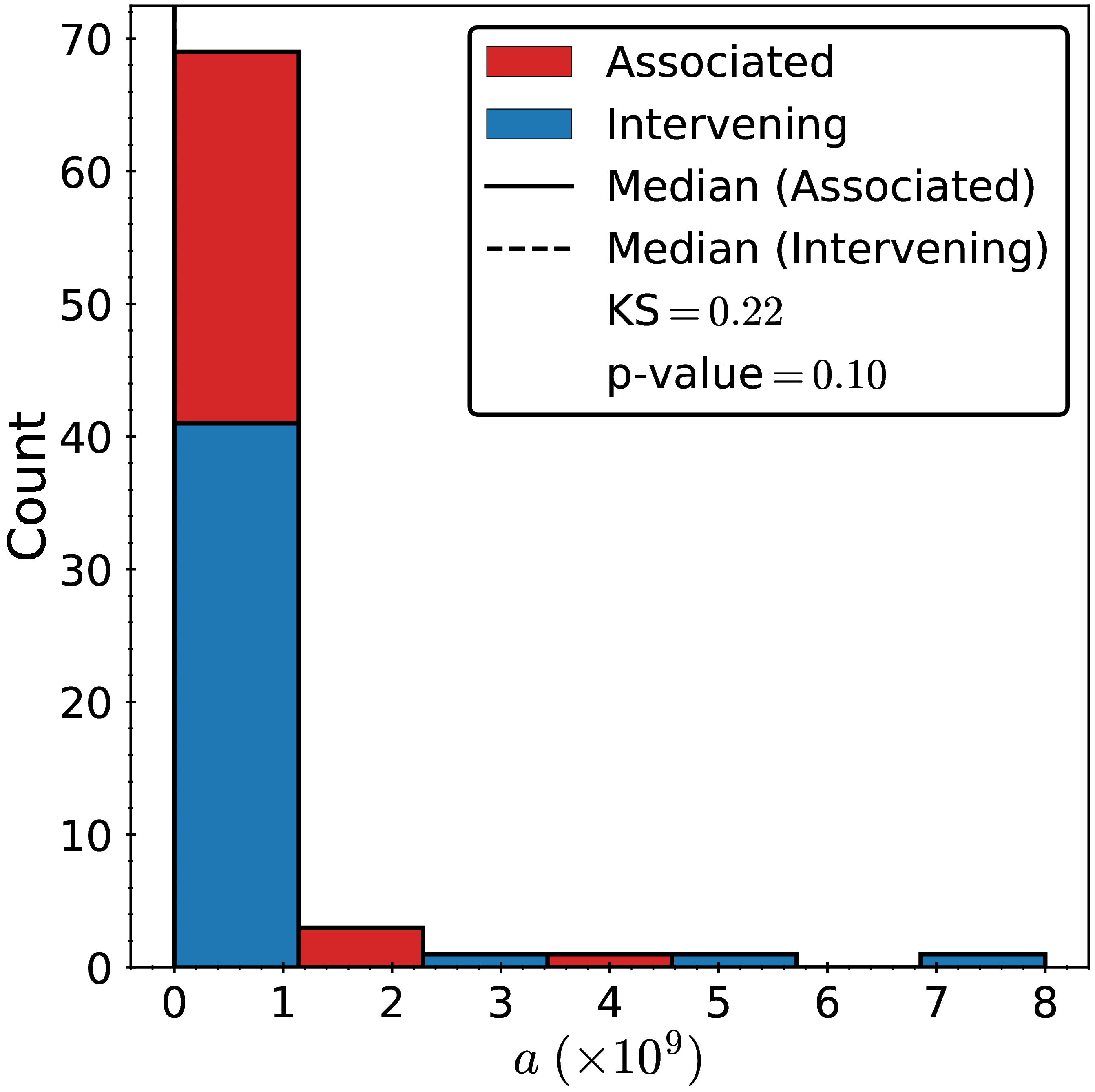}}
\hfill
\subfloat[\label{fig:2c}]{\includegraphics[width=0.24\textwidth]{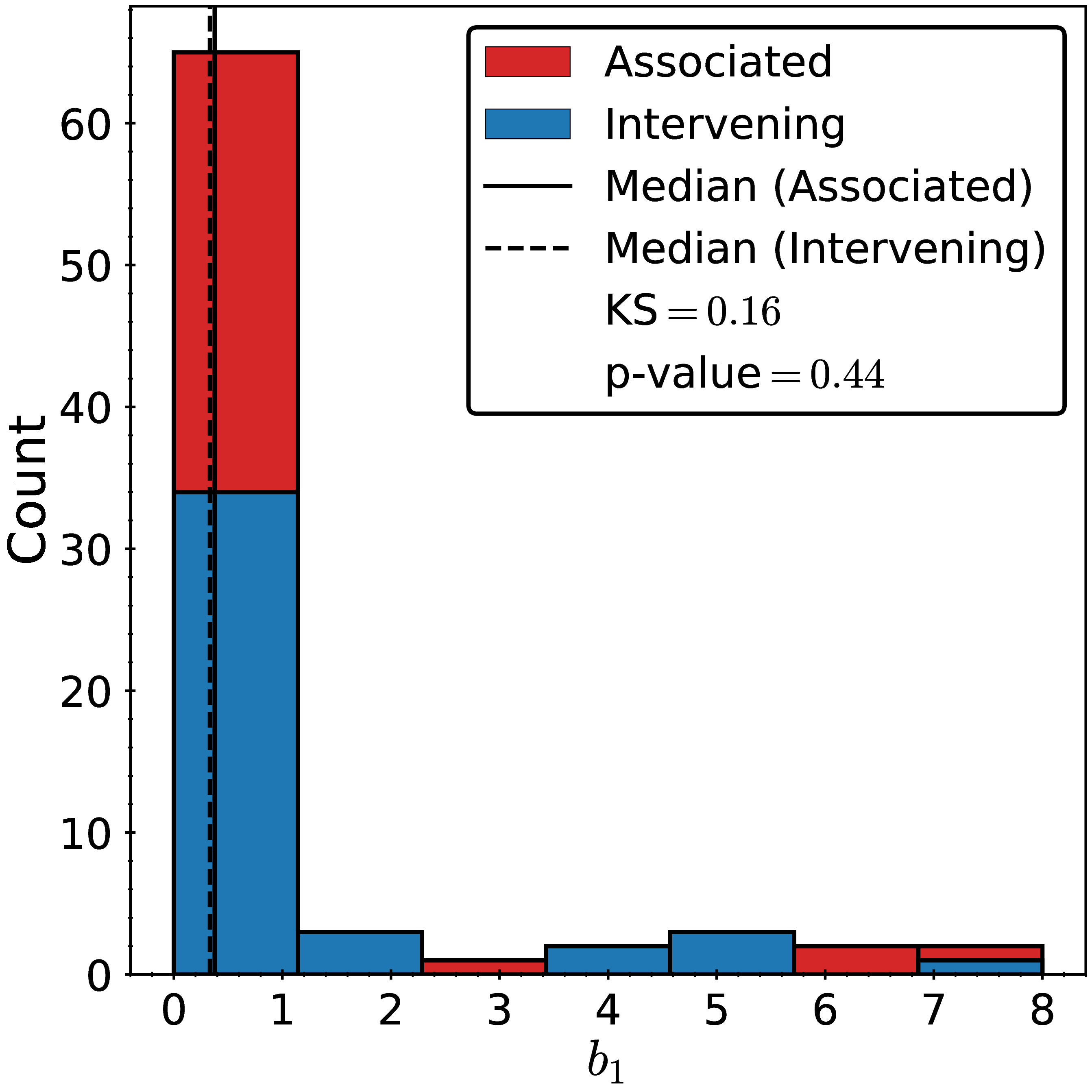}}
\hfill
\subfloat[\label{fig:2d}]{\includegraphics[width=0.24\textwidth]{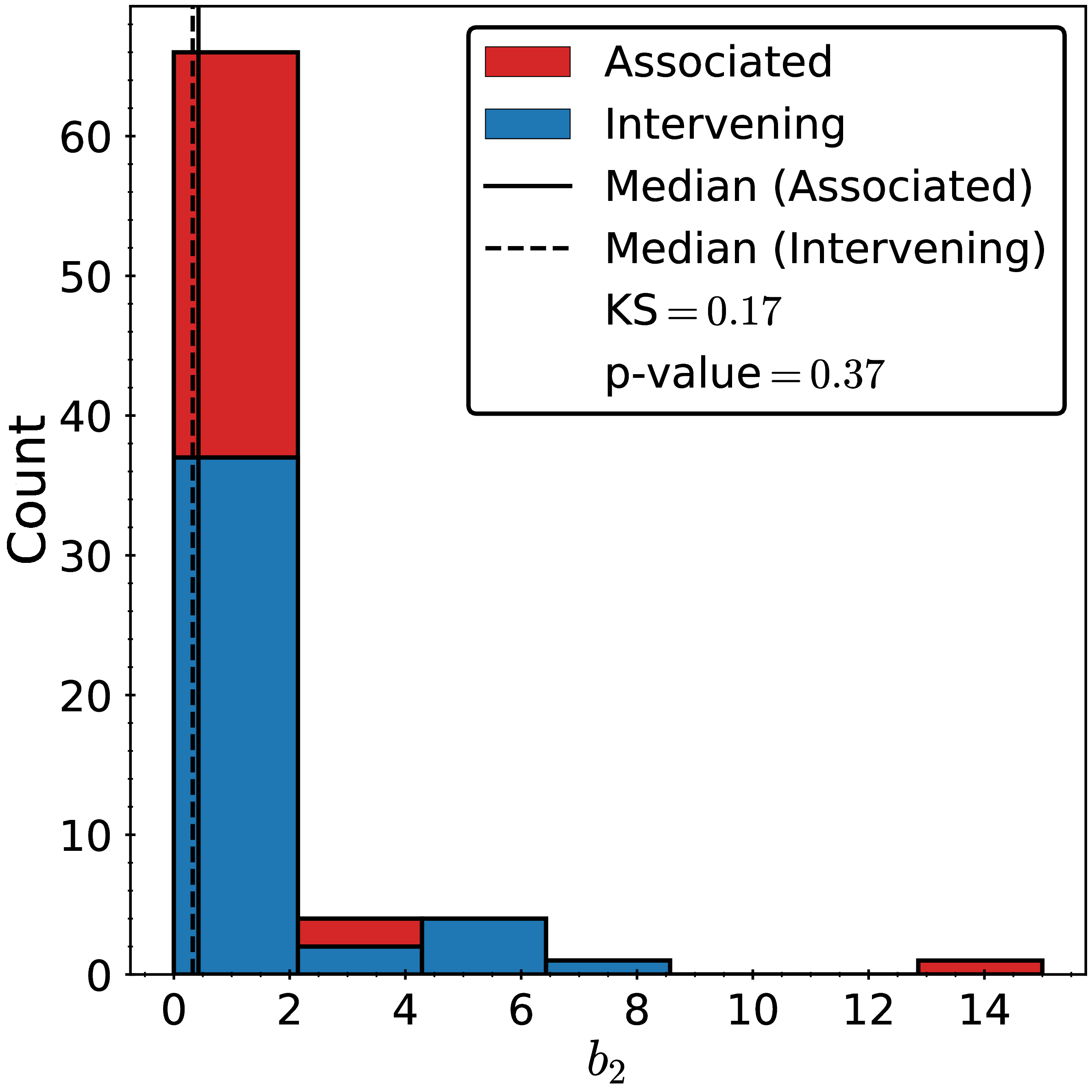}}
\hfill
\subfloat[\label{fig:2e}]{\includegraphics[width=0.24\textwidth]{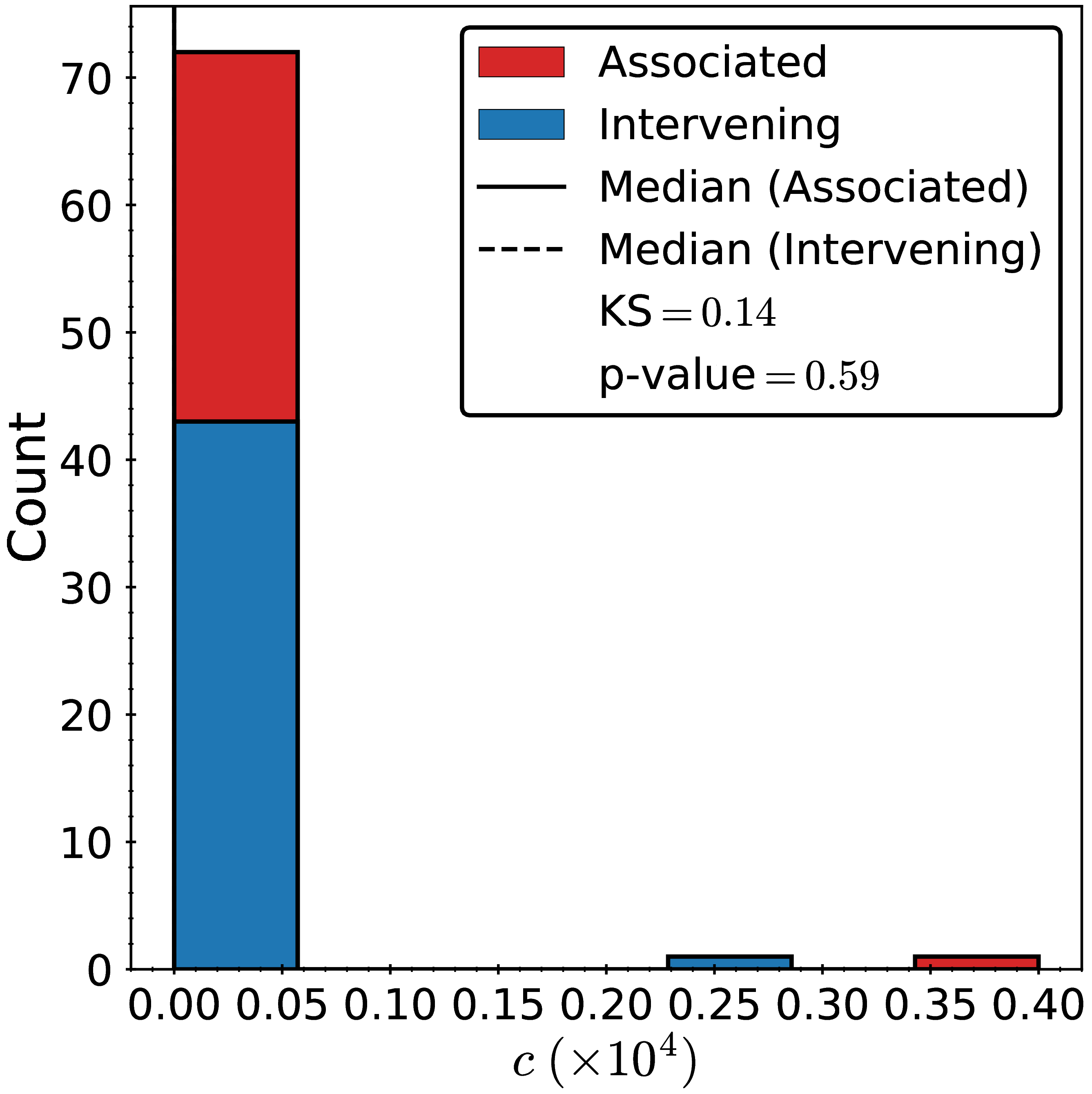}}
\hfill
\subfloat[\label{fig:2f}]{\includegraphics[width=0.24\textwidth]{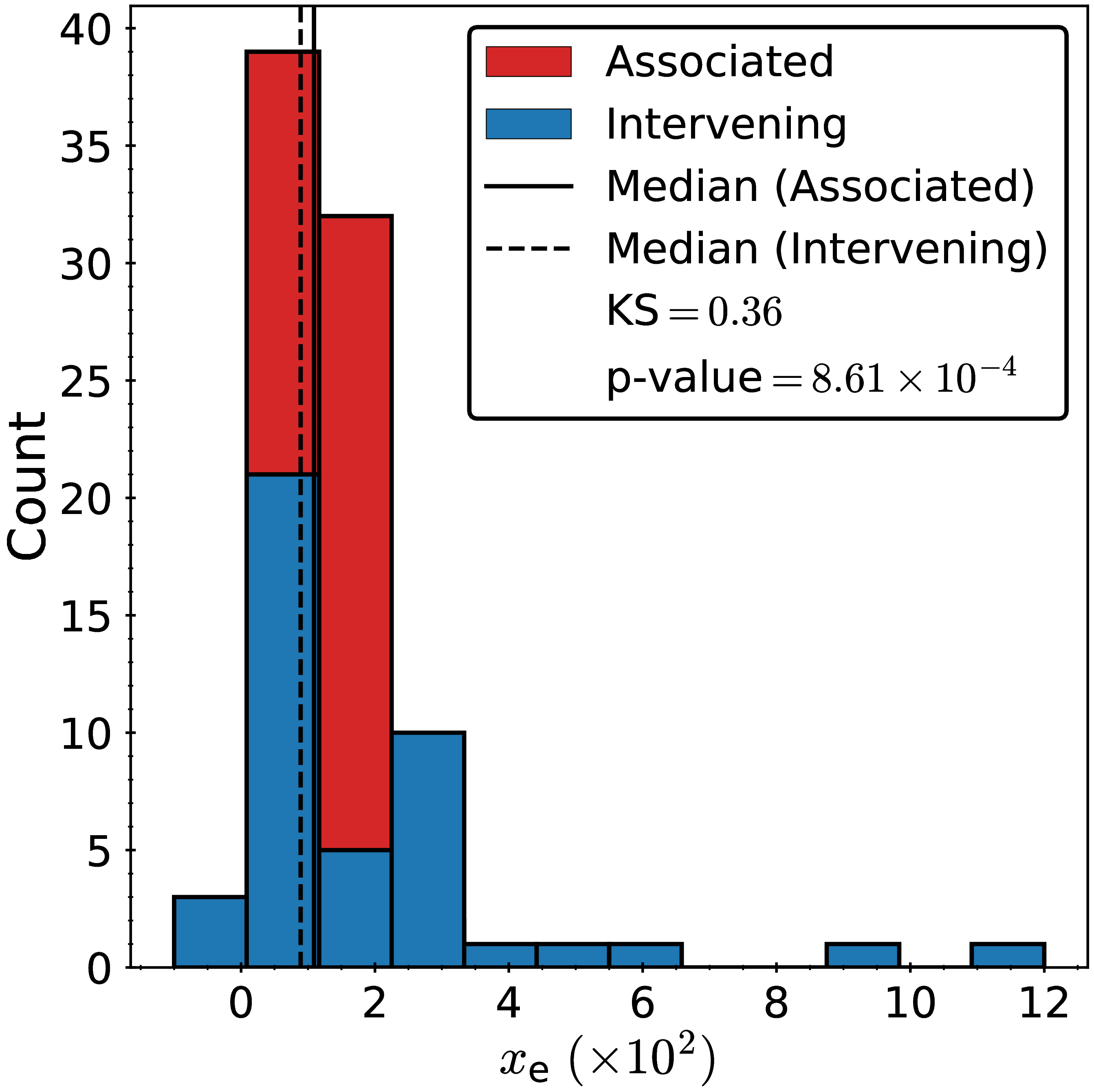}}
\hfill
\subfloat[\label{fig:2g}]{\includegraphics[width=0.24\textwidth]{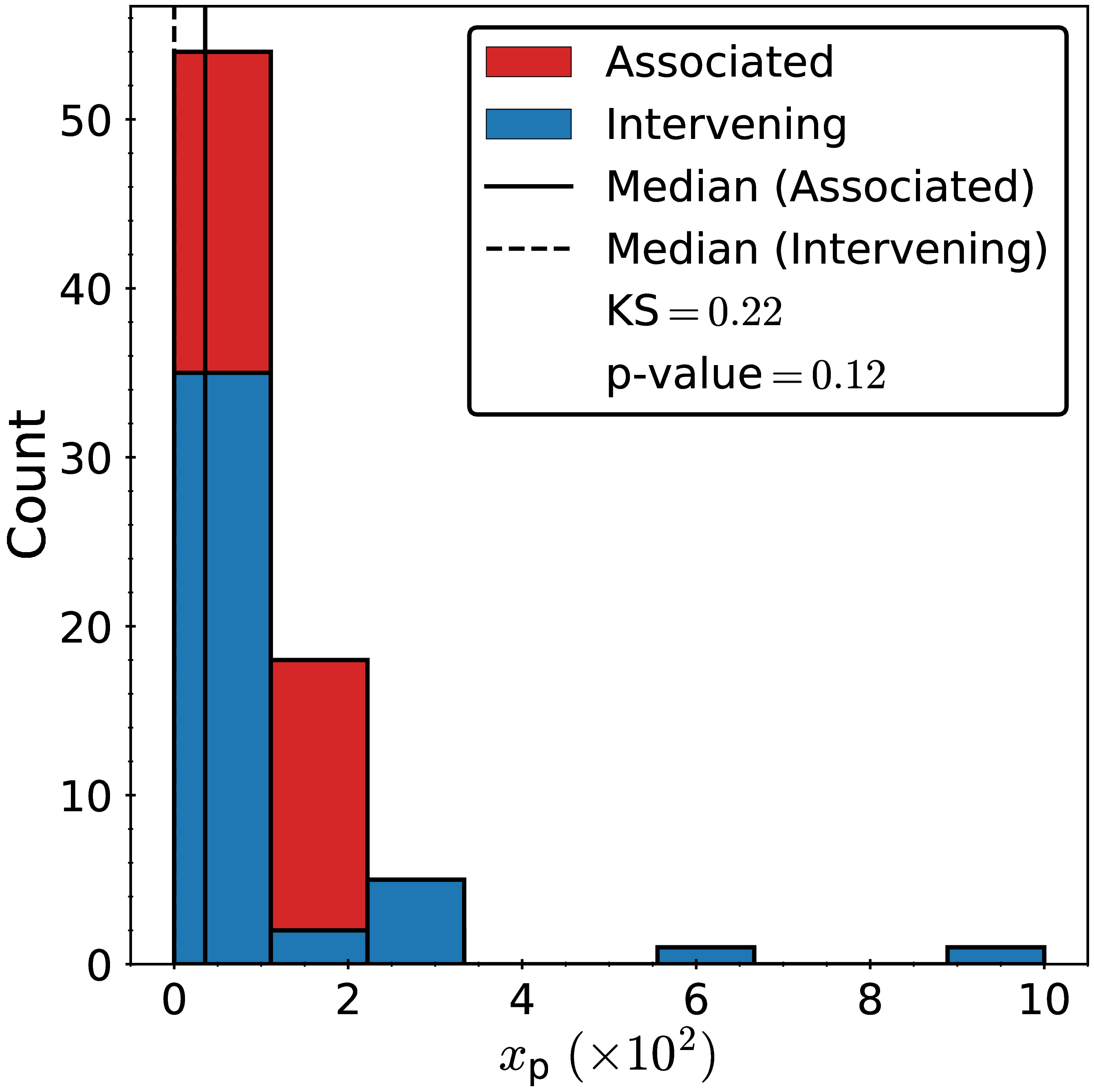}}
\hfill
\subfloat[\label{fig:2h}]{\includegraphics[width=0.24\textwidth]{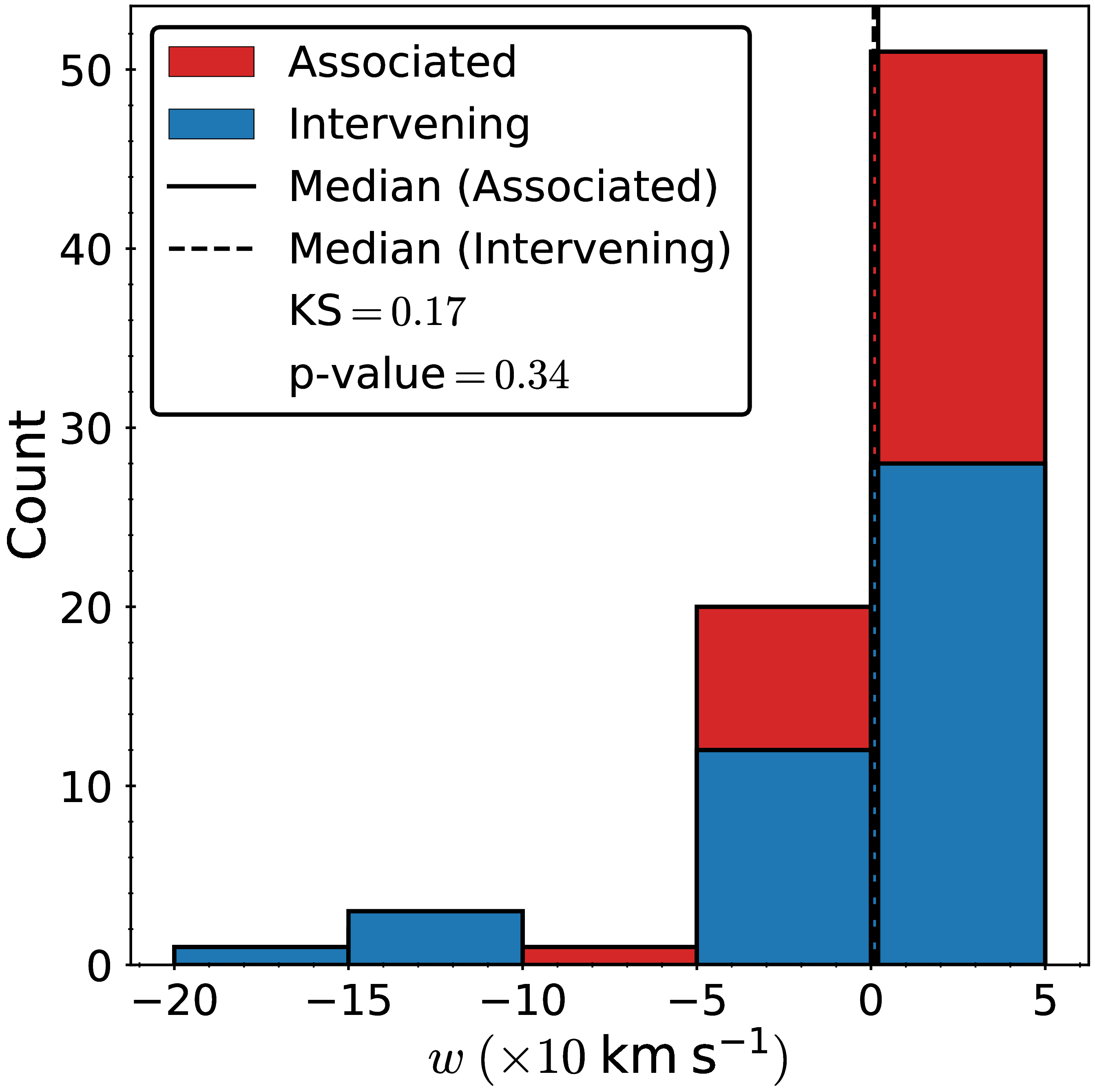}}
\hfill
\subfloat[\label{fig:2i}]{\includegraphics[width=0.24\textwidth]{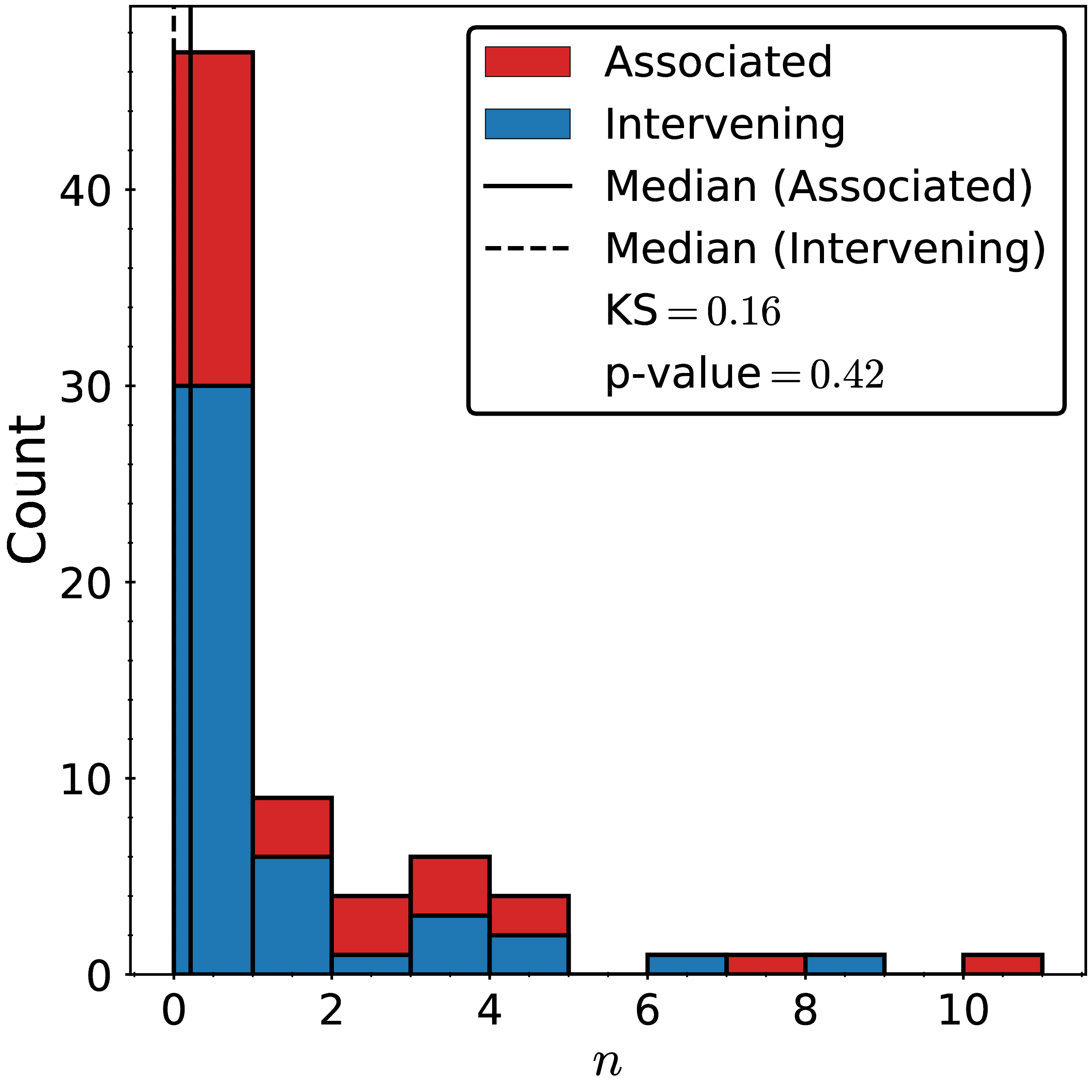}}
\hfill
\subfloat[\label{fig:2j}]{\includegraphics[width=0.24\textwidth]{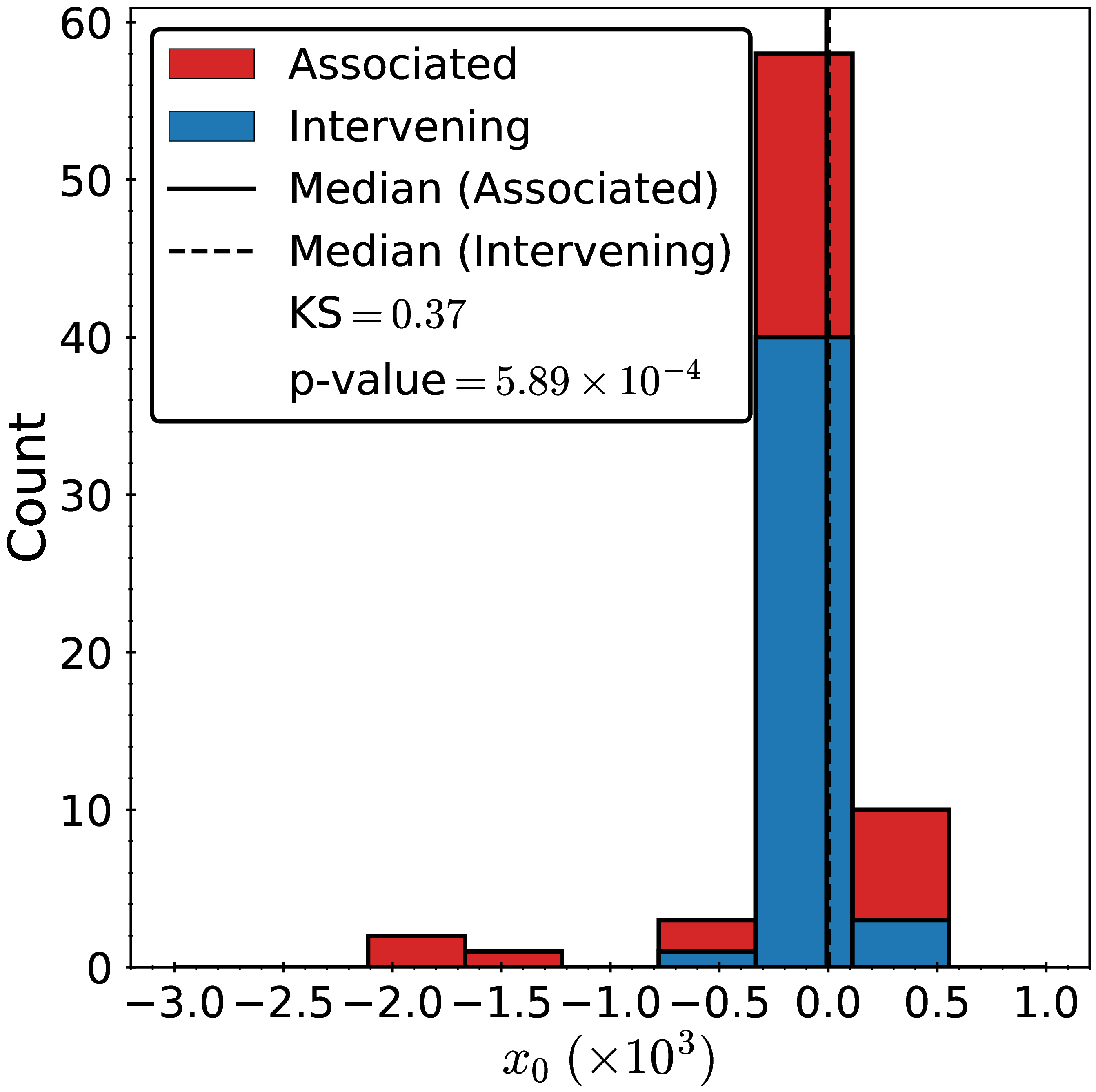}}
\hfill
\subfloat[\label{fig:2k}]{\includegraphics[width=0.24\textwidth]{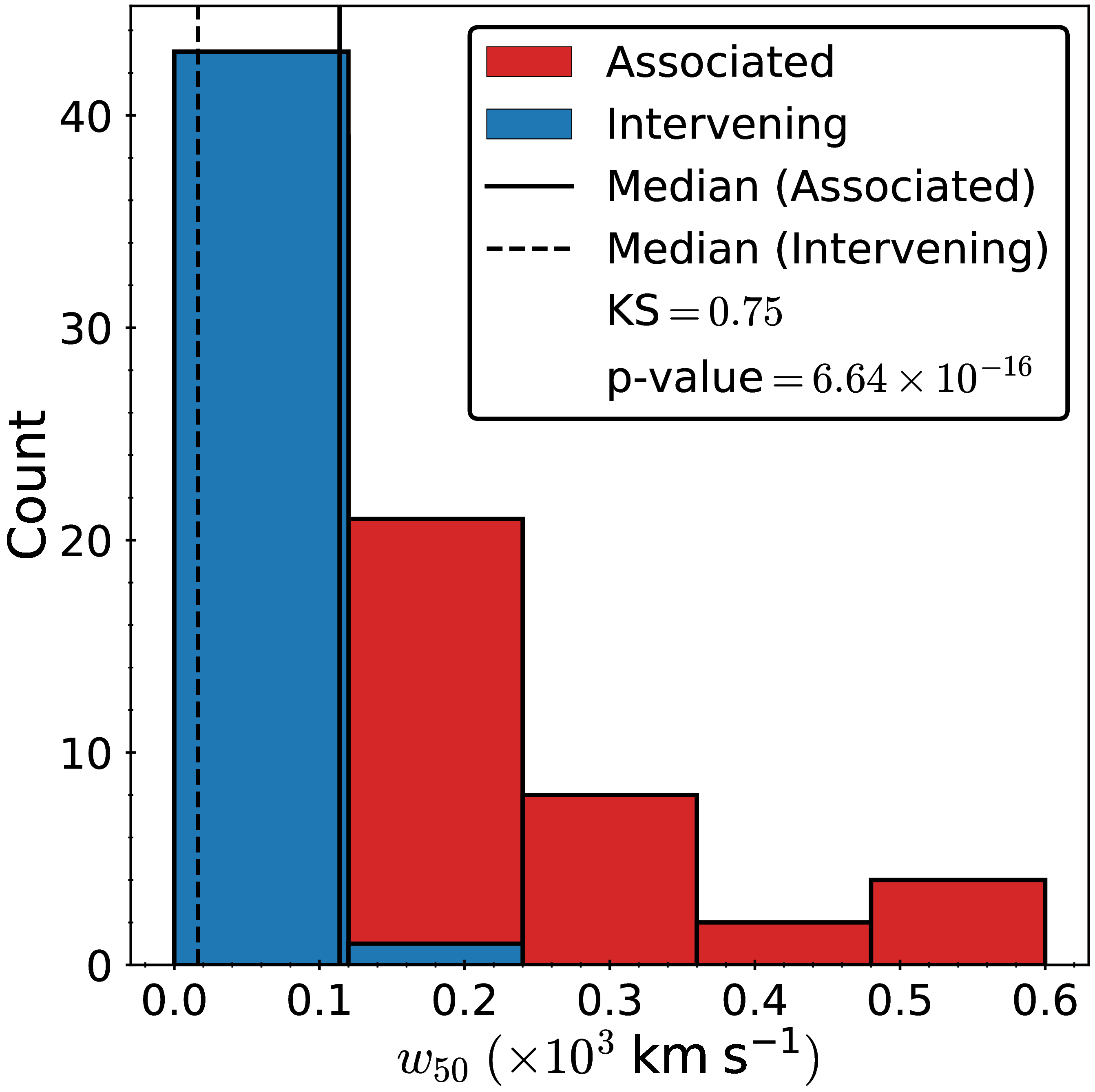}}
\hfill
\subfloat[\label{fig:2l}]{\includegraphics[width=0.24\textwidth]{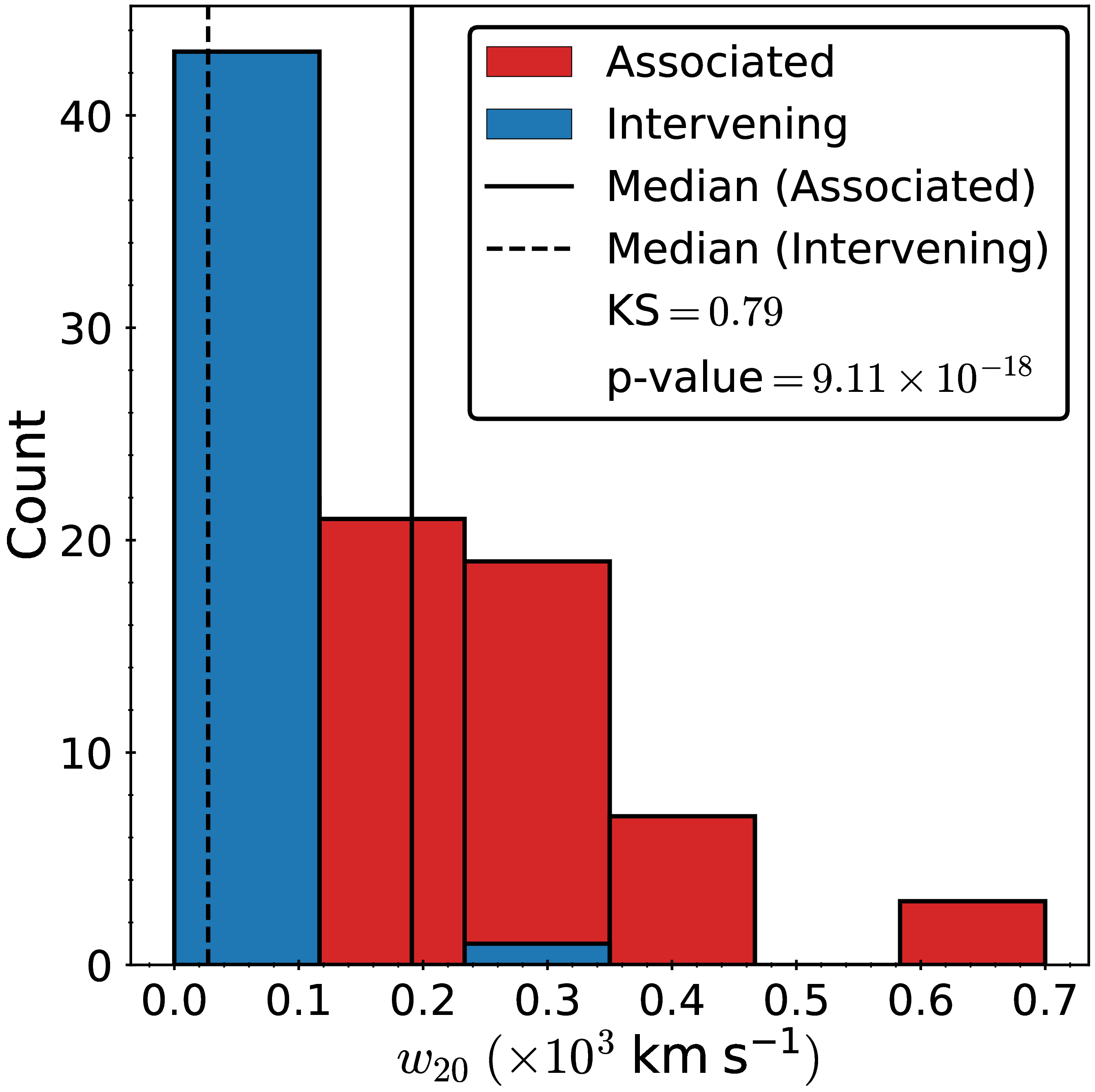}}
\hfill
\subfloat[\label{fig:2m}]{\includegraphics[width=0.24\textwidth]{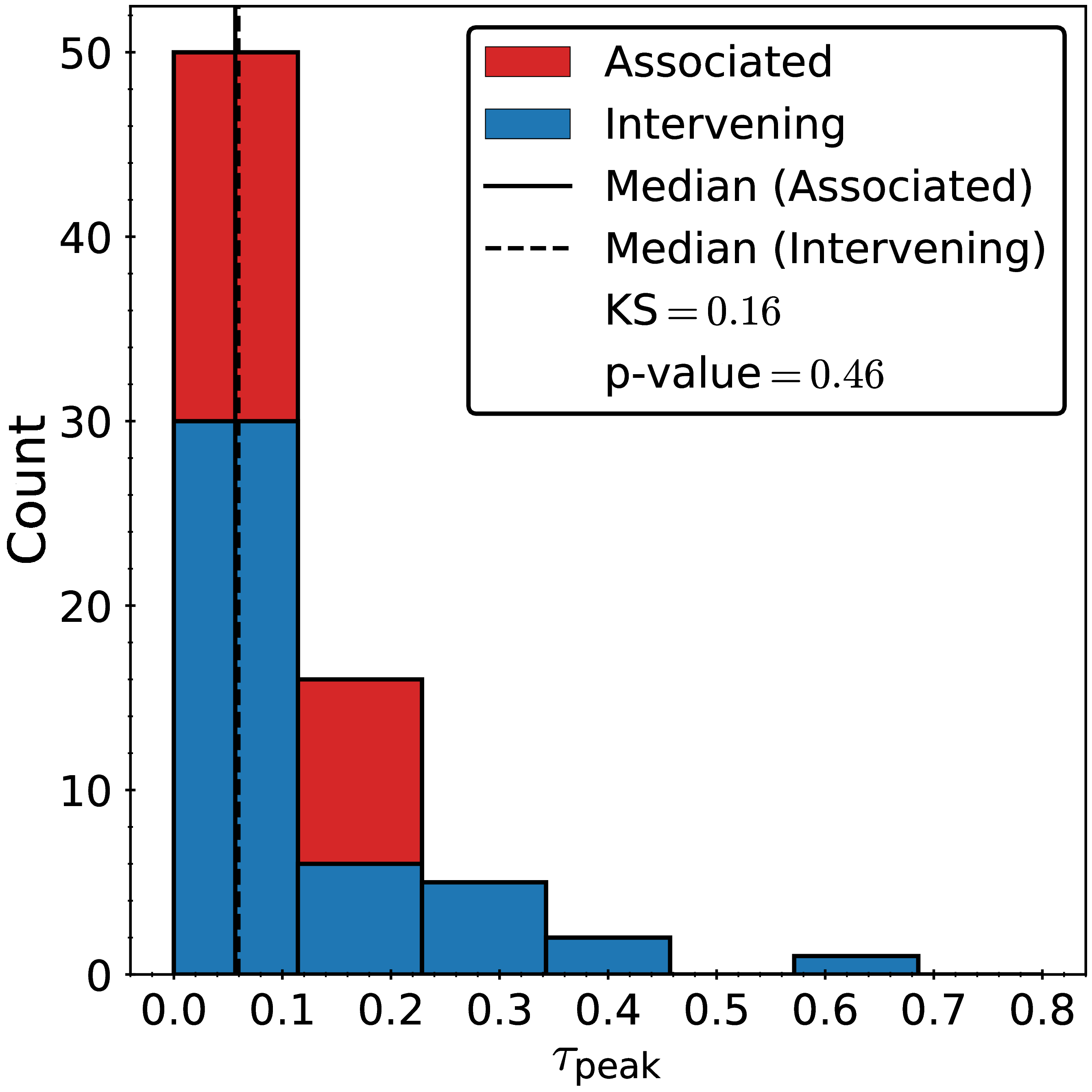}}
\subfloat[\label{fig:2n}]{\includegraphics[width=0.24\textwidth]{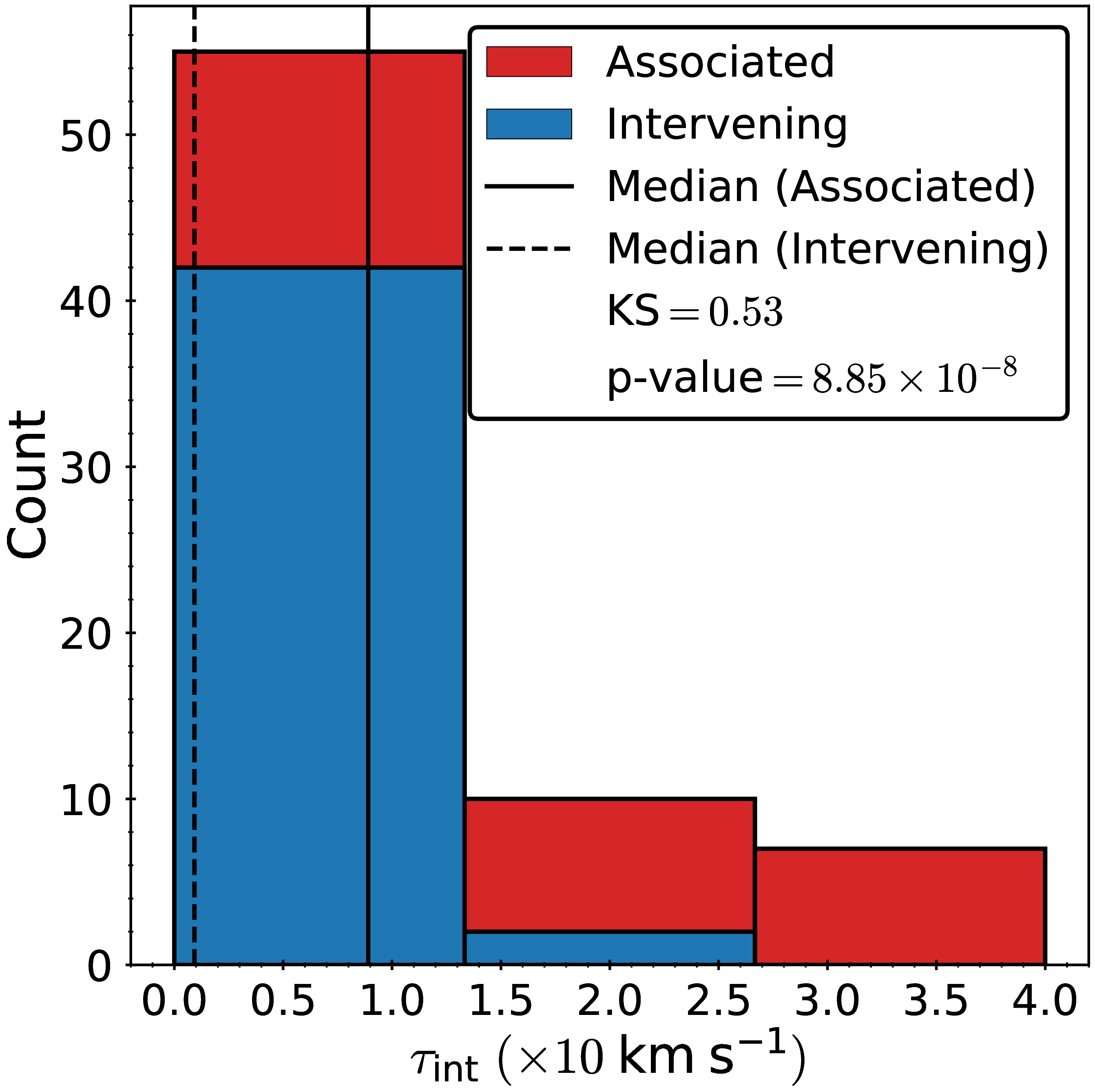}}
\subfloat[\label{fig:2o}]{\includegraphics[width=0.24\textwidth]{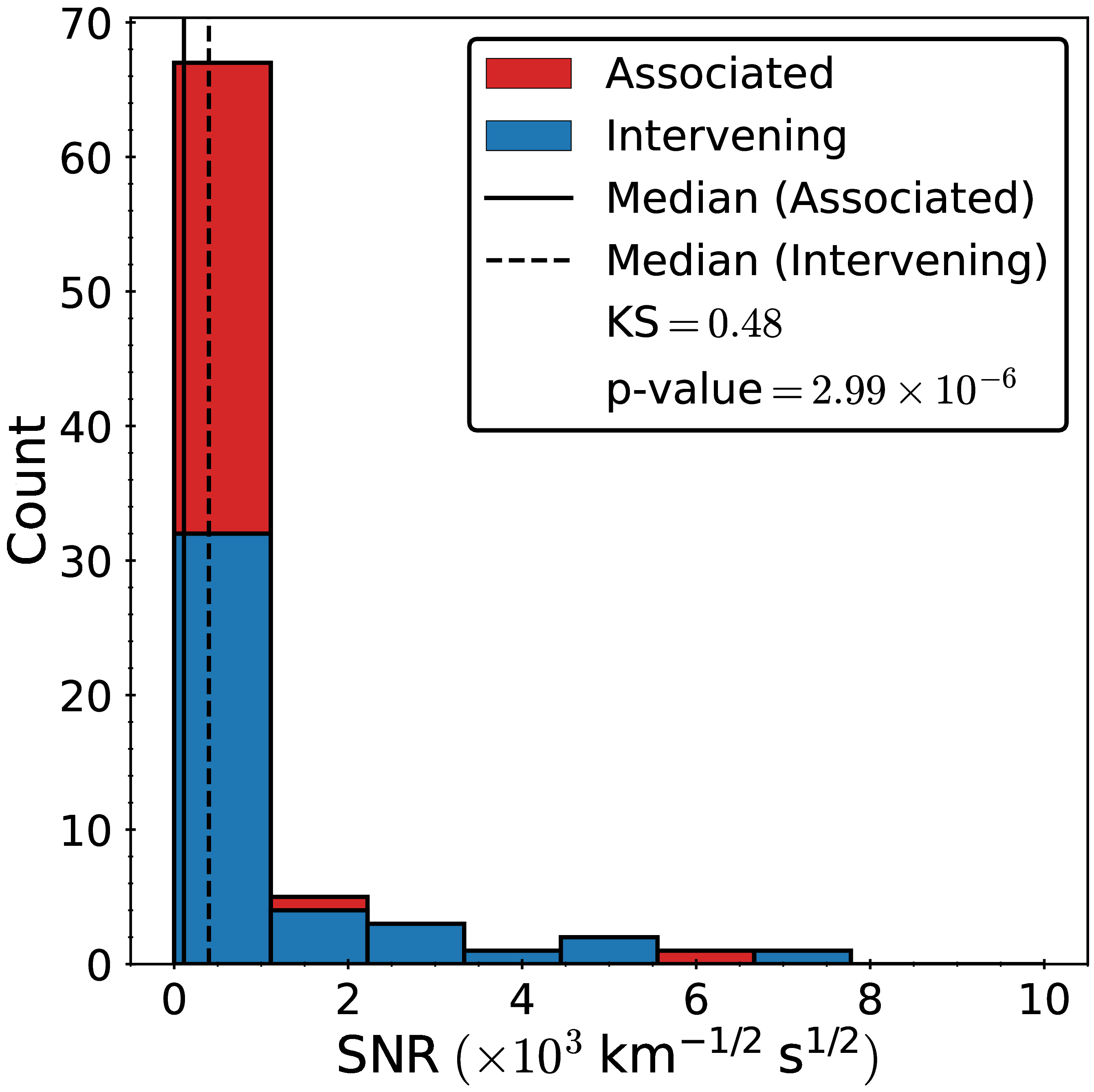}}
\caption{The histograms of the 13 spectral parameters extracted from Busy function fitting, the absorber redshift ($z_\text{abs}$) and the SNR for all the associated (red) and intervening (blue) absorbers of our data sample. The solid and dashed lines denote the median values for the associated and intervening absorbers, respectively. The two-sample KS statistic and corresponding $\mathrm{p-value}$ for each parameter are mentioned in the legend.}
\label{fig:2}
\end{figure*}

As per \citet{Curran2021}, we show the histograms and the Kolmogorov–Smirnov (KS) of the underlying distributions of spectral parameters. The histograms of all the 13 best-fit spectral parameters, including the absorber redshift ($z_\text{abs}$) and the SNR for the associated and intervening absorber samples, are shown in Fig. \ref{fig:2}. We performed the two-sample KS test on each parameter to compare the underlying distributions of the parameters between the associated and intervening samples. It helps to quantify the discriminating power of each parameter regarding the classification of the absorber type. Also, $\mathrm{p-values}$ corresponding to KS values are calculated to check their significance level. A parameter with a high KS statistic could have a strong influence on the absorber type classification task. All the KS statistic and associated $\mathrm{p-values}$ are also given with respective parameter distributions in Fig. \ref{fig:2}. From these KS statistic values, it is evident that the linewidth parameters ($w_{20}$ and $w_{50}$) and the integrated optical depth parameter ($\tau_{\text{int}}$) could have more substantial influence on the absorber type classification task (see Fig. \ref{fig:2}). The distributions of $w_{20}$, $w_{50}$ and $\tau_{\text{int}}$ show the most significant difference between the associated and intervening samples, with medians values of $w_{20}$ = $190.921 \; \mathrm{km \; s^{-1}}$, $27.214 \; \mathrm{km \; s^{-1}}$, $w_{50}$ = $113.85 \;\mathrm{km \; s^{-1}}$, $16.3 \; \mathrm{km \; s^{-1}}$, $\tau_{\text{int}}$ = $8.909 \; \mathrm{km \; s^{-1}}$, $0.926 \; \mathrm{km \; s^{-1}}$ for the associated and intervening absorber samples, respectively. Similarly, \citet{Curran2016} and \citet{Curran2021} found the linewidth to be the dominant factor. Since the peak optical depth shows no correlation, the integrated optical depth differences are only due to the linewidths. We explore the implication of this further in Section \ref{sec:4.3}.

Fig. \ref{fig:2} also shows the distributions of the absorber redshifts, which are quite different between the associated and intervening samples, with median values of $z_\text{abs}$ being 0.097 and 0.833, respectively. This difference arises mainly due to observational limitations \citep{Curran2016}. In Section \ref{sec:4.2}, we investigated whether the difference in redshift distributions between the two samples affects the classification results.

\section{Machine Learning Classification}
\label{sec:3}

The following six supervised ML classification algorithms -- Gaussian naive Bayes, logistic regression, decision tree, random forest, SVM and XGBoost are used for model training on our labelled dataset. 

We have started with the simplest one, i.e., naive Bayes, which assumes that all features are conditionally independent given the class label. Then we checked with the logistic regression, which uses the Sigmoid function to infer the probabilities of class labels. After that, we moved to a more complex type, i.e. tree-based models. We first checked with a simple tree-based model, i.e., a decision tree. Then, we checked for a more complex tree-based model, i.e., a random forest, which uses an ensemble of decision trees to reduce the possible chances of overfitting. Next, we checked with a more advanced type, i.e. support vector machine, which uses nonlinear kernel functions to classify the data. Finally, we checked with the most advanced model, i.e., XGBoost, which utilises sequential weak models to establish a strong classification model with low bias. We have excluded neural networks since they are like black-box models, thus making it hard to interpret the features. Also, we have very limited data to train such networks. A brief overview of each of the models we have used is given in Section \ref{sec:a1}. 

The scheme we followed to execute an binary classification task using any of the above ML algorithms on our dataset (comprising the best Busy function fit spectral parameters, the SNR and the category of each absorber) is as follows:

\begin{enumerate}[label = (\roman*), labelwidth = 0pt]
\item First, we identified the predictor and response variables. In our dataset, the predictor variables are the 13 Busy function fitted spectral parameters (as mentioned in Section \ref{sec:2.2}) plus the SNR, and the response variable is the absorber type.

\item Secondly, we checked whether the predictor variables are strongly correlated with each other or not. We keep only those variables in the dataset with the least correlations with others and discard the rest. This helps to robustly interpret the model's performance and also to reduce the dimensionality of the dataset rendering the model training process faster. We have conducted Pearson and Spearman correlation tests between each pair of the spectral parameters to identify the correlated predictors. Both tests give similar results. In Fig. \ref{fig:3}, we have shown Pearson's correlation values between every pair of spectral parameters (including the absorber redshift and the SNR) in the form of a matrix. From this, it is evident that $w_{50}$ and $w_{20}$, are highly correlated with each other as expected, with the Pearson's correlation coefficient $= 0.95$ and $\mathrm{p-value} \sim 2\times10^{-60}$. Thus, we keep the $w_{20}$ parameter in the dataset and exclude the $w_{50}$ parameter as a redundant variable. Now, the reduced dataset has the remaining 12 spectral parameters and the SNR as predictor variables, and the absorber type as the response variable. The model training process and predictions have done on this dimensionally reduced dataset.

\begin{figure*}
\centering
\includegraphics[width=\textwidth]{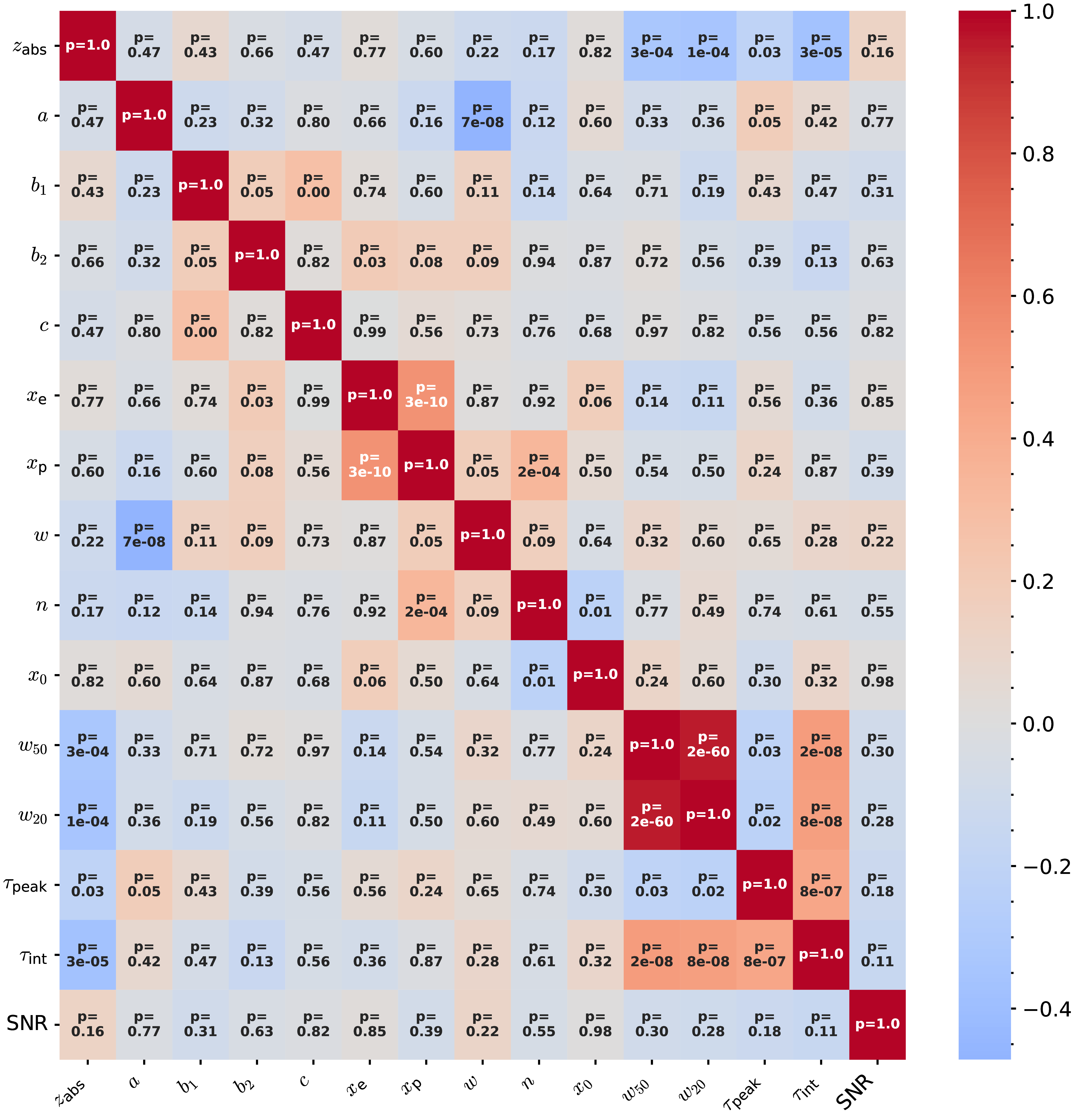}
\caption{Pearson's correlation matrix for the spectral parameters obtained from Busy function fitting, the absorber redshift and the SNR. The annotated numbers are corresponding $\mathrm{p-values}$}.
\label{fig:3}
\end{figure*}

\item After that, we divided the dataset into training and testing parts. Our sample size is relatively small for any ML model training. Our data sample consists of 74 associated and 44 intervening absorbers (labelled `0' and `1' respectively in the column `Class' of Table \ref{tab:b1}). This imbalance in the number of absorber types may introduce bias in the classification results towards any specific absorber type. Thus, we selected 44 associated absorbers randomly from 74 available and added them with 44 intervening absorbers to make a dataset of 88 absorbers to ensure an equal representation of both absorber types. Then, we divided the spectral parameter dataset of this reduced sample into $80:20$ ratio. During this train-test division, we have used the random stratified sampling technique to keep the absorber type ratio fixed across the train and test samples. The overall approach helps to minimize bias in the training data towards any specific absorber type, leading to a more accurate and generalizable model. We implemented this using the \texttt{train\_test\_split} module of the \texttt{scikit-learn Python} library.

\item Then, we standardise the training and test sets individually using the standard scaling (Z-score normalization) technique to ensure smooth and quick convergence of the gradient descent algorithm while minimising the loss function of a classification task \citep{Ruder2016}. We implemented this using the \texttt{StandardScaler} module of the \texttt{scikit-learn Python} library.

\item Next, we tuned the model hyperparameters. To infer an unbiased estimate for our small sample of 118 \hi\ 21-cm absorption lines, model hyperparameter tuning has been done by using the leave-one-out technique \citep{Geroldinger2023} on top of the grid search cross-validation (CV) technique \citep{Liashchynskyi2019}. This process helps to find the optimal model hyperparameters from a given grid of hyperparameter values by using the whole dataset for training, excluding only one data point, and then iterating for every data point. We implemented these using the \texttt{GridSearchCV} and \texttt{LeaveOneOut} modules of the \texttt{scikit-learn Python} library. Each model's hyperparameter grid and associated optimal values are given in Table \ref{tab:1}. 

\item Thereafter, we trained the model using the optimal hyperparameter values and used the result to make predictions on the test data.

\begin{table*}
\centering
\scriptsize
\caption{For the all spectral parameter sample, the description of each of the six ML classification models' hyperparameter grid. Each grid comprises sets of trial hyperparameter values. Details about each model's hyperparameters are available in their respective \texttt{Python} modules (see appendix \ref{sec:a1}).}
\label{tab:1}
\begin{tabular}{c c c c c}
\hline 
ML classification model & Trial hyperparameter values & Total number of fits & Most preferred optimal hyperparameter & Total run time\\ 
 &  & in each run & value over 1000 runs \\
\hline
\hline
Gaussian naive Bayes &                                    not applicable &     -- & -- & 4 sec 962 ms \\
\hline
 Logistic regression & $C: [0.0001, 0.01, 0.1, 1, 10, 100, 1000, 10000]$ &   2256 &  1 & 53 min 17 sec 787 ms \\
                     &          penalty$: [l_1, l_2, \text{elasticnet}]$ &        & $l_1$ & \\
\hline
                     &        max\_depth$: [2, 3, 5, 7, 10, 15, 17, 20]$ &        &  2 & 1 hr 3 min 17 sec 367 ms \\
       Decision tree &    min\_samples\_leaf$: [5, 10, 20, 50, 75, 100]$ &   9024 & 10 & \\
                     &        criterion$: [\text{gini}, \text{entropy}]$ &        & gini & \\
\hline
                     &        max\_depth$: [2, 3, 5, 7, 10, 15, 17, 20]$ &        & 2 \\
       Random forest & min\_samples\_leaf$: [5, 10, 20, 50, 75, 100]$ &  9024 & 5 & 19 hr 41 min 8 sec 118 ms \\
                     &        criterion$: [\text{gini}, \text{entropy}]$ &        & gini \\
\hline
                     & $C: [0.0001, 0.01, 0.1, 1, 10, 100, 1000, 10000]$ &        & 1 & \\
                 SVM &       gamma$: [0.0001, 0.001, 0.01, 0.1, 1]$ & 15040 &   1 & 24 hr 34 min 45 sec 701 ms \\
                     & kernel$: [\text{rbf}, \text{poly}, \text{sigmoid}, \text{linear}]$ &  & sigmoid & \\
\hline
             XGBoost &    max\_depth$: [2, 3, 5, 7, 10, 15, 17, 20]$ &  752 &  2 &  24 min 45 sec 94 ms \\
\hline
\end{tabular}
\end{table*}

\item Finally, we evaluated the model classification performance for both the training and the testing data using the metrics -- accuracy, $F1$-score and AUC score \citep{Gonccalves2014,Wardhani2019}. These can be calculated from a confusion matrix, which summarizes all the predicted and actual labels. Brief overviews of these three performance metrics are given below.

\begin{itemize}[leftmargin=*]
\item Accuracy: It is defined as,
$$\text{accuracy} = \frac{\text{number of true positives} + \text{true negatives}}{\text{total number of predictions}},$$
where positive and negative signify intervening (i.e. `1') and associated (i.e. `0') absorber categories respectively. A high accuracy means that ML model predictions are much closer to the actual values.\\

\item $F_1$-score: It is defined as,
$$F_1\text{-score} = 2\times \frac{\text{precision} \times \text{recall}}{\text{precision} + \text{recall}},$$
$$\text{where,} \; \text{precision} = \frac{\text{number of true positives}}{\text{number of true and false positives}},$$ 
$$\text{and} \; \text{recall} = \frac{\text{number of true positives}}{\text{number of true positives and false negatives}}.$$ 
A high $F_1$-score means that the ML model is good at identifying both true positive and true negative predictions from all predictions. \\

\item AUC score: It is defined as the area under the ROC (receiver operating characteristic) curve and its $x$-axis (i.e., recall). A ROC curve is a trade-off between the false positive rate and true positive rate (also known as recall). A higher AUC score signifies a better ML classification model. 
\end{itemize}
We have evaluated all these performance indicators using the \texttt{metrics} module of the \texttt{scikit-learn Python} library.

\item To get an unbiased estimate of these metrics, we repeat all the model training processes (including hyperparameter tuning) 1000 times. In each trial run, we randomly divided the data into $80:20$ training and test ratio. For each ML classification model, the average values of all these metrics are provided in Table \ref{tab:2}. Additionally, for the best-performing ML model, we provided the distribution of test accuracy scores, as shown in Fig. \ref{fig:4}. Moreover, we have shown the corresponding mean confusion matrix and mean ROC curve for the test data in Figs. \ref{fig:5a} and \ref{fig:5b} respectively. 

\item Finally, we assigned a feature importance score to each of the predictor variables based on their contribution to the model's predictive power. For the best-performing ML model, we have shown the mean feature importance plot (e.g. Fig. \ref{fig:5c}). We implemented this using the \texttt{permutation\_importance} module of the \texttt{scikit-learn Python} library.
\end{enumerate}
 
The best ML model should have the highest test accuracy, test $F_1$-score and test AUC score than others. Additionally, the difference between the train and test accuracy values for each ML model should not be high; otherwise, the model will be prone to severe overfitting. We trained each ML model on a 16-core \texttt{Intel 13th Gen i7} processor with 32 GB RAM and 2 TB HDD. The runtime of each model is provided in Table \ref{tab:1}.

\section{Results and Discussion}
\label{sec:4}

A set of 13 Busy function fitted spectral parameters ($a$, $b_1$, $b_2$, $c$, $x_{\text{e}}$, $x_{\text{p}}$, $w$, $n$, $x_0$, $w_{50}$, $w_{20}$, $\tau_{\text{peak}}$ and $\tau_{\text{int}}$) are extracted for each of the 118 \hi\ 21-cm absorption spectra of our data sample as described in Section \ref{sec:2}. After that, $w_{50}$ is dropped from the set of the spectral parameters as it is strongly correlated with $w_{20}$. We have used the remaining 12 spectral parameters, the SNR and the absorber type to train all six ML models. The results are discussed below for the following three cases.

\subsection{Results for all spectral parameters}
\label{sec:4.1}

In the first case, we trained all six ML models using all 12 spectral parameters (excluding $w_{50}$) and the SNR as predictor variables (the concerned dataset is termed as the all spectral parameter sample) over 1000 runs. The predictive performance of each model is tabulated in Table \ref{tab:2} and the key results are discussed below.

\begin{figure*}
\centering
\subfloat[All spectral parameter sample\label{fig:4a}]{\includegraphics[width=0.49\textwidth]{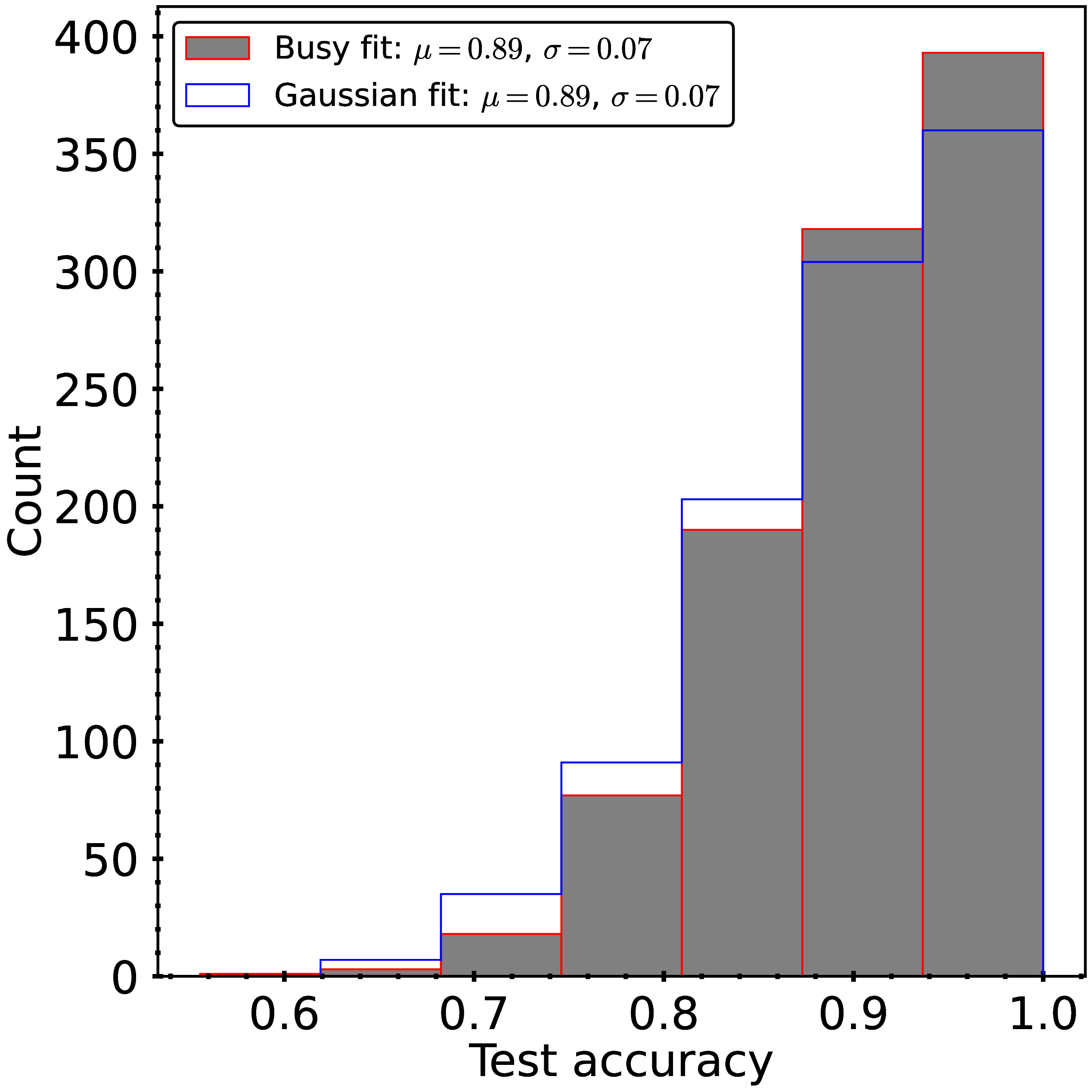}}
\hfill
\subfloat[Redshift cut sample\label{fig:4b}]{\includegraphics[width=0.49\textwidth]{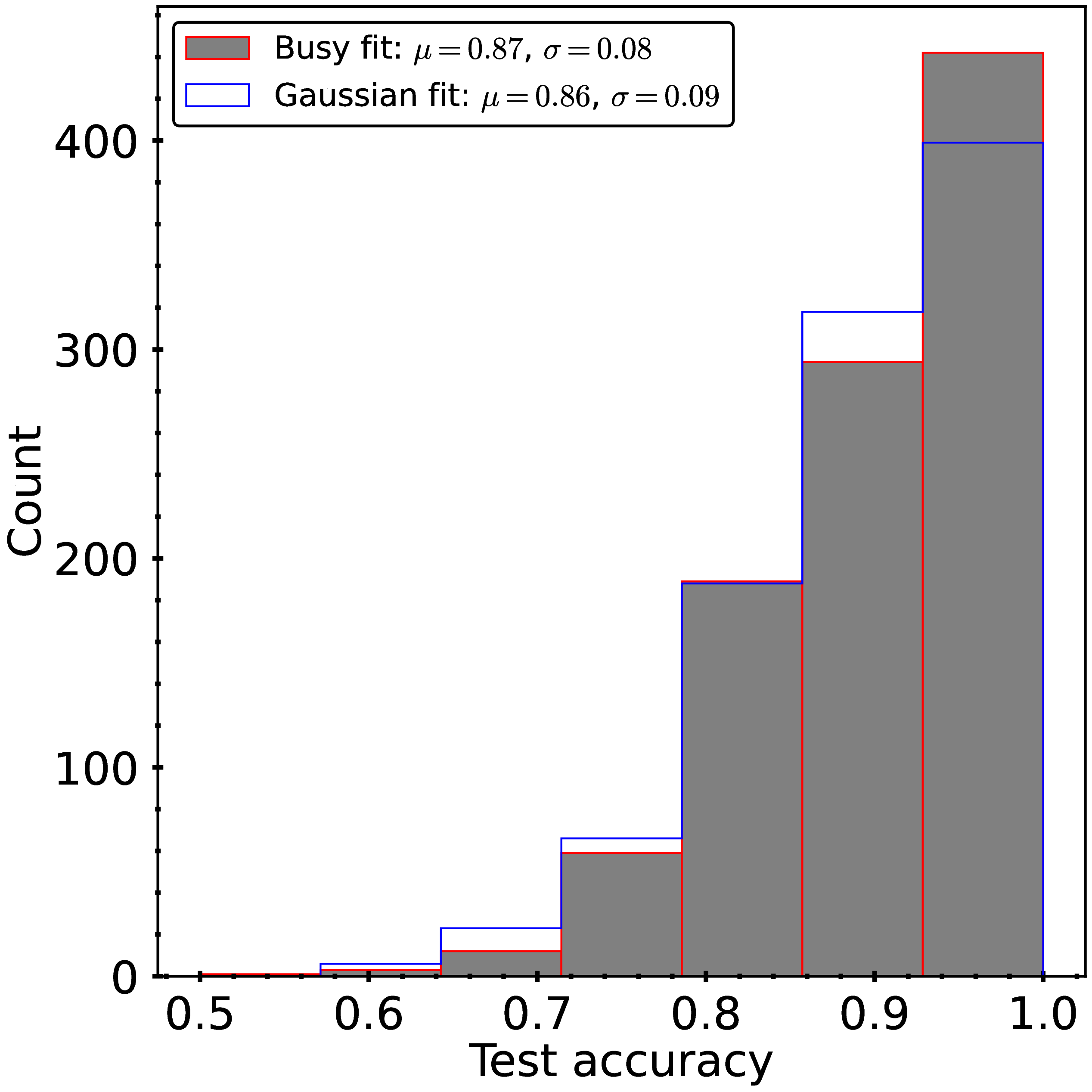}}
\caption{Busy function vs. multi-Gaussian function fit -- the test accuracy evaluation of the random forest model over 1000 runs, where $\mu$ and $\sigma$ denote the average values of the mean and the standard deviations.}
\label{fig:4}
\end{figure*}

\begin{figure*}
\centering
\subfloat[Confusion matrix\label{fig:5a}]{\includegraphics[width=0.33\textwidth]{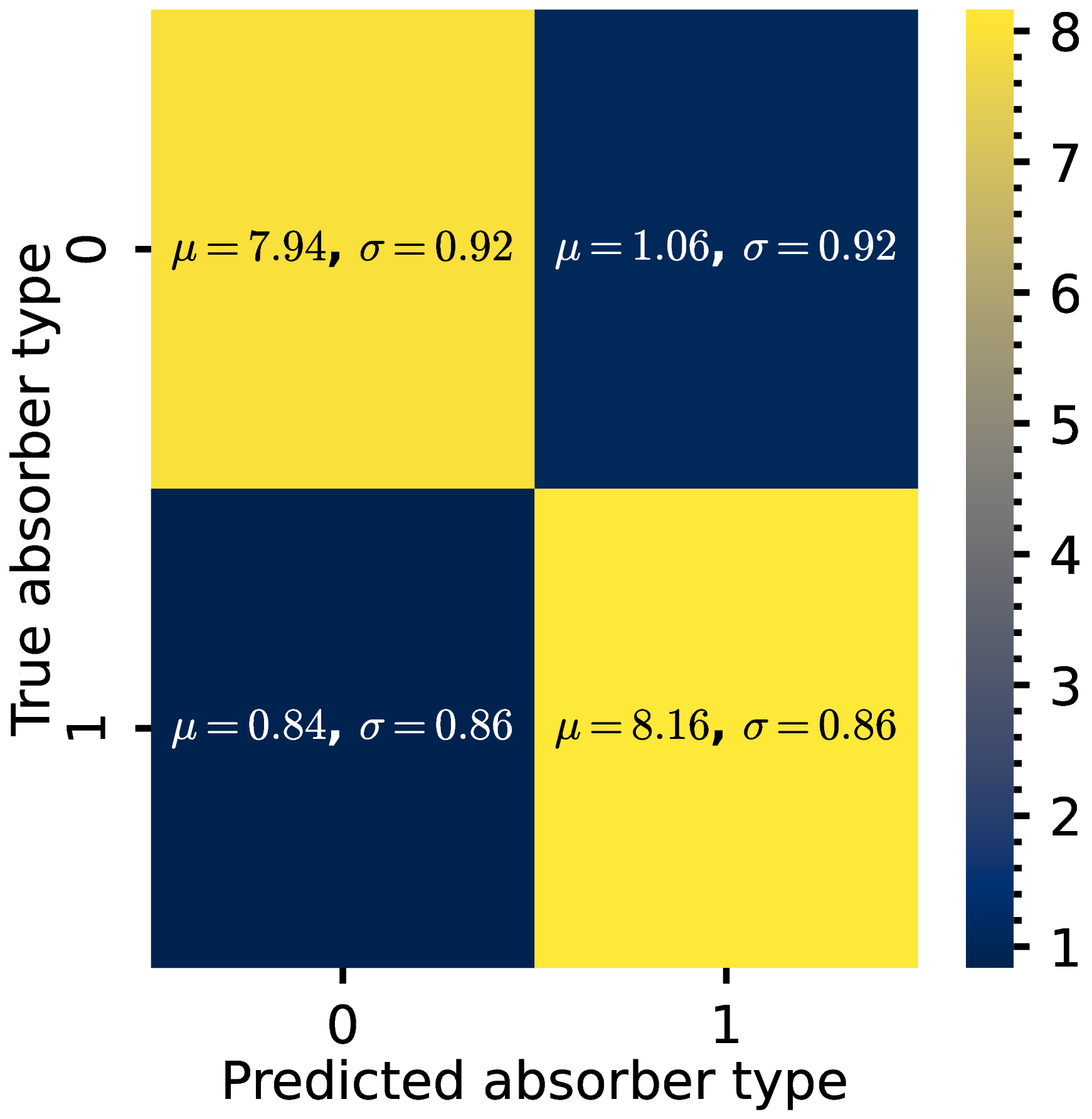}}
\hfill
\subfloat[ROC curve\label{fig:5b}]{\includegraphics[width=0.33\textwidth]{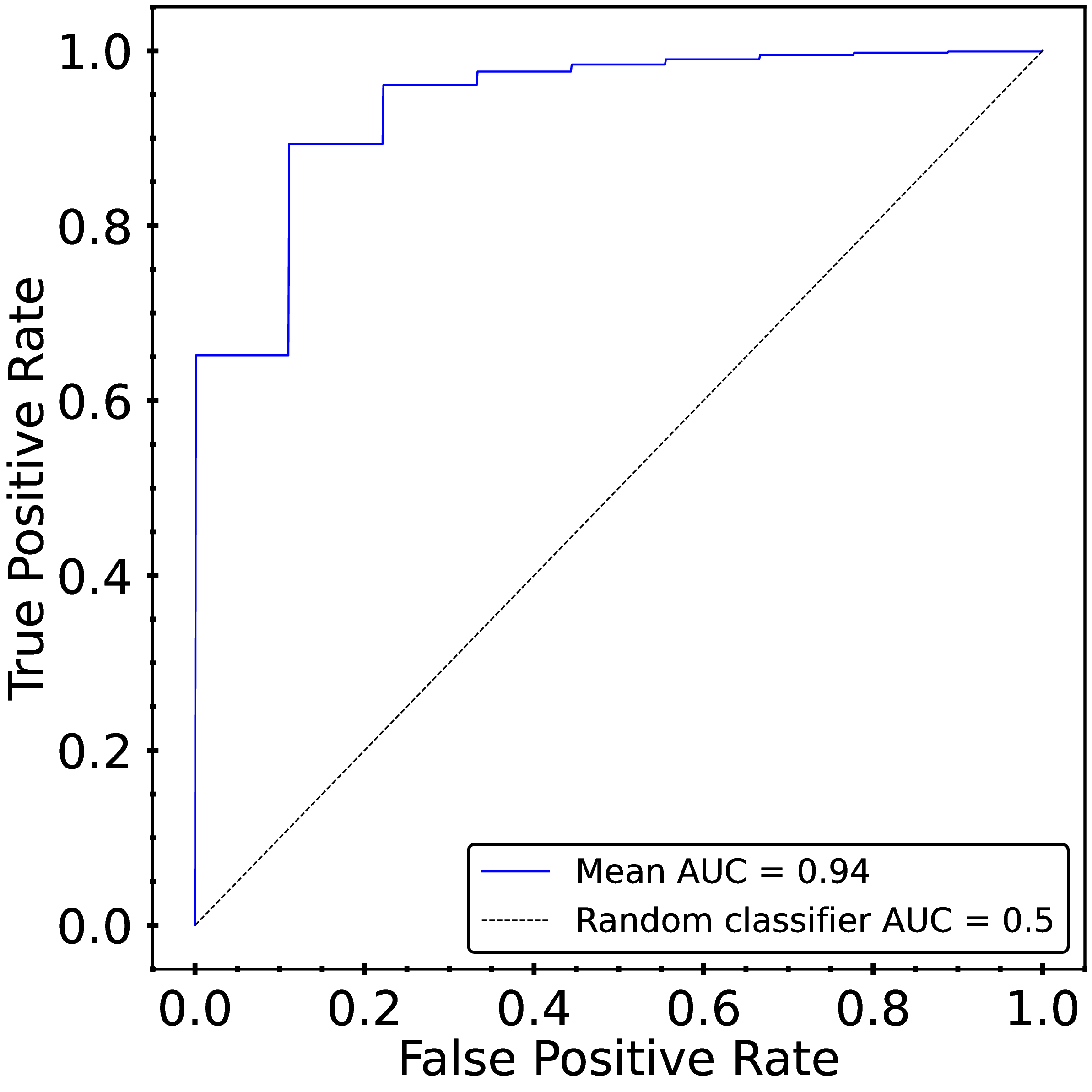}}
\hfill
\subfloat[Feature importance\label{fig:5c}]{\includegraphics[width=0.33\textwidth]{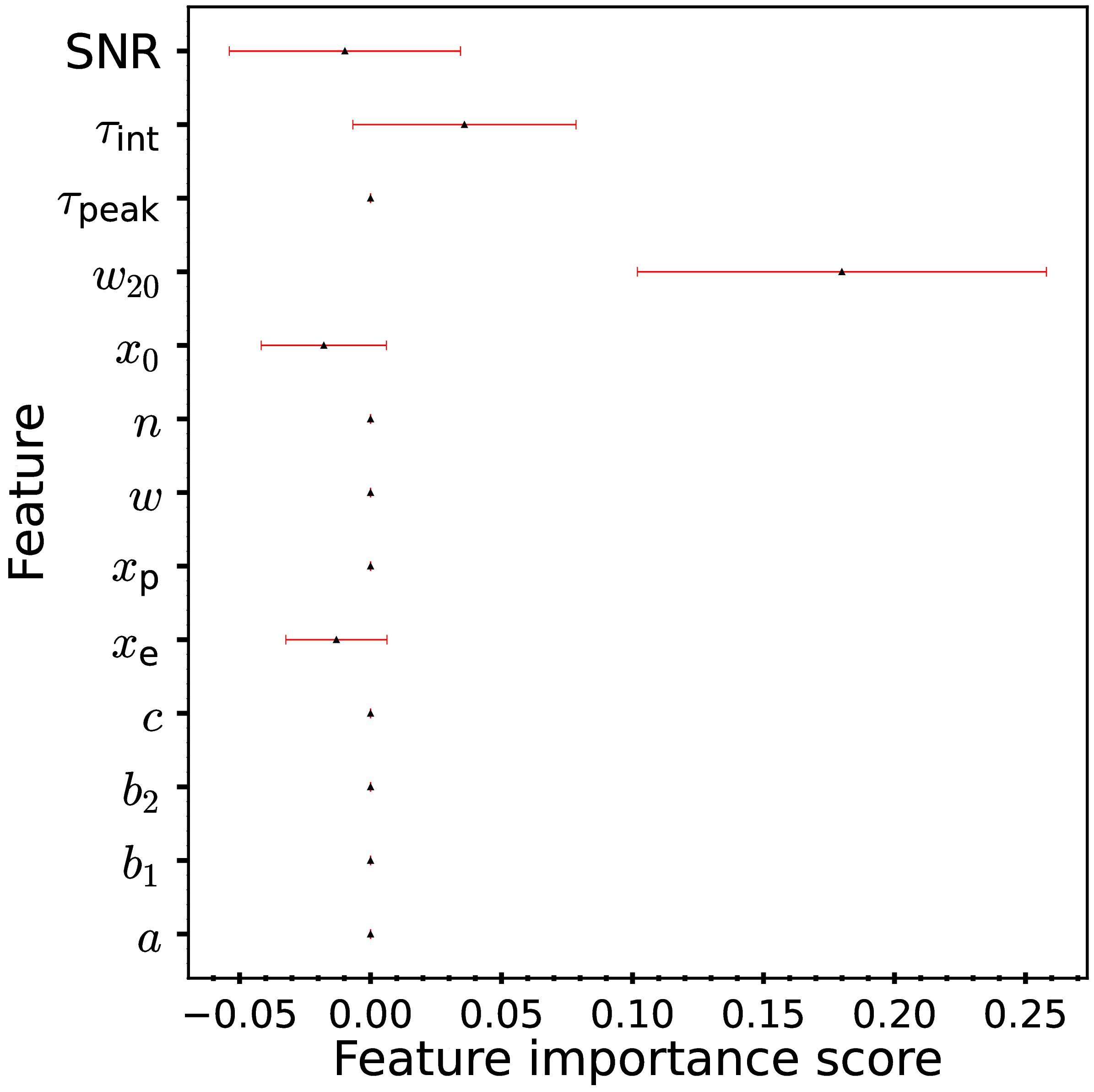}}
\caption{For the all spectral parameter sample, the predictive performance of the random forest model on the test data, where in (a) $\mu$ and $\sigma$ denote the average values of the mean and the standard deviations, and in (c) error bars are shown in red.}
\label{fig:5}
\end{figure*}

\begin{enumerate}[label = (\roman*), labelwidth = 0pt]
\item Among all the ML models, the random forest emerged as the best classification model with the highest average accuracy of 89\%, the highest average $F_1$-score of 0.90 and the highest average AUC score of 0.94 on test data. Also, the difference between the model's average training and test accuracies is 6\% (see Table \ref{tab:2}), which is a little high but acceptable because we have a small data sample. Thus, this random forest model is not prone to severe overfitting. Fig. \ref{fig:5} displays the mean confusion matrix, the mean ROC curve, and the associated mean feature importance weight graph for this random forest model. From Fig. \ref{fig:5c}, it is evident that $w_{20}$ is the most significant spectral parameter influencing the random forest's predictive performance, and $\tau_{\text{int}}$ is the next significant spectral parameter. \\

\item Also, we have fitted multiple Gaussian functions to each absorber in our sample to check whether using a Busy function fitting method offers any advantages over Gaussian fitting for absorber type classification. Multiple Gaussian functions have been fitted using the \texttt{curve\_fit} module of the \texttt{SciPy} Python library. The following spectral parameters are extracted from each best-fitted multi-Gaussian profile: the number of Gaussian components, the maximum, minimum and average FWHM (full width at half maximum) values of the fitted Gaussians, the maximum, minimum and average values of the peak optical depth of the fitted Gaussians, integrated optical depth, FWZI (full width at zero intensity), the average velocity offset and the average velocity offset over FWZI. The number of Gaussian components in the best-fit is determined using the Bayesian information criterion (BIC). This Gaussian fitting process has been performed in accordance with \citet{Curran2021}. We again trained the random forest model over 1000 runs using these multi-Gaussian function fitted spectral parameters and the SNR as predictor variables. This retrained random forest model yields an average accuracy of 89\%, an average $F_1$-score of 0.89 and an average AUC score of 0.95 on test data. Also, its average training and test accuracy difference is 3\% only. Thus, model does not suffer from severe overfitting. \\

\item For the all spectral parameter sample, the random forest model trained on the Busy function fitted spectral parameters provides an equal average test accuracy as its multi-Gaussian function fitted counterpart. Fig. \ref{fig:4a} provides a comparison between the histograms of test accuracy scores of the random forest models trained using the Busy function fitted and the multi-Gaussian function fitted spectral parameters. \\

\item The random forest algorithm performs well on our dataset compared to others. Because it captures the training data patterns efficiently and handles the overfitting issue simultaneously, as it uses different subsets of the training data and features for its constituent decision trees. However, more complex models like SVM and XGBoost lead to severe overfitting for our data sample. In such cases, we need more training data to prevent overfitting. \\
\end{enumerate}

\begin{table*}
\centering
\caption{The average classification metrics values of all ML models over 1000 runs for the all spectral parameter and the redshift cut samples. Random forest (indicated by \textdagger) is the best-performing ML classification model for both cases.} 
\label{tab:2}
\begin{tabular}{c|c c c|c c c}
\hline
& \multicolumn{3}{c|}{All spectral parameter sample} & \multicolumn{3}{c}{Redshift cut sample}\\
\hline
ML classification model &  Accuracy & $F_1$-score &       AUC &  Accuracy & $F_1$-score & AUC \\ 
                        & (average) &   (average) & (average) & (average) &   (average) & (average) \\ 
\hline
\hline
     Gaussian naive Bayes (Training) & 0.82 & 0.81 & 0.93 & 0.84 & 0.84 & 0.92\\
         Gaussian naive Bayes (Test) & 0.74 & 0.73 & 0.82 & 0.76 & 0.76 & 0.81\\
\hline    
      Logistic regression (Training) & 0.91 & 0.91 & 0.95 & 0.89 & 0.90 & 0.93\\
          Logistic regression (Test) & 0.86 & 0.87 & 0.91 & 0.83 & 0.84 & 0.86\\
\hline 
            Decision tree (Training) & 0.91 & 0.91 & 0.96 & 0.91 & 0.91 & 0.95\\
                Decision tree (Test) & 0.86 & 0.85 & 0.89 & 0.84 & 0.85 & 0.88\\
\hline
            Random forest (Training) & 0.95 & 0.95 & 0.98 & 0.92 & 0.93 & 0.98\\
\textdagger     Random forest (Test) & 0.89 & 0.90 & 0.94 & 0.87 & 0.88 & 0.92\\
\hline
                      SVM (Training) & 0.90 & 0.90 & 0.93 & 0.89 & 0.90 & 0.92\\
                          SVM (Test) & 0.82 & 0.83 & 0.88 & 0.80 & 0.80 & 0.84\\ 
\hline
                  XGBoost (Training) & 1.00 & 1.00 & 1.00 & 1.00 & 1.00 & 1.00\\
               XGBoost forest (Test) & 0.87 & 0.87 & 0.92 & 0.85 & 0.86 & 0.89\\

\hline
\end{tabular}
\end{table*}

\subsection{Results for all spectral parameters with redshift cut}
\label{sec:4.2}

Earlier works, viz. \citet{Curran2016} and \citet{Curran2021} had conducted similar ML classification of \hi\ 21-cm absorption lines, where Gaussian profiles were used to fit each spectrum. The logistic regression emerged as their best ML classification model that efficiently determined the type of \hi\ 21-cm absorber into associated or intervening with accuracy $\approx 80\%$. Notably, they had applied an absorber redshift cut-off ($z_\text{abs} \geq 0.1$) on their data sample for the following reasons: (i) to prevent resolved sightlines at $z \lesssim 0.1$ \citep[e.g.][]{Dutta2016,Reeves2016} from introducing any systematic difference between the samples at high and low redshifts, (ii) to reduce the dilution by \hi\ 21-cm emission, which could be significant at $z \lesssim 0.1$, and (iii) finally, to keep the sample sizes for both the intervening and associated absorber categories equal, and thus prevent the ML models from favouring one over the other. \\

We checked that there are no resolved sightlines or discernible effects of \hi\ 21-cm emission in our sample. From the correlation matrix presented in Fig. \ref{fig:3}, we find no strong correlation (i.e., $|$correlation coefficient$|$ $>0.5$) of any of the spectral parameters with redshift. Nevertheless, to check whether any intrinsic evolution of spectral parameters with redshift affects the predictive performance of all six ML models on our data sample, we limit our dataset to the $z_\text{abs} \geq 0.1$ range similar to the study of \citet{Curran2016}. This reduces our \hi\ 21-cm absorber sample size to 74 (34 associated and 40 intervening; the concerned dataset is termed as the redshift cut sample). For this redshift cut sample, we again trained all six ML models over 1000 runs as per the model training scheme outlined in Section \ref{sec:3}. The predictive performance of each model is tabulated in Table \ref{tab:2} and discussed below.

\begin{figure*}
\centering
\subfloat[Confusion matrix\label{fig:6a}]{\includegraphics[width=0.33\textwidth]{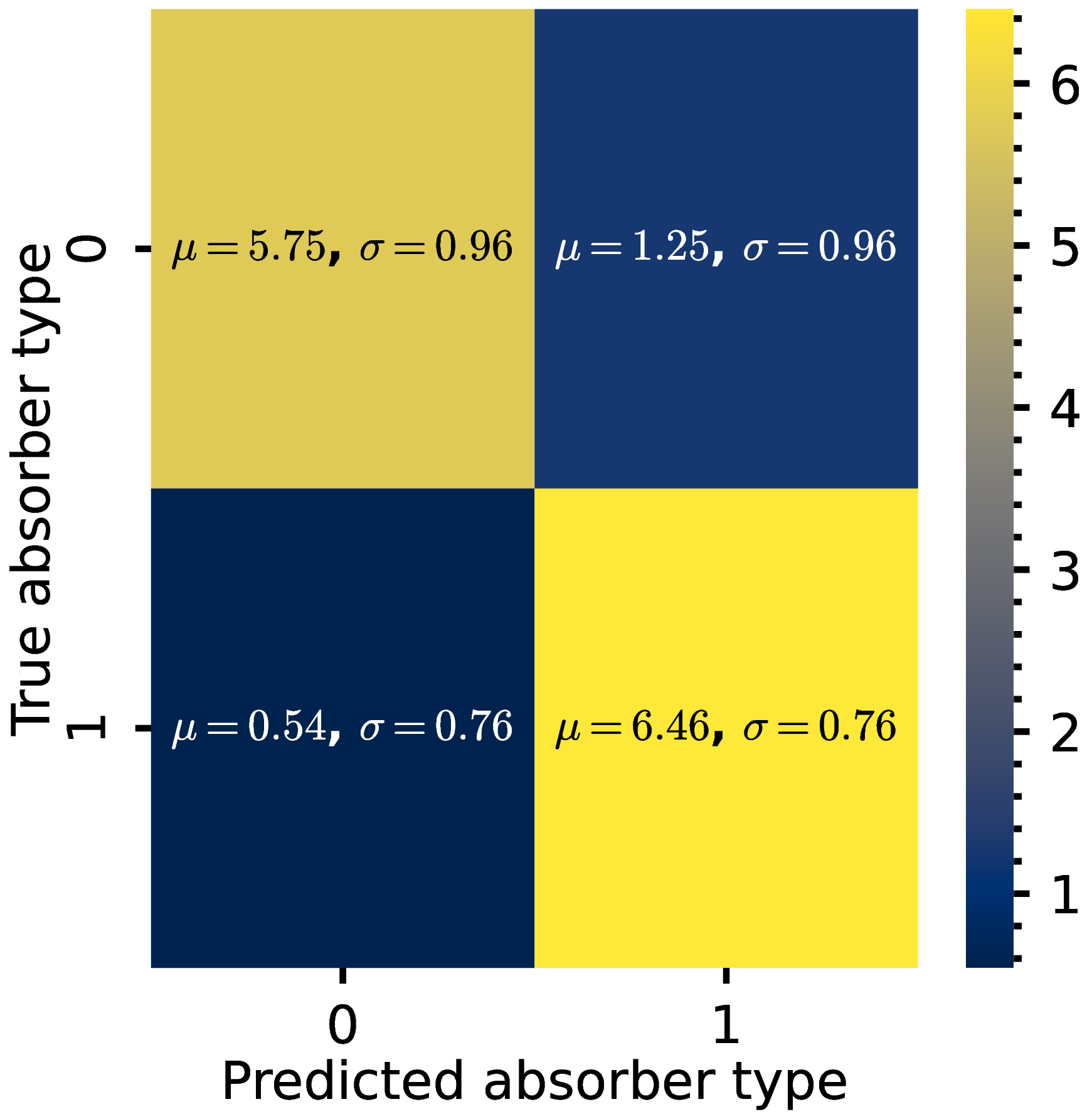}}
\hfill
\subfloat[ROC curve\label{fig:6b}]{\includegraphics[width=0.33\textwidth]{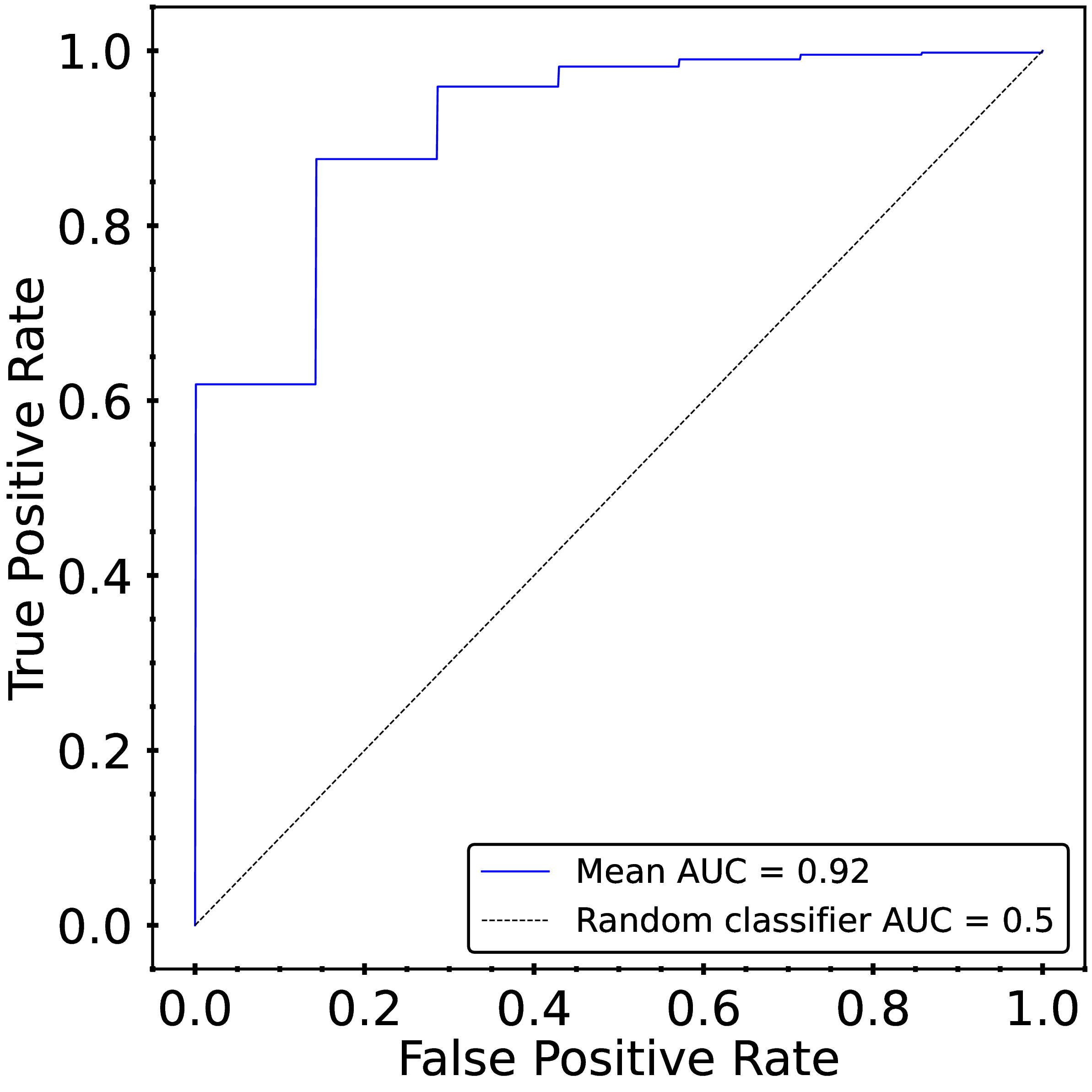}}
\hfill
\subfloat[Feature importance\label{fig:6c}]{\includegraphics[width=0.33\textwidth]{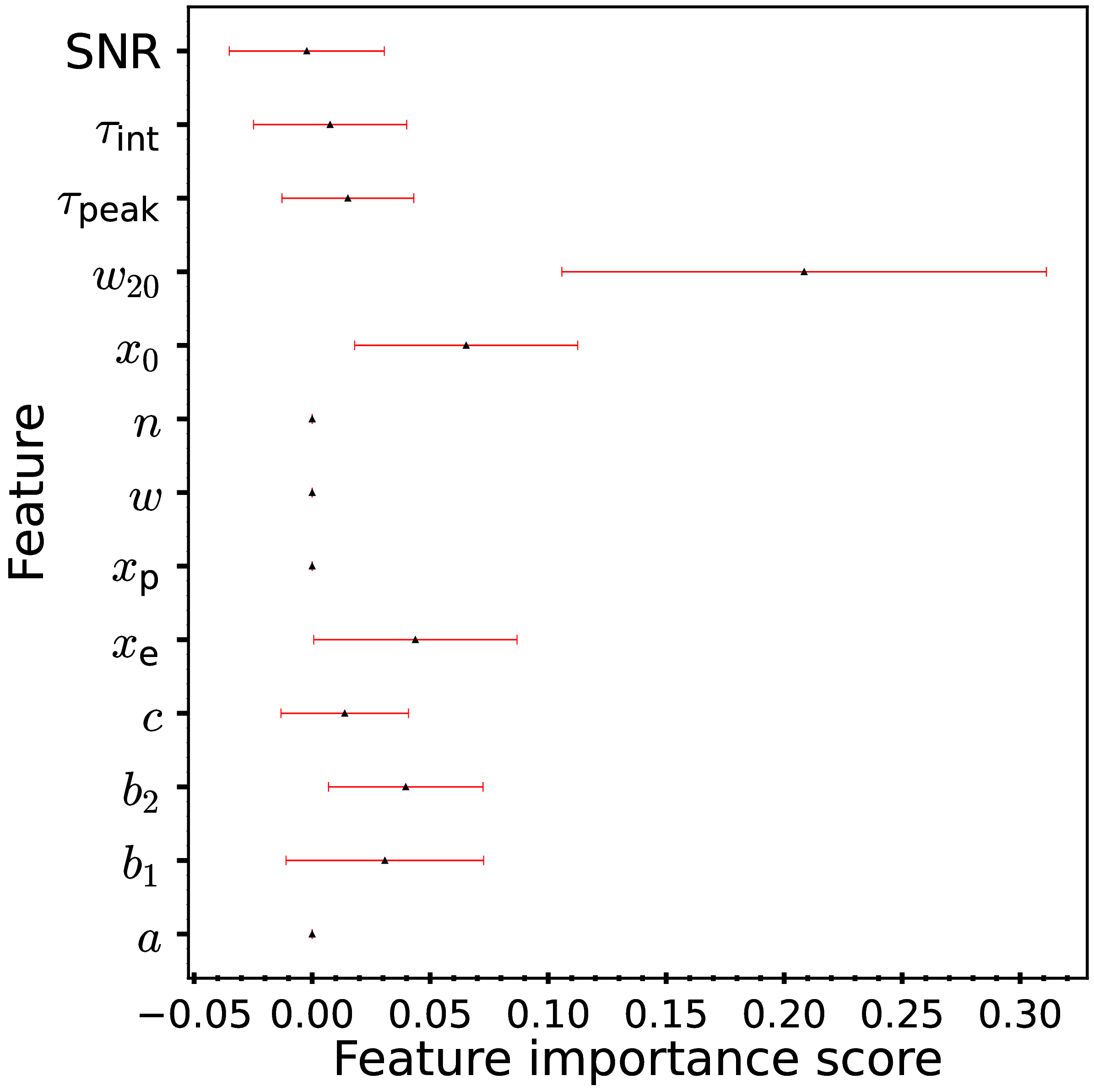}}
\caption{For the redshift cut sample, the predictive performance of the random forest model on the test data, where in (a) $\mu$ and $\sigma$ denote the average values of the mean and the standard deviations, and in (c) error bars are shown in red.}
\label{fig:6}
\end{figure*}

\begin{enumerate}[label = (\roman*), labelwidth = 0pt]
\item For this redshift cut sample, the random forest emerged as the best ML classification model with the average accuracy of 87\%, the average $F_1$-score of 0.88 and the average AUC score of 0.92 on test data. Moreover, the difference between the model's average training and test accuracies is 5\% (see Table \ref{tab:2}), which signifies the model does not suffer from severe overfitting. The mean confusion matrix, the mean ROC curve and the corresponding mean feature importance graph for this random forest model are shown in Fig. \ref{fig:6}. Like the all spectral parameter case (see Section \ref{sec:4.1}), here also, $w_{20}$ is the most significant spectral parameter that influences the predictive performance of this random forest model, and $x_0$ is the next significant spectral parameter (see Fig. \ref{fig:6c}). \\

\item In this case, the random forest's average test accuracy is 2\% (= 89 - 87\%) less compared to the earlier all spectral parameter case (see Table \ref{tab:2}). Also, the values of other performance metrics ($F_1$ and AUC scores) do not deviate much either. Hence, we can conclude that imposing a redshift cut of $z_\text{abs} \geq 0.1$ has a little impact on this random forest model's predictive power, despite the original sample size being reduced significantly after applying the redshift cut. Also, this random forest model yields a better classification performance (accuracy $\approx$ 87\%) than \citet{Curran2016} and \citet{Curran2021} (accuracy $\approx$ 80\%). A possible reason could be using Busy function fitted spectral features in model training. \\

\item Also, for this redshift cut sample, we again trained the random forest model using multi-Gaussian function fitted spectral parameters and the SNR as predictor variables. On test data, this random forest model's average accuracy is 86\%, with an average $F_1$-score of 0.87 and an average AUC score of 0.94. The difference between the model's average training and test accuracy is 6\%, so it suffers from some overfitting but not severely.

\item For the redshift cut sample, the random forest model trained using Busy function fitted spectral parameters only provides little better ML classification performance, 1\% (= 87 - 86\%) more average test accuracy. A comparison between the histograms of test accuracy scores of the random forest models trained using Busy function fitted and multi-Gaussian function fitted spectral parameters is given in Fig. \ref{fig:4b}.
\end{enumerate}

\subsection{Results for two spectral parameters}
\label{sec:4.3}

As indicated in Section \ref{sec:2.2} and as seen in Sections \ref{sec:4.1} and \ref{sec:4.2}, $w_{20}$ is the most significant spectral parameter that influences the predictive performance of the best ML classification models. This is consistent with the results of \citet{Curran2021}, who found that the linewidth of the absorption lines was the most important feature in their ML classification models. The stronger and broader associated absorption lines arise from AGN, whereas the weaker and narrower intervening absorption lines arise from normal, star-forming galaxies. Fast outflowing gas, accreting gas onto the black hole, and rotating or disturbed gas in the circumnuclear region could give rise to the stronger and/or broader absorption lines in the case of associated absorbers.

FLASH is a wide-area radio survey being conducted using the ASKAP radio telescope to study the cold neutral gas in and around galaxies using the \hi\ 21-cm absorption line in the intermediate redshift range, $0.4 < z < 1.0$ \citep{Allison2022}. \citet{Yoon2025} reported the absorber type of these 30 new \hi\ 21-cm absorption lines using the ML classification model (logistic regression) of \citet{Curran2021} trained on the spectral parameters extracted using Gaussian profile fitting. Among these Gaussian function fitted parameters `Linewidth' (corresponds to the FWHM from a single Gaussian fit) and $\tau_{\text{int}}$ (corresponds to the integrated optical depth of the spectral line) (see Table \ref{tab:b2}) are similar to our dataset's $w_{20}$ and $\tau_{\text{int}}$ spectral parameters (see Table \ref{tab:b1}). This motivates us to train a random forest model (which is the best ML classification model in both the previous cases we discussed till now), using only the Busy function fitted spectral parameters $w_{20}$ and $\tau_{\text{int}}$. We aim to use this newly trained random forest model to predict the absorber type of 30 new \hi\ 21-cm absorbers detected blindly in the First Large Absorption Survey in \hi\ (FLASH) pilot surveys \citep{Yoon2025}. This also helps to ensure whether these two parameters are sufficient for classifying \hi\ 21-cm absorbers or not.

In accordance with this, we first limit our all spectral parameter dataset to only $w_{20}$, $\tau_{\text{int}}$ and the absorber type (the concerned dataset is termed as the two spectral parameter sample). Then, we trained the random forest model over 1000 runs using it. The model's average predictive performance is tabulated in Table \ref{tab:3} and discussed below.

\begin{figure*}
\centering
\subfloat[Confusion matrix\label{fig:7a}]{\includegraphics[width=0.33\textwidth]{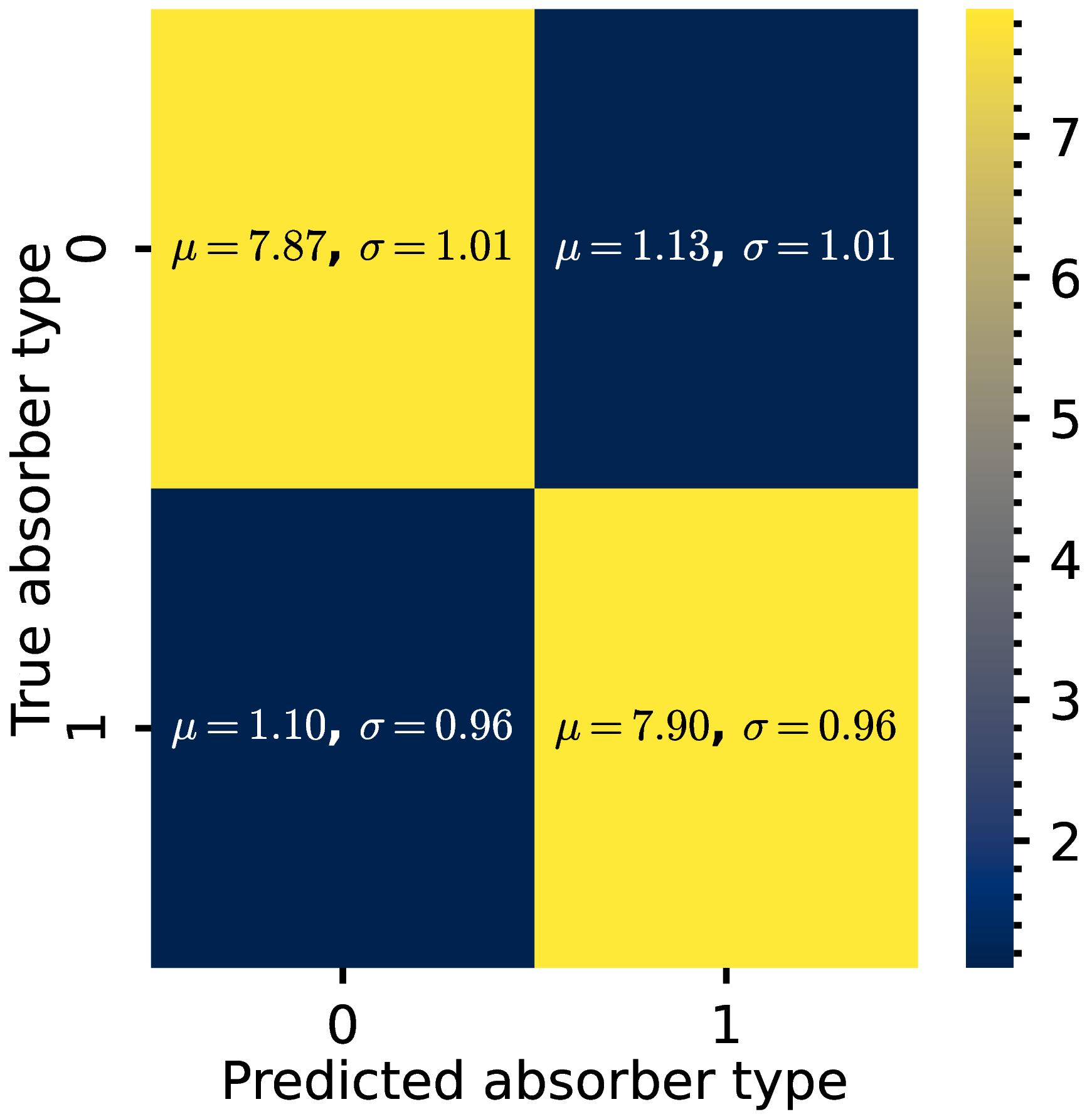}}
\subfloat[ROC curve\label{fig:7b}]{\includegraphics[width=0.33\textwidth]{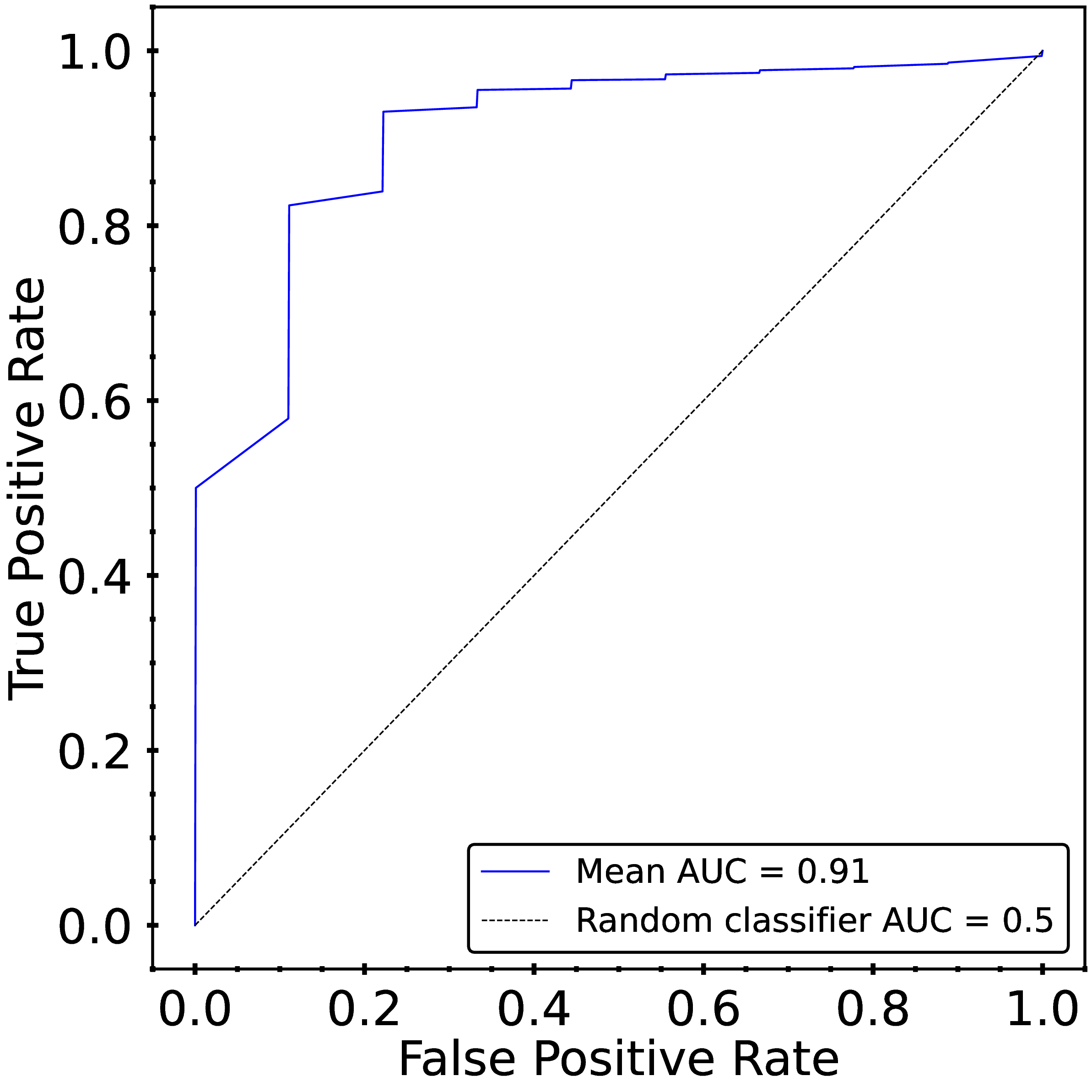}}
\caption{For the two spectral parameter sample, the predictive performance of the random forest model on the test data, where in (a) $\mu$ and $\sigma$ denote the average values of the mean and the standard deviations.}
\label{fig:7}
\end{figure*}

\begin{enumerate}[label = (\roman*), labelwidth = 0pt]
\item For the two spectral parameter sample, the random forest model achieved an average accuracy of 88\%, an average $F_1$-score of 0.88 and an average AUC score of 0.91 on test data. Also, the training and test accuracy difference of this random forest model is 3\% (see Table \ref{tab:3}), which is quite less, thus the model is not prone to severe overfitting. The histogram of the accuracy scores, the mean confusion matrix and the mean ROC curve of this random forest model on test data are shown in Fig. \ref{fig:7}. \\

\item The average test accuracy of this newly trained random forest model is only 1\% (= 89 - 88\%) less compared to the case for all spectral parameters. The values of other performance metrics are also consistent with those obtained from the all spectral parameter case and those obtained from the all spectral parameter with the redshift cut case (see Tables \ref{tab:2} and \ref{tab:3}). Thus, the predictive performance of the random forest model does not affect much despite being trained only on the two spectral parameters. The use of the most prominent spectral parameter $w_{20}$ might be one of the reasons behind this.

\item We identified the most preferred model hyperparameter values over 1000 runs during this random forest model training. We used those to predict the absorber type of above-mentioned 30 new \hi\ 21-cm absorbers reported in \citet{Yoon2025}. The ML classification labels predicted by the random forest model with the most preferred hyperparameters are in $\approx 80\%$ agreement (24 out of 30) with the ML classification labels predicted by the logistic regression model of \citet{Curran2021}, which has $\approx 80\%$ accuracy on the absorber type prediction. The description of this 30 new \hi\ 21-cm absorption line data sample, along with their ML classification labels as per this newly trained random forest model and their ML classification label agreement with \citet{Curran2021} are given in Table \ref{tab:b2}.
\end{enumerate} 

\begin{table}
\centering
\caption{For the two spectral parameter sample, the average classification metrics values of the random forest model over 1000 runs.}
\label{tab:3}
\begin{tabular}{c c c c c}
\hline
 ML classification model &  Accuracy & $F_1$-score & AUC \\
                         & (average) &   (average) & (average) \\
\hline
\hline
Random forest (Training) &      0.91 &        0.91 & 0.95 \\
    Random forest (Test) &      0.88 &        0.88 & 0.91 \\
\hline
\end{tabular}
\end{table}

\section{Conclusions}
\label{sec:5}

This work aims to develop an efficient ML classification model to predict the origin of \hi\ 21-cm absorption lines, i.e., whether they arise in `intervening' galaxies or are `associated' with the radio AGN. Such an ML model would be extremely useful to accurately predict the type of absorbers in upcoming large, blind \hi\ 21-cm surveys using SKA pre-cursors in an automated manner without the need for follow-up spectroscopy. To this end, we used the Busy function on a data sample of 118 \hi\ 21-cm absorption line spectra (74 associated and 44 intervening) to extract spectral features. We trained six ML classification models -- Gaussian naive Bayes, logistic regression, decision tree, random forest, XGBoost and SVM -- on the dataset of these features and the absorber type. First, we have used the random stratified sampling technique during the train-test splitting of the dataset, such that trained models do not favour a specific absorber type. Additionally, to make unbiased model hyperparameters for this small data sample, the leave-one-out CV technique is used on top of the grid search CV during the training process. We ran each ML model 1000 times to get an unbiased estimate of the classification metrics. The main results from this analysis are outlined below.

\begin{enumerate}[label = (\roman*), labelwidth = 0pt]
\item Among all the models, the random forest emerges as the most efficient ML classification model, achieving a test accuracy of 89\%, a test $F_1$-score of 0.9 and a test AUC score of 0.94 when considering all the spectral parameters and the SNR as predictor variables. \\

\item $w_{20}$ is the most significant spectral parameter influencing the absorber type classification task of this random forest model. Associated absorbers have stronger and broader absorption profiles on average compared to intervening absorbers, which is also evident in \citet{Curran2016}. The reason could be due to fast rotation, outflows and accretion in the circumnuclear region of AGN. We retrained this random forest model on the dataset of two spectral parameters ($w_{20}$ and $\tau_{\text{int}}$), the SNR and the absorber type. This yields a test accuracy of 88\% along with a test $F_1$-score of 0.88 and a test AUC score of 0.91. Comparing these values with the random forest model of the all spectral parameter case, we can conclude that these two parameters could be sufficient for predicting the origin of the \hi\ 21-cm absorption lines. \\

\item The random forest with $w_{20}$ as its most significant spectral parameter also emerges as the most reliable ML model when trained on the redshift cut sample, where a $z_\text{abs} \geq 0.1$ cut has been applied to mitigate any effect of redshift evolution on the spectral parameters. This retrained random forest model gives a test accuracy of 87\%, a test $F_1$-score of 0.88 and a test AUC score of 0.92. Overall, the classification results of this random forest model are close to the classification results obtained by the random forest model trained on all spectral parameter sample without applying any redshift cut. It indicates that the redshift cut does not significantly affect the absorber-type ML classification task. \\

\item Compared to Gaussian fitting technique employed in previous similar studies \citep{Curran2016,Curran2021}, this work is the first to use Busy function fitting to extract spectral parameters of \hi\ 21-cm absorption lines and use these to train ML models. Moreover, this work explores a broader range of six different ML classification models to compare their predictive power in classifying the type of \hi\ 21-cm absorbers. In addition, we have provided a detailed breakdown of the training and test accuracy values, $F_1$ and AUC scores, and optimal hyperparameter values averaged over 1000 runs for each ML model. In particular, \citet{Curran2021} had trained four different ML models, and obtained the highest accuracy of $\approx 80\%$ for the logistic regression model (they did not provide $F_1$ and AUC scores). Overall, this work provides an efficient ML classification framework in greater detail for low data samples like ours. \\

\item We have also fitted multi-Gaussian functions on our data samples and used the extracted spectral parameters to train ML models to classify the absorber type. We found that the random forest yields the same 89\% test accuracy for the all-parameter sample and little (1\%) less 86\% test accuracy for the redshift cut sample, compared to the Busy function fitted counterpart. Moreover, the spectral linewidth emerges as the most robust feature in both ML classification tasks, using the Busy function fitted and the multi-Gaussian function fitted spectral parameters. Thus, we conclude that the absorber type classification results obtained by training ML models using the Busy function's extracted spectral parameters are as good as those obtained using multi-Gaussian function fitted spectral parameters for a similar classification task. Also, we found that for our data sample, we required between 1 to 5 Gaussians to fit the spectral profile, and each Gaussian has three free parameters (mean, variance and amplitude). Thus, we must estimate the values of 15 free parameters in some cases. Now, compared to this, for the Busy function, we need to estimate a fixed number of eight free parameters regardless of the spectra. Moreover, the Busy function parameters are physically more interpretable than the Gaussian parameters. Thus, for such ML-based classification tasks, the Busy function can also be used as a suitable and efficient alternative to Gaussian functions (see discussion in Section \ref{sec:2.2} to check how the Busy function parameters are physically related to the double-horn profiles of \hi\ 21-cm absorption lines). \\

\item To demonstrate the applicability of our work for future large \hi\ 21-cm absorption surveys, the random forest model retrained on the Busy function fitted spectral parameters $w_{20}$ and $\tau_{\text{int}}$ are used to predict the absorber type of a new data sample of 30 \hi\ 21-cm absorption lines detected in the FLASH pilot surveys \citep{Yoon2025}. The absorber type prediction by this model is in $\approx 80\%$ agreement with that of \citet{Curran2021}. \\
\end{enumerate}

As more \hi\ 21-cm absorption spectra become available in the next few years for training the ML models, the predictive power of such models will increase. Therefore, the techniques developed in this work are likely to be of significant value in statistical studies of large samples of \hi\ 21-cm absorbers in the SKA-era.

\section*{Acknowledgements}

The authors gratefully acknowledge Nissim Kanekar, Filippo Marcello Maccagni and Stephen J. Curran for providing the necessary spectra of \hi\ 21-cm absorption lines for our data sample. We also thank Rajeshwari Dutta and Debasish Koner for their valuable suggestions and inputs, which have greatly improved the quality of the paper. Moreover, authors would like to acknowledge the Science and Engineering Research Board (SERB), Department of Science and Technology (DST), Government of India, for providing financial support through a research grant (SERB/CRG/2022/005963). Also, the author DM acknowledges the Council of Scientific \& Industrial Research (CSIR), Government of India, for providing a research grant via an associateship (File No. 09/1178(25506)/2025-EMR-I). We also thank the anonymous referee for insightful remarks, which significantly improved the manuscript.

\section*{Data Availability}

The data used in this study is available within this article and codes will be made available with a reasonable request to the corresponding author.



\bibliographystyle{mnras}
\bibliography{paper_bib} 

@article{Westmeier2014,
title = {The busy function: a new analytic function for describing the integrated 21-cm spectral profile of galaxies},
author = {Westmeier, Tobias and Jurek, Russell and Obreschkow, Danail and Koribalski, B{\"a}rbel S. and Staveley-Smith, Lister},
journal = {\mnras},
year = {2014},
volume = {438},
pages = {1176},
doi = {https://doi.org/10.1093/mnras/stt2266}
}

@article{Cramer2002,
title = {The origins of logistic regression},
author = {Cramer, Jan Salomon},
journal = {Tinbergen Institute Working Paper},
year = {2002},
volume = {4},
pages = {119},
doi = {http://dx.doi.org/10.2139/ssrn.360300}
}

@article{Utgoff1989,
title = {Incremental induction of decision trees},
author = {Utgoff, Paul E.},
journal = {Mach. Learn.},
year = {1989},
volume = {4},
pages = {161},
doi = {https://doi.org/10.1023/A:1022699900025}
}

@article{Ho1998,
title = {The random subspace method for constructing decision forests},
author = {Ho, Tin Kam},
journal = {IEEE Trans. Pattern Anal. Mach. Intell.},
year = {1998},
volume = {20},
pages = {832},
doi = {https://doi.org/10.1109/34.709601}
}

@article{Chen2016,
title = {XGBoost: A scalable tree boosting system},
author = {Chen, Tianqi and Guestrin, Carlos},
journal = {Proc. ACM SIGKDD Int. Conf. Knowl. Discov. Data Min.},
year = {2016},
pages = {785},
doi = {https://doi.org/10.1145/2939672.2939785}
}

@article{Cortes1995,
title = {Support-vector networks},
author = {Cortes, Corinna and Vapnik, Vladimir},
journal = {Mach. Learn.},
year = {1995},
volume = {20},
pages = {273},
doi = {https://doi.org/10.1007/BF00994018}
}

@article{Aditya2018b,
title = {A Giant Metrewave Radio Telescope survey for associated HI 21 cm absorption in the Caltech--Jodrell flat-spectrum sample},
author = {Aditya, J. N. H. S. and Kanekar, Nissim},
journal = {\mnras},
year = {2018},
volume = {481},
pages = {1578},
doi = {https://doi.org/10.1093/mnras/sty2184}
}

@article{Ostorero2017,
title = {Correlation between X-ray and radio absorption in compact radio galaxies},
author = {Ostorero, Luisa and Morganti, Raffaella and Diaferio, Antonaldo and Siemiginowska, Aneta and Stawarz, {\L}ukasz and Moderski, Rafal and Labiano, Alvaro},
journal = {\apj},
year = {2017},
volume = {849},
pages = {34},
doi = {https://doi.org/10.3847/1538-4357/aa8ef6}
}

@article{Aditya2019,
title = {uGMRT detections of HI 21-cm absorption associated with intermediate redshift galaxies},
author = {Aditya, J. N. H. S.},
journal = {\mnras},
year = {2019},
volume = {482},
pages = {5597},
doi = {https://doi.org/10.1093/mnras/sty3062}
}

@article{kanekar2009a,
title = {The covering factor of high-redshift damped Lyman-$\alpha$ systems},
author = {Kanekar, Nissim and Lane, W. M. and Momjian, E. and Briggs, F. H. and Chengalur, Jayaram N. },
journal = {\mnras},
year = {2009},
volume = {394},
pages = {L61},
doi = {https://doi.org/10.1111/j.1745-3933.2008.00610.x}
}

@article{Kanekar2009b,
title = {A search for HI 21 cm absorption in strong Mg II absorbers in the redshift desert},
author = {Kanekar, Nissim and Prochaska, J. X. and Ellison, S. L. and Chengalur, Jayaram N. },
journal = {\mnras},
year = {2009},
volume = {396},
pages = {385},
doi = {https://doi.org/10.1111/j.1365-2966.2009.14661.x}
}

@article{Aditya2018a,
title = {A giant metrewave radio telescope search for associated HI 21 cm absorption in GHz-peaked-spectrum sources},
author = {Aditya, J. N. H. S. and Kanekar, Nissim},
journal = {\mnras},
year = {2018},
volume = {473},
pages = {59},
doi = {https://doi.org/10.1093/mnras/stx2325}
}

@article{Carilli1998,
title = {Redshifted neutral hydrogen 21 centimeter absorption toward red quasars},
author = {Carilli, C. L. and Menten, Karl M. and Reid, Mark J. and Rupen, M. P. and Yun, Min Su},
journal = {\apj},
year = {1998},
volume = {494},
pages = {175},
doi = {https://doi.org/10.1086/305191}
}

@article{Dutta2017a,
title = {HI 21-cm absorption survey of quasar-galaxy pairs: Distribution of cold gas around z < 0.4 galaxies},
author = {Dutta, Rajeshwari and Srianand, Raghunathan and Gupta, Neeraj and Momjian, E. and Noterdaeme, P. and Petitjean, P. and Rahmani, H.},
journal = {\mnras},
year = {2017},
volume = {465},
pages = {588},
doi = {https://doi.org/10.1093/mnras/stw2689}
}

@article{Dutta2017b,
title = {Incidence of HI 21-cm absorption in strong Fe II systems at 0.5 < z < 1.5},
author = {Dutta, Rajeshwari and Srianand, Raghunathan and Gupta, Neeraj and Joshi, R. and Petitjean, P. and Noterdaeme, P. and Ge, J. and Krogager, J. K.},
journal = {\mnras},
year = {2017},
volume = {465},
pages = {4249},
doi = {https://doi.org/10.1093/mnras/stw3040}
}

@article{Dutta2017c,
title = {HI 21-cm absorption from z ~ 0.35 strong Mg II absorbers},
author = {Dutta, Rajeshwari and Srianand, Raghunathan and Gupta, Neeraj and Joshi, Ravi},
journal = {\mnras},
year = {2017},
volume = {468},
pages = {1029},
doi = {https://doi.org/10.1093/mnras/stx538}
}

@article{Dutta2020,
title = {uGMRT search for cold gas at z~ 1--1.4 towards red quasars},
author = {Dutta, Rajeshwari and Raghunathan, S. and Gupta, Neeraj and Joshi, Ravi},
journal = {\mnras},
year = {2020},
volume = {491},
pages = {838},
doi = {https://doi.org/10.1093/mnras/stz3084}
}

@article{Gupta2009,
title = {A complete sample of 21-cm absorbers at z~ 1.3: Giant Metrewave Radio Telescope survey using Mg ii systems},
author={Gupta, Neeraj and Srianand, Raghunathan and Petitjean, Patrick and Noterdaeme, P. and Saikia, D. J.},
journal = {\mnras},
year = {2009},
volume = {398},
pages = {201},
doi = {https://doi.org/10.1111/j.1365-2966.2009.14933.x}
}

@article{Kanekar2014b,
title = {Giant metrewave radio telescope detection of two new HI 21 cm absorbers at z≈2},
author = {Kanekar, Nissim},
journal = {\apjl},
year = {2014},
volume = {797},
pages = {L20},
doi = {https://doi.org/10.1088/2041-8205/797/2/L20}
}

@article{Kanekar2013,
title = {A search for HI 21 cm absorption towards a radio-selected quasar sample--II. A new low spin temperature DLA at high redshift},
author = {Kanekar, Nissim and Ellison, Sara L. and Momjian, Emmanuel and York, Brian A. and Pettini, Max},
journal = {\mnras},
year = {2013},
volume = {428},
pages = {532},
doi = {https://doi.org/10.1093/mnras/sts058}
}

@article{Kanekar2006,
title = {HI 21 cm absorption at z~2.347 towards PKS B0438-436},
author = {Kanekar, Nissim and Subrahmanyan, Ravi and Ellison, S. L. and Lane, W. M. and Chengalur, Jayaram N.},
journal = {\mnras},
year = {2006},
volume = {370},
pages = {L46},
doi = {https://doi.org/10.1111/j.1745-3933.2006.00186.x}
}

@article{Kanekar2007,
title = {HI 21-cm absorption at z~3.39 towards PKS 0201+113},
author = {Kanekar, Nissim and Chengalur, Jayaram N. and Lane, W M},
journal = {\mnras},
year = {2007},
volume = {375},
pages = {1528},
doi = {https://doi.org/10.1111/j.1365-2966.2007.11430.x}
}

@article{Maccagni2017,
title = {Kinematics and physical conditions of HI in nearby radio sources-The last survey of the old Westerbork Synthesis Radio Telescope},
author = {Maccagni, Filippo M. and Morganti, Raffaella and Oosterloo, Tom A. and Ger{\'e}b, Katinka and Maddox, Natasha},
journal = {\aap},
year = {2017},
volume = {604},
pages = {A43},
doi = {https://doi.org/10.1051/0004-6361/201730563}
}

@article{Curran2021,
title = {Intervening or associated? Machine learning classification of redshifted HI 21-cm absorption},
author = {Curran, Stephen J.},
journal = {\mnras},
year = {2021},
volume = {506},
pages = {1548},
doi = {https://doi.org/10.1093/mnras/stab1865}
}

@article{Curran2016,
title = {A comparative study of intervening and associated H i 21-cm absorption profiles in redshifted galaxies},
author = {Curran, Stephen J. and Duchesne, S. W. and Divoli, A. and Allison, J. R.},
journal = {\mnras},
year = {2016},
volume = {462},
pages = {4197},
doi = {https://doi.org/10.1093/mnras/stw1938}
}

@article{Demleitner2001,
title = {ADS’s Dexter Data Extraction Applet},
author = {Demleitner, M. and Accomazzi, A. and Eichhorn, G. and Grant, C. S. and Kurtz, M. J. and Murray, S. S.},
journal = {ASP Conf. Proc.},
year = {2001},
volume = {238},
pages = {321}
}

@article{Gereb2015,
title = {The HI absorption “Zoo”},
author = {Ger{\'e}b, K. and Maccagni, F. M. and Morganti, R. and Oosterloo, T. A.},
journal = {\aap},
year = {2015},
volume = {575},
pages = {A44},
doi = {https://doi.org/10.1051/0004-6361/201424655}
}

@article{Kanekar2001a,
title = {HI 21 cm absorption in low z damped Lyman-$\alpha$ systems},
author = {Kanekar, Nissim and Chengalur, Jayaram N.},
journal = {\aap},
year = {2001},
volume = {369},
pages = {42},
doi = {https://doi.org/10.1051/0004-6361:20010096}
}

@article{Kanekar2001b,
title = {Detection of a multi-phase ISM at z = 0.2212},
author = {Kanekar, Nissim and Ghosh, Tapasi and Chengalur, Jayaram N.},
journal = {\aap},
year = {2001},
volume = {373},
pages = {394},
doi = {https://doi.org/10.1051/0004-6361:20010545}
}

@article{Kanekar2002,
title = {A new 21-cm absorber identified with an $L ~ L^*$ galaxy},
author = {Kanekar, Nissim and Athreya, R. M. and Chengalur, Jayaram N.},
journal = {\aap},
year = {2002},
volume = {382},
pages = {838},
doi = {https://doi.org/10.1051/0004-6361:20011674}
}

@article{Kanekar2003,
title = {A deep search for 21-cm absorption in high redshift damped Lyman-$\alpha$ systems},
author = {Kanekar, Nissim and Chengalur, Jayaram N.},
journal = {\aap},
year = {2003},
volume = {399},
pages = {857},
doi = {https://doi.org/10.1051/0004-6361:20021922}
}

@article{Ellison2012,
title = {HI content, metallicities and spin temperatures of damped and sub-damped Ly$\alpha$ systems in the redshift desert (0.6 < $z_abs$ < 1.7)},
author = {Ellison, Sara L. and Kanekar, Nissim and Prochaska, J. Xavier and Momjian, Emmanuel and Worseck, Gabor},
journal = {\mnras},
year = {2012},
volume = {424},
pages = {293},
doi = {https://doi.org/10.1111/j.1365-2966.2012.21194.x}
}

@article{Darling2004,
title = {Detection of 21 centimeter HI absorption at z = 0.78 in a survey of radio continuum sources},
author = {Darling, Jeremy and Giovanelli, Riccardo and Haynes, Martha P. and Bolatto, Alberto D. and Bower, Geoffrey C.},
journal = {\apj},
year = {2004},
volume = {613},
pages = {L101},
doi = {https://doi.org/10.1086/425143}
}

@article{Allison2016,
title = {Tracing the neutral gas environments of young radio AGN with ASKAP},
author = {Allison, J R and Sadler, E M and Moss, V A and Harvey-Smith, Lisa and Heywood, Ian and Indermuehle, B T and McConnell, D. and Sault, R. J. and Whiting, M. T.},
journal = {Astron. Nachr.},
year = {2016},
volume = {337},
pages = {175},
doi = {https://doi.org/10.1002/asna.201512288}
}

@article{Weltman2020, 
title = {Fundamental physics with the Square Kilometre Array}, 
author = {Weltman, A. and Bull, P. and Camera, S. and Kelley, K. and Padmanabhan, H. and Pritchard, J. and Raccanelli, A. and Riemer-Sørensen, S. and Shao, L. and Andrianomena, S. and et al.},
journal = {PASA},
year = {2020}, 
volume = {37},
doi = {https://doi.org/10.1017/pasa.2019.42}
}

@article{Ruder2016,
title = {An overview of gradient descent optimization algorithms},
author = {Ruder, Sebastian},
journal = {preprint (arXiv:1609.04747)},
year = {2016},
doi = {https://doi.org/10.48550/arXiv.1609.04747}
}

@article{Dutta2019,
title = {Cold neutral hydrogen gas in galaxies},
author = {Dutta, Rajeshwari},
journal = {J. Astrophys. Astron.},
year = {2019},
volume = {40},
pages = {41},
doi = {https://doi.org/10.1007/s12036-019-9610-5}
}

@article{Yan2012,
title = {Invisible Active Galactic Nuclei. I. Sample Selection and Optical/Near-IR Spectral Energy Distributions},
author = {Yan, Ting and Stocke, John T. and Darling, Jeremy and Hearty, Fred},
journal = {\apj},
year = {2012},
volume = {144},
pages = {124},
doi = {https://doi.org/10.1088/0004-6256/144/4/124}
}

@article{Yan2016,
title = {Invisible Active Galactic Nuclei. II. Radio Morphologies and Five New HI 21-cm Absorption Line Detectors},
author = {Yan, Ting and Stocke, John T. and Darling, Jeremy and Momjian, Emmanuel and Sharma, Soniya and Kanekar, Nissim},
journal = {\aj},
year = {2016},
volume = {151},
pages = {74},
doi = {https://doi.org/10.3847/0004-6256/151/3/74}
}

@article{Beck2021,
title = {PS1-STRM: neural network source classification and photometric redshift catalogue for PS1 3$\pi$ DR1},
author = {Beck, R{\'o}bert and Szapudi, Istv{\'a}n and Flewelling, Heather and Holmberg, Conrad and Magnier, Eugene and Chambers, Kenneth C.},
journal = {\mnras},
year = {2021},
volume = {500},
pages = {1633},
doi = {https://doi.org/10.1093/mnras/staa2587}
}

@article{Henghes2022,
title = {Deep learning methods for obtaining photometric redshift estimations from images},
author = {Henghes, Ben and Thiyagalingam, Jeyan and Pettitt, Connor and Hey, Tony and Lahav, Ofer},
journal = {\mnras},
year = {2022},
volume = {512},
pages = {1696},
doi = {https://doi.org/10.1093/mnras/stac480}
}

@article{Liashchynskyi2019,
title = {Grid search, random search, genetic algorithm: a big comparison for NAS},
author = {Liashchynskyi, Petro and Liashchynskyi, Pavlo},
journal = {preprint (arXiv:1912.06059)},
year = {2019},
doi = {https://doi.org/10.48550/arXiv.1912.06059}
}

@article{Peroux2020,
title = {The Cosmic Baryon and Metal Cycles},
author = {P{\'e}roux, C{\'e}line and Howk, J. Christopher},
journal = {\araa},
year = {2020},
volume = {58},
pages = {363},
doi = {https://doi.org/10.1146/annurev-astro-021820-120014}
}

@article{McClure-Griffiths2023,
title = {Atomic Hydrogen in the Milky Way: A Stepping Stone in the Evolution of Galaxies},
author = {McClure-Griffiths, Naomi M. and Stanimirovi{\'c}, Sne{\v{z}}ana and Rybarczyk, Daniel R.},
journal = {\araa},
year = {2023},
volume = {61},
pages = {19},
doi = {https://doi.org/10.1146/annurev-astro-052920-104851}
}

@article{Dutta2022,
title = {Probing galaxy evolution through HI 21-cm emission and absorption: current status and prospects with square kilometre array},
author = {Dutta, Rajeshwari and Kurapati, Sushma and Aditya, J. N. H. S. and Bait, Omkar and Das, Mousumi and Dutta, Prasun and Indulekha, K. and Nandakumar, Meera and Patra, Narendra Nath and Roy, Nirupam and Roychowdhury, Sambit},
journal = {J. Astrophys. Astron.},
year = {2022},
volume = {43},
pages = {103},
doi = {https://doi.org/10.1007/s12036-022-09875-y}
}

@article{Fernandez2016,
title = {Highest Redshift Image of Neutral Hydrogen in Emission: A CHILES Detection of a Starbursting Galaxy at z = 0.376},
author = {Fern{\'a}ndez, Ximena and Gim, Hansung B. and van Gorkom, J. H. and Yun, Min S. and Momjian, Emmanuel and Popping, Attila and Chomiuk, Laura and Hess, Kelley M. and Hunt, Lucas and Kreckel, Kathryn and Lucero, Danielle and Maddox, Natasha and Oosterloo, Tom and Pisano, D. J. and Verheijen, M. A. W. and Hales, Christopher A. and Chung, Aeree and Dodson, Richard and Golap, Kumar and Gross, Julia and Henning, Patricia and Hibbard, John and Jaff{\'e}, Yara L. and Donovan Meyer, Jennifer and Meyer, Martin and Sanchez-Barrantes, Monica and Schiminovich, David and Wicenec, Andreas and Wilcots, Eric and Bershady, Matthew and Scoville, Nick and Strader, Jay and Tremou, Evangelia and Salinas, Ricardo and Ch{\'a}vez, Ricardo},
journal = {\apjl},
year = {2016},
volume = {824},
pages = {L1},
doi = {https://doi.org/10.3847/2041-8205/824/1/L1}
}

@ARTICLE{Morganti2018,
title = {The interstellar and circumnuclear medium of active nuclei traced by HI 21 cm absorption},
author = {Morganti, Raffaella and Oosterloo, Tom},
journal = {\aapr},
year = {2018},
volume = {26},
pages = {4},
doi = {https://doi.org/10.1007/s00159-018-0109-x}
}

@article{Murthy2021,
title = {The HI absorption zoo: JVLA extension to z {\ensuremath{\sim}} 0.4},
author = {Murthy, Suma and Morganti, Raffaella and Oosterloo, Tom and Maccagni, Filippo M.},
journal = {\aap},
year = {2021},
volume = {654},
pages = {A94},
doi = {https://doi.org/10.1051/0004-6361/202141566}
}

@article{Jonas2009,
title = {MeerKAT - The South African Array With Composite Dishes and Wide-Band Single Pixel Feeds},
author = {Jonas, J. L.},
journal = {Proc. IEEE},
year = {2009},
volume = {97},
pages = {1522},
doi = {https://doi.org/10.1109/JPROC.2009.2020713}
}

@article{Johnston2008,
title = {Science with ASKAP. The Australian square-kilometre-array pathfinder},
author = {Johnston, S. and Taylor, R. and Bailes, M. and Bartel, N. and Baugh, C. and Bietenholz, M. and Blake, C. and Braun, R. and Brown, J. and Chatterjee, S. and Darling, J. and Deller, A. and Dodson, R. and Edwards, P. and Ekers, R. and Ellingsen, S. and Feain, I. and Gaensler, B. and Haverkorn, M. and Hobbs, G. and Hopkins, A. and Jackson, C. and James, C. and Joncas, G. and Kaspi, V. and Kilborn, V. and Koribalski, B. and Kothes, R. and Landecker, T. and Lenc, E. and Lovell, J. and Macquart, J. P. and Manchester, R. and Matthews, D. and McClure-Griffiths, N. and Norris, R. and Pen, U. L. and Phillips, C. and Power, C. and Protheroe, R. and Sadler, E. and Schmidt, B. and Stairs, I. and Staveley-Smith, L. and Stil, J. and Tingay, S. and Tzioumis, A. and Walker, M. and Wall, J. and Wolleben, M.},
journal = {Exp. Astron.},
year = {2008},
volume = {22},
pages = {151},
doi = {https://doi.org/10.1007/s10686-008-9124-7}}

@article{Gupta2021,
title = {Blind HI and OH Absorption Line Search: First Results with MALS and uGMRT Processed Using ARTIP},
author = {Gupta, Neeraj and Jagannathan, P. and Srianand, Raghunathan and Bhatnagar, S. and Noterdaeme, P. and Combes, F. and Petitjean, P. and Jose, J. and Pandey, S. and Kaski, C. and Baker, A. J. and Balashev, S. A. and Boettcher, E. and Chen, H. W. and Cress, C. and Dutta, Rajeshwari and Goedhart, S. and Heald, G. and J{\'o}zsa, G. I. G. and Kamau, E. and Kamphuis, P. and Kerp, J. and Kl{\"o}ckner, H. R. and Knowles, K. and Krishnan, V. and Krogager, J. K. and Kulkarni, V. P. and Momjian, E. and Moodley, K. and Passmoor, S. and Schr{\"o}eder, A. and Sekhar, S. and Sikhosana, S. and Wagenveld, J. and Wong, O. I.},
journal = {\apj},
year = {2021},
volume = {907},
pages = {11},
doi = {https://doi.org/10.3847/1538-4357/abcb85}
}

@article{Allison2022,
title = {The First Large Absorption Survey in HI (FLASH): I. Science goals and survey design},
author = {Allison, James R. and Sadler, E. M. and Amaral, A. D. and An, T. and Curran, S. J. and Darling, J. and Edge, A. C. and Ellison, S. L. and Emig, K. L. and Gaensler, B. M. and Garratt-Smithson, L. and Glowacki, M. and Grasha, K. and Koribalski, B. S. and Lagos, C. del P. and Lah, P. and Mahony, E. K. and Mao, S. A. and Morganti, R. and Moss, V. A. and Pettini, M. and Pimbblet, K. A. and Power, C. and Salas, P. and Staveley-Smith, L. and Whiting, M. T. and Wong, O. I. and Yoon, H. and Zheng, Z. and Zwaan, M. A.},
journal = {\pasa},
year = {2022},
volume = {39},
pages = {e010},
doi = {https://doi.org/10.1017/pasa.2022.3}
}

@article{Yoon2025,
title = {The First Large Absorption Survey in HI (FLASH): II. Pilot Survey data release and first results},
author = {Yoon, Hyein and Sadler, Elaine M. and Mahony, Elizabeth K. and Aditya, J. N. H. S. and Allison, James R. and Glowacki, Marcin and Kerrison, Emily F. and Moss, Vanessa A. and Su, Renzhi and Weng, Simon and et al.},
journal = {\pasa},
year = {2025},
volume = {42},
pages = {e088},
doi = {https://doi.org/10.1017/pasa.2025.10046}
}

@article{Geroldinger2023,
title = {Leave-one-out cross-validation, penalization, and differential bias of some prediction model performance measures—A simulation study},
author = {Geroldinger, Angelika and Lusa, Lara and Nold, Mariana and Heinze, Georg},
journal = {Diagn. Progn. Res.},
year = {2023},
volume = {7},
pages = {9},
doi = {https://doi.org/10.1186/s41512-023-00146-0}
}

@article{Reeves2016,
title = {HI emission and absorption in nearby, gas-rich galaxies-II. Sample completion and detection of intervening absorption in NGC 5156},
author = {Reeves, Sarah N. and Sadler, Elaine M. and Allison, James R. and Koribalski, Barbel S. and Curran, Stephen J. and Pracy, Michael B. and Phillips, Christopher J. and Bignall, Hayley E. and Reynolds, Cormac},
journal = {\mnras},
year = {2016},
volume = {457},
pages = {2613},
doi = {https://doi.org/10.1093/mnras/stv3011}
}

@article{Wardhani2019,
title = {Cross-validation metrics for evaluating classification performance on imbalanced data},
author = {Wardhani, Ni Wayan Surya and Rochayani, Masithoh Yessi and Iriany, Atiek and Sulistyono, Agus Dwi and Lestantyo, Prayudi},
journal = {IEEE IC3INA},
year = {2019},
pages = {14},
doi = {https://doi.org/10.1109/IC3INA48034.2019.8949568}
}

@article{Ellison2002,
title = {The CORALS survey. II. Clues to galaxy clustering around QSOs from z$_{abs}$ \raisebox{-0.5ex}\textasciitilde z$_{em}$ damped Lyman alpha systems},
author = {Ellison, S. L. and Yan, L. and Hook, I. M. and Pettini, M. and Wall, J. V. and Shaver, P.},
journal = {\aap},
year = {2002},
volume = {383},
pages = {91},
doi = {https://doi.org/10.1051/0004-6361:20011738}
}

@article{Prochaska2008,
title = {The SDSS-DR5 Survey for Proximate Damped Ly{\ensuremath{\alpha}} Systems},
author = {Prochaska, Jason X. and Hennawi, Joseph F. and Herbert-Fort, St{\'e}phane},
journal = {\apj},
year = {2008},
volume = {675},
pages = {1002},
doi = {https://doi.org/10.1086/526508}
}

@article{Roy2013,
title = {The temperature of the diffuse H I in the Milky Way - II. Gaussian decomposition of the H I-21 cm absorption spectra},
author = {Roy, Nirupam and Kanekar, Nissim and Chengalur, Jayaram N.},
journal = {\mnras},
year = {2013},
volume = {436},
pages = {2366},
doi = {https://doi.org/10.1093/mnras/stt1746}
}

@article{Heiles2003,
title = {The Millennium Arecibo 21 Centimeter Absorption-Line Survey. I. Techniques and Gaussian Fits},
author = {Heiles, Carl and Troland, T. H.},
journal = {\apjs},
year = {2003},
volume = {145},
pages = {329},
doi = {https://doi.org/10.1086/367785}
}

@article{Field1959,
title = {The Spin Temperature of Intergalactic Neutral Hydrogen},
author = {Field, George B.},
journal = {\apj},
year = {1959},
volume = {129},
pages = {536},
doi = {https://doi.org/10.1086/146653}
}

@book{Kulkarni1988,
title = {Neutral hydrogen and the diffuse interstellar medium},
author = {Kulkarni, Shrinivas R. and Heiles, Carl},
journal = {Galactic and Extragalactic Radio Astronomy},
year = {1988},
pages = {95},
doi = {https://doi.org/10.1007/978-1-4612-3936-9_3},
publisher = {Springer New York}
}

@article{Dutta2016,
title = {Mapping kiloparsec-scale structures in the extended H I disc of the galaxy UGC 000439 by H I 21-cm absorption},
author = {Dutta, Rajeshwari and Gupta, Neeraj and Srianand, Raghunathan and O'Meara, J. M.},
journal = {\mnras},
year = {2016},
volume = {456},
pages = {4209},
doi = {https://doi.org/10.1093/mnras/stv2980}
}

@article{Anand2022,
title = {Gaussian Na{\"\i}ve Bayes algorithm: a reliable technique involved in the assortment of the segregation in cancer},
author = {Anand, M. Vijay and KiranBala, B. and Srividhya, S. R. and Kavitha, C. and Younus, Mohammed and Rahman, Md Habibur},
journal = {Mob. Inf. Syst.},
year = {2022},
volume = {2022},
pages = {2436946},
doi = {https://doi.org/10.1155/2022/2436946}
}

@article{Curran2006,
title = {A survey for redshifted molecular and atomic absorption lines--I. The Parkes half-Jansky flat-spectrum red quasar sample},
author = {Curran, Stephen J and Whiting, M T and Murphy, M T and Webb, J K and Longmore, S N and Pihlstr{\"o}m, Y M and Athreya, R and Blake, Chris},
journal = {\mnras},
year = {2006},
volume = {371},
pages = {431},
doi = {https://doi.org/10.1111/j.1365-2966.2006.10677.x}
}

@article{Curran2008,
title = {A survey for redshifted molecular and atomic absorption lines--II. Associated H I, OH and millimetre lines in the z≳ 3 Parkes quarter-Jansky flat-spectrum sample},
author = {Curran, Stephen J and Whiting, M T and Wiklind, Tommy and Webb, J K and Murphy, M T and Purcell, C R},
journal = {\mnras},
year = {2008},
volume = {391},
pages = {765},
doi = {https://doi.org/10.1111/j.1365-2966.2008.13925.x}
}

@article{Allison2015,
title = {Discovery of H I gas in a young radio galaxy at z = 0.44 using the Australian Square Kilometre Array Pathfinder},
author = {Allison, J R and Sadler, Elaine M and Moss, V A and Whiting, M T and Hunstead, R W and Pracy, M B and Curran, Stephen J and Croom, S M and Glowacki, M and Morganti, R and et al.},
journal = {\mnras},
year = {2015},
volume = {453},
pages = {1249},
doi = {https://doi.org/10.1093/mnras/stv1532}
}

@article{Gonccalves2014,
title = {ROC curve estimation: An overview},
author = {Gon{\c{c}}alves, Luzia and Subtil, Ana and Oliveira, M Ros{\'a}rio and de Zea Bermudez, Patricia},
journal = {Revstat Stat. J.},
year = {2014},
volume = {12},
pages = {1},
doi = {https://doi.org/10.57805/revstat.v12i1.141}
}

@article{Stewart2014,
title = {A simple model for global H I profiles of galaxies},
author = {Stewart, I M and Blyth, S L and de Blok, W J G},
journal ={\aap},
year = {2014},
volume = {567},
pages = {A61},
doi = {https://doi.org/10.1051/0004-6361/201423602}
}

@article{Koch2021,
title = {A lack of constraints on the cold opaque H I mass: H I spectra in M31 and M33 prefer multicomponent models over a single cold opaque component},
author = {Koch, Eric W and Rosolowsky, Erik W and Leroy, Adam K and Chastenet, J{\'e}r{\'e}my and Chiang, I-Da and Dalcanton, Julianne and Kepley, Amanda A and Sandstrom, Karin M and Schruba, Andreas and Stanimirovi{\'c}, Sne{\v{z}}ana and et al.},
journal = {\mnras},
year = {2021},
volume = {504},
pages = {1801},
doi = {https://doi.org/10.1093/mnras/stab981}
}




\appendix

\section{Machine Learning Classification Algorithms}
\label{sec:a1}

\begin{enumerate}[label = (\roman*), labelwidth = 0pt]
\item Gaussian naive Bayes: It is a probabilistic classification algorithm based on the Bayes theorem, assuming features are conditionally independent given the class and follow a Gaussian distribution \citep{Anand2022}. It is available under the \texttt{GaussianNB} module of the \texttt{scikit-learn Python} library.

\item Logistic regression: It is a regression-based parametric classification algorithm \citep{Cramer2002} and available under the \texttt{LogisticRegression} module of the \texttt{scikit-learn Python} library.

\item Decision tree: It is a tree-based non-parametric classification algorithm \citep{Utgoff1989} and available under the \texttt{DecisionTreeClassifier} module of the \texttt{scikit-learn Python} library.

\item Random forest: It is an ensemble-based non-parametric classification algorithm that leverages the collective decision-making of multiple decision trees to enhance the accuracy and robustness of the model \citep{Ho1998}. It is available under the \texttt{RandomForestClassifier} module of the \texttt{scikit-learn Python} library.

\item XGBoost: It is an ensemble-based non-parametric classification algorithm that combines the predictions of multiple weak learners to create a strong learner \citep{Chen2016}. It is available under the \texttt{XGBClassifier} module of the \texttt{xgboost Python} library. 

\item SVM: It is a non-parametric classification algorithm that excels at separating data into distinct classes by finding the optimal hyperplane that maximizes the margin or the decision boundary that separates them \citep{Cortes1995}. It is available under the \texttt{SVC} module from the \texttt{scikit-learn Python} library.
\end{enumerate}

\section{Data Samples}
\label{sec:a2}

\onecolumn
\begin{landscape}
\centering
\scriptsize
\begin{longtable}{c c c c c c c c c c c c c c c c c c c c}
\caption{Detailed description of our data sample. The column $S_\nu$ denotes the background source flux density and $\frac{\chi^2}{\mathrm{d.o.f.}}$ denotes the reduced chi-square goodness of fit. Associated and intervening absorbers are labelled as `0' and `1' respectively in the column `Class'. Spectra marked with $\ast$ in the column `Spectra name' are obtained in digitized versions using ADS’s \texttt{Dexter Data Extraction Applet} \citep{Demleitner2001}, and the others obtained in \texttt{ASCII} format from their respective literature as given in the `Reference' column, where the following nomenclatures are used: A18a $\rightarrow$ \citet{Aditya2018a}, A18b $\rightarrow$ \citet{Aditya2018b}, A19 $\rightarrow$ \citet{Aditya2019}, C98 $\rightarrow$ \citet{Carilli1998}, D04 $\rightarrow$ \citet{Darling2004}, D17a $\rightarrow$ \citet{Dutta2017a}, D17b $\rightarrow$ \citet{Dutta2017b}, D17c $\rightarrow$ \citet{Dutta2017c}, D20 $\rightarrow$ \citet{Dutta2020}, E12 $\rightarrow$ \citet{Ellison2012}, G09 $\rightarrow$ \citet{Gupta2009}, G15 $\rightarrow$ \citet{Gereb2015}, K01a $\rightarrow$ \citet{Kanekar2001a}, K01b $\rightarrow$ \citet{Kanekar2001b}, K02 $\rightarrow$ \citet{Kanekar2002}, K03 $\rightarrow$ \citet{Kanekar2003}, K06 $\rightarrow$ \citet{Kanekar2006}, K07 $\rightarrow$ \citet{Kanekar2007}, K09a $\rightarrow$ \citet{Kanekar2009a}, K09b $\rightarrow$ \citet{Kanekar2009b}, K13 $\rightarrow$ \citet{Kanekar2013}, K14 $\rightarrow$ \citet{Kanekar2014b}, M17 $\rightarrow$ \citet{Maccagni2017}, O17 $\rightarrow$ \citet{Ostorero2017}.} 
\label{tab:b1}\\
\hline
Spectra name & $S_\nu$ & $z_\text{abs}$ & $a$ & $b_1$ & $b_2$ & $c$ & $x_{\text{e}}$ & $x_{\text{p}}$ & $w$ & $n$ & $x_0$ & $w_{50}$ & $w_{20}$ & $\tau_{\text{peak}}$ & $\tau_{\text{int}}$ & SNR & $\frac{\chi^2}{\mathrm{d.o.f.}}$ & Class & Reference \\ 
& (Jy) &  &  &  &  &  &  &  & (km $\text{s}^{-1}$) &  &  & (km $\text{s}^{-1}$) & (km $\text{s}^{-1}$) &  & (km $\text{s}^{-1}$) & ($\text{km}^{-1/2}$ $\text{s}^{1/2}$) &  &  & \\ 
\endfirsthead
\multicolumn{19}{c}%
{{\tablename\ \thetable{} -- continued from previous page}}\\ \\
\endhead
\hline 
$\ast$TXS 0003+380 & 0.547 & 0.229 & 0.07 & 0.158 & 0.315 & 0 & 17.92 & 0.1 & 4.062 & 0 & -46.47 & 0 & 0 & 0.056 & 0 & 267.876 & 0.597 & 0 & A18b \\ 
$\ast$0035+227 & 0.583 & 0.096 & 0.008 & 3.618 & 3.433 & 0.449 & 43.199 & 45.003 & 3.378 & 0.895 & 425.947 & 169.005 & 183.261 & 0.021 & 2.507 & 548.881 & 0.065 & 0 & O17 \\ 
0105-008 & 1.26 & 1.371 & 0.244 & 0.139 & 0.248 & 0 & 89.449 & 0.1 & -0.263 & 0 & -26.323 & 15.332 & 23.949 & 0.059 & 0.987 & 1561.174 & 0.072 & 0 & K09b \\ 
$\ast$0941-080 & 2.58 & 0.228 & 0.018 & 0.375 & 14.91 & 0 & 29.756 & 0.1 & 4.979 & 0 & -501.49 & 271.151 & 316.234 & 0.018 & 4.929 & 124.663 & 0.237 & 0 & O17 \\ 
$\ast$SDSS J101301.60+244837.3 & 0.892 & 0.95 & 0.005 & 0.05 & 0.23 & 0 & 83.426 & 0.1 & 34.667 & 0 & -525.438 & 201.073 & 243.043 & 0.005 & 1.076 & 1651.321 & 0.023 & 0 & A19 \\ 
$\ast$SDSS J104830.37+353800.8 & 0.553 & 0.846 & 1.219 & 0.008 & 0.192 & 0 & 216.327 & 0.1 & -85.765 & 0 & -26.071 & 208.543 & 409.57 & 0.03 & 7.764 & 597.04 & 0.271 & 0 & A19 \\ 
1142+052\footnote{We have found one duplicate in our sample corresponding to this spectra. For this, we only considered the best data collected from the Giant Metrewave Radio Telescope (GMRT) observations \citep{Kanekar2009b}.} & 1.01 & 1.343 & 0.003 & 0.362 & 0.285 & 0.074 & 52.754 & 49.021 & 13.725 & 0.998 & -15.646 & 107.432 & 122.156 & 0.006 & 0.562 & 2066.999 & 0.014 & 0 & K09b \\ 
$\ast$TXS 1200+045 & 1.675 & 1.226 & 0.004 & 0.64 & 0.05 & 0.194 & 14.188 & 20.748 & 3.296 & 1.303 & -2092.86 & 18.276 & 97.686 & 0.01 & 0.408 & 841.837 & 0.018 & 0 & A18a \\ 
$\ast$TXS 1245-197 & 8.302 & 1.275 & 0.083 & 0.139 & 0.155 & 0.001 & 36.585 & 38.423 & -2.904 & 4.236 & -6.266 & 100.362 & 203.863 & 0.02 & 2.586 & 1292.803 & 0.008 & 0 & A18a \\ 
$\ast$1504+377 & 1 & 0.673 & 0 & 0.337 & 0.306 & 3550.72 & 82.512 & 81.523 & 4.339 & 0.353 & 334.214 & 79.96 & 105.758 & 0.413 & 29.84 & 96.621 & 2.989 & 0 & C98 \\ 
SDSS J014652.79-015721.2 & 1.804 & 0.959 & 0.024 & 0.051 & 0.026 & 0 & 59.03 & 0.1 & 10.057 & 0 & 39.73 & 0 & 0 & 0.012 & 0 & 496.016 & 0.047 & 0 & A19 \\ 
SDSS J075756.71+395936.1 & 0.091 & 0.066 & 0.067 & 0.339 & 0.122 & 0 & 91.009 & 0.1 & 2.082 & 0 & -0.761 & 150.632 & 240.913 & 0.036 & 5.955 & 216.778 & 0.16 & 0 & G15, M17 \\ 
SDSS J080601.51+190614.7 & 0.142 & 0.098 & $1.152 \times 10^9$ & 0.193 & 0.017 & 0 & -35.136 & 0.1 & -131.494 & 0 & 133.327 & 220.748 & 366.912 & 0.033 & 8.31 & 148.952 & 0.217 & 0 & G15, M17 \\ 
SDSS J080938.88+345537.2 & 0.142 & 0.082 & 0.009 & 0.668 & 1.341 & 0 & 80.72 & 0.1 & 2.147 & 0 & -213.66 & 76.776 & 100.658 & 0.008 & 0.653 & 310.802 & 0.059 & 0 & G15, M17 \\ 
SDSS J082133.60+470237.3 & 1.24 & 0.128 & 0.01 & 4.835 & 4.541 & 0 & 129.971 & 0.1 & 1 & 0 & -207.428 & 37.172 & 42.595 & 0.01 & 0.363 & 852.284 & 0.017 & 0 & M17 \\ 
SDSS J083548.14+151717.0 & 0.045 & 0.168 & 0.157 & 0.331 & 1.058 & 0 & 115.169 & 112.987 & 0.405 & 8.952 & -276.114 & 58.873 & 209.356 & 0.117 & 9.145 & 40.04 & 15.719 & 0 & M17 \\ 
SDSS J083637.84+440109.6 & 0.134 & 0.055 & 0.043 & 0.417 & 0.318 & 0.032 & 90.956 & 90.992 & -0.85 & 4.823 & 37.597 & 62.498 & 181.383 & 0.016 & 1.421 & 370.049 & 0.132 & 0 & G15, M17 \\ 
SDSS J084307.11+453742.8 & 0.331 & 0.192 & $7.735 \times 10^9$ & 0.073 & 0.593 & 0 & 125.576 & 0.1 & -30.925 & 0 & 52.382 & 70.713 & 113.563 & 0.278 & 21.973 & 310.929 & 0.3 & 0 & G15, M17 \\ 
SDSS J090325.54+162256.0 & 0.048 & 0.182 & 0.017 & 0.707 & 0.452 & 2.491 & 128.719 & 129.229 & 1.938 & 1.676 & 17.209 & 118.199 & 160.788 & 0.101 & 8.464 & 47.025 & 6.769 & 0 & M17 \\ 
SDSS J090734.91+325722.9 & 0.045 & 0.049 & 0.023 & 2.62 & 0.427 & 10.11 & 127.834 & 126.232 & 2.616 & 0.477 & 16.425 & 102.47 & 128.709 & 0.323 & 28.772 & 34.684 & 11.786 & 0 & M17 \\ 
SDSS J090937.44+192808.2 & 0.063 & 0.028 & 0.075 & 0.249 & 0.873 & 0.182 & 95.729 & 96.231 & 1.753 & 2.372 & -30.568 & 106.122 & 185.032 & 0.119 & 14.365 & 101.039 & 0.496 & 0 & G15 \\ 
SDSS J093551.59+612111.3 & 0.148 & 0.039 & 0.074 & 0.047 & 0.102 & 0 & 90.413 & 0.1 & 12.299 & 0 & -74.197 & 527.342 & 806.61 & 0.057 & 31.856 & 263.781 & 0.207 & 0 & G15, M17 \\ 
SDSS J102053.67+483124.3 & 0.082 & 0.053 & 0.019 & 0.89 & 0.867 & 0.064 & 90.032 & 85.982 & 3.925 & 1.355 & -53.959 & 126.271 & 156.028 & 0.033 & 3.763 & 99.28 & 0.626 & 0 & G15, M17 \\ 
SDSS J102400.53+511248.1 & 0.047 & 0.214 & 569.267 & 0.031 & 0.737 & 0 & 157.408 & 0.1 & -40.378 & 0 & -321.02 & 124.044 & 239.057 & 0.08 & 12.468 & 60.269 & 4.514 & 0 & M17 \\ 
SDSS J102544.22+102230.4 & 0.093 & 0.046 & 0.145 & 0.861 & 2.922 & 0.012 & 127.055 & 124.001 & 1.896 & 0.427 & -15.125 & 65.187 & 80.908 & 0.148 & 9.651 & 91.882 & 1.048 & 0 & M17 \\ 
SDSS J103932.12+461205.3 & 0.031 & 0.186 & 0.202 & 0.565 & 0.597 & 0 & 128.596 & 0.1 & 1.675 & 0 & 13.914 & 75.19 & 111.499 & 0.169 & 13.3 & 15.595 & 9.942 & 0 & M17 \\ 
SDSS J110017.98+100256.8 & 0.126 & 0.036 & 0.147 & 0.351 & 0.273 & 0.026 & 124.083 & 122.619 & 2.583 & 3.327 & -9.169 & 100.082 & 195.363 & 0.369 & 46.185 & 113.409 & 1.009 & 0 & M17 \\ 
SDSS J111113.19+284147.0 & 0.036 & 0.029 & 0.125 & 0.11 & 0.716 & 0 & 133.268 & 0.1 & 3.284 & 0 & 51.133 & 149.387 & 240.749 & 0.093 & 15.072 & 33.055 & 5.211 & 0 & M17 \\ 
SDSS J111916.54+623925.7 & 0.032 & 0.11 & $1.68 \times 10^9$ & 0.384 & 0.058 & 0 & 85.118 & 0.1 & -37.617 & 0 & -91.844 & 96.812 & 153.253 & 0.28 & 29.866 & 15.576 & 16.62 & 0 & M17 \\ 
SDSS J112030.04+273610.7 & 0.177 & 0.113 & 0.361 & 0.465 & 0.554 & 0 & 88.576 & 0.1 & 0.563 & 0 & -89.214 & 58.535 & 90.155 & 0.156 & 9.777 & 259.679 & 0.368 & 0 & G15, M17 \\ 
SDSS J112332.04+235047.8 & 0.143 & 0.207 & 0.011 & 0.874 & 0.68 & 0 & 32.45 & 32.975 & 1.559 & 10.925 & -1906.4 & 33.132 & 180.289 & 0.026 & 1.759 & 112.452 & 0.714 & 0 & M17 \\ 
SDSS J120231.12+163741.8 & 0.082 & 0.12 & 0.328 & 0.35 & 0.062 & 0 & 74.352 & 0.1 & -5.46 & 0 & -123.367 & 159.459 & 273.954 & 0.036 & 6.533 & 79.017 & 1.035 & 0 & G15, M17 \\ 
SDSS J122513.09+321401.6 & 0.051 & 0.059 & 0.041 & 0.677 & 0.963 & 0.422 & 139.15 & 139.973 & 2.235 & 2.347 & 175.964 & 51.518 & 120.752 & 0.142 & 9.491 & 99.05 & 1.225 & 0 & M17 \\ 
SDSS J123200.55+331747.6 & 0.094 & 0.079 & 0.009 & 0.879 & 0.476 & 0.729 & 91.114 & 91.413 & 2.893 & 1.734 & -39.897 & 133.22 & 165.267 & 0.034 & 3.17 & 351.678 & 0.118 & 0 & G15, M17 \\ 
SDSS J123905.13+174457.5 & 0.066 & 0.065 & 0.036 & 1.236 & 0.653 & 0 & 130.329 & 0.1 & 2.113 & 0 & 45.241 & 74.641 & 99.121 & 0.035 & 2.661 & 84.452 & 1.196 & 0 & M17 \\ 
SDSS J124707.32+490017.9 & 1.14 & 0.207 & 0.024 & 0.244 & 0.087 & 0.22 & 69.204 & 75.629 & -6.808 & 1.792 & -262.929 & 290.412 & 404.55 & 0.001 & 0.351 & 2009.654 & 0.001 & 0 & G15, M17 \\ 
SDSS J130125.26+291849.5 & 0.036 & 0.023 & 0.007 & 0.627 & 1.388 & 1.284 & 134.606 & 137.668 & 7.547 & 0.804 & 84.083 & 251.861 & 279.221 & 0.053 & 8.672 & 44.337 & 1.727 & 0 & M17 \\ 
SDSS J130556.95+395621.5 & 0.036 & 0.153 & 1.56 & 7.145 & 0.424 & 7.192 & 54.03 & 56 & -1.15 & 0.061 & -1363.67 & 13.386 & 29.323 & 0.802 & 13.941 & 11.842 & 15.813 & 0 & M17 \\ 
SDSS J131535.10+620728.4 & 0.045 & 0.031 & 0.07 & 0.692 & 0.127 & 0.049 & 127.424 & 130.947 & -0.552 & 3.362 & 146.905 & 274.583 & 331.669 & 0.058 & 11.393 & 67.915 & 1.247 & 0 & M17 \\ 
SDSS J131739.20+411545.6 & 0.246 & 0.066 & $2.108 \times 10^9$ & 0.019 & 0.398 & 0 & 224.735 & 0.1 & -121.927 & 0 & 75.942 & 134.38 & 241.888 & 0.022 & 3.613 & 735.139 & 0.029 & 0 & G15, M17 \\ 
SDSS J132035.40+340821.7 & 0.097 & 0.023 & 0.15 & 0.193 & 0.123 & 0 & 95.293 & 0.1 & 7.525 & 0 & 41.807 & 274.393 & 399.718 & 0.135 & 38.198 & 253.446 & 0.324 & 0 & G15, M17 \\ 
SDSS J132513.37+395553.2 & 0.037 & 0.076 & 0.036 & 0.797 & 0.023 & 0 & 94.836 & 0.1 & 10.983 & 0 & 260.909 & 573.697 & 973.273 & 0.027 & 16.832 & 51.592 & 1.208 & 0 & G15, M17 \\ 
SDSS J133455.94+134431.7 & 0.026 & 0.023 & 0.176 & 0.929 & 1.245 & 0 & 121.888 & 0.1 & 2.093 & 0 & -101.289 & 70.609 & 89.639 & 0.176 & 12.424 & 14.745 & 25.293 & 0 & M17 \\ 
SDSS J133817.24+481629.7 & 0.079 & 0.028 & 0.013 & 0.466 & 2.942 & 0.005 & 137.631 & 138.077 & 4.26 & 4.152 & 132.12 & 185.079 & 205.138 & 0.03 & 3.762 & 74.714 & 1.109 & 0 & M17 \\ 
SDSS J134035.20+444817.3 & 0.036 & 0.065 & 0.434 & 0.361 & 2.389 & 0 & 92.404 & 0.1 & 0.502 & 0 & -13.573 & 38.203 & 64.365 & 0.249 & 10.62 & 111.694 & 1.214 & 0 & G15, M17 \\ 
SDSS J134111.14+302241.3 & 0.039 & 0.04 & 0.171 & 0.215 & 0.075 & 0 & 124.977 & 0.1 & -0.517 & 0 & 39.032 & 197.09 & 318.785 & 0.047 & 10.279 & 71.594 & 3.063 & 0 & M17 \\ 
SDSS J134442.16+555313.5 & 0.132 & 0.037 & 0.067 & 0.239 & 0.304 & 0.005 & 99.61 & 104.324 & 16.199 & 1.556 & 90.195 & 562.21 & 639.378 & 0.09 & 44.382 & 390.852 & 0.083 & 0 & G15, M17 \\ 
SDSS J134649.45+142401.7 & 0.162 & 0.022 & 0.017 & 0.946 & 6.985 & 0 & 115.73 & 0.1 & 1.335 & 0 & -205.957 & 45.087 & 57.266 & 0.017 & 0.762 & 150.515 & 0.264 & 0 & M17 \\ 
SDSS J135217.88+312646.4 & 3.53 & 0.045 & 0.029 & 0.314 & 0.456 & 0 & 84.163 & 81.852 & 4.313 & 6.662 & -130.463 & 105.418 & 223.623 & 0.057 & 7.645 & 5829.725 & 0.018 & 0 & G15, M17 \\ 
SDSS J135646.10+102609.0 & 0.061 & 0.123 & 0.059 & 6.237 & 1.267 & 0 & 137.793 & 0.1 & 6.752 & 0 & 184.09 & 250.194 & 261.609 & 0.059 & 14.816 & 21.264 & 10.145 & 0 & M17 \\ 
SDSS J135806.05+214021.1 & 0.061 & 0.066 & 0.023 & 0.747 & 0.624 & 0.519 & 130.546 & 131.506 & 2.432 & 2.653 & 24.671 & 45.62 & 72.875 & 0.162 & 9.724 & 82.203 & 1.702 & 0 & M17 \\ 
SDSS J142210.81+210554.1 & 0.084 & 0.191 & $0.493 \times 10^9$ & 0.046 & 0.356 & 0 & 134.178 & 0.1 & -48.709 & 0 & -168.21 & 116.382 & 186.48 & 0.033 & 4.302 & 148.478 & 0.334 & 0 & G15, M17 \\ 
SDSS J143521.67+505122.9 & 0.141 & 0.1 & 0.318 & 0.591 & 0.103 & 0.036 & 75.775 & 83.014 & -6.428 & 3.172 & -69.643 & 225.865 & 300.9 & 0.008 & 1.577 & 321.335 & 0.112 & 0 & G15, M17 \\ 
SDSS J144921.58+631614.0 & 2.5 & 0.042 & 0.008 & 0.059 & 0.181 & 0.12 & 102.638 & 91.291 & -4.547 & 1.664 & 27.051 & 160.325 & 455.355 & 0.004 & 0.864 & 2399.019 & 0.002 & 0 & G15, M17 \\ 
SDSS J150034.56+364845.1 & 0.061 & 0.066 & $3.878 \times 10^9$ & 0.364 & 0.05 & 0 & 48.087 & 0.1 & -45.536 & 0 & 34.592 & 100.2 & 159.618 & 0.163 & 18.163 & 123.555 & 0.836 & 0 & G15, M17 \\ 
SDSS J150721.87+101844.8 & 0.403 & 0.078 & 0.007 & 0.239 & 0.938 & 0.018 & 128.985 & 127.867 & 2.861 & 3.422 & -15.428 & 68.066 & 241.934 & 0.011 & 1.404 & 231.334 & 0.056 & 0 & M17 \\ 
SDSS J151319.23+343133.7 & 0.035 & 0.127 & 0.003 & 0.371 & 0.448 & 109.99 & 129.281 & 130.564 & 3.03 & 0.071 & 22.981 & 123.592 & 174.778 & 0.299 & 37.122 & 30.342 & 6.717 & 0 & M17 \\ 
SDSS J152446.01+230723.5 & 0.041 & 0.216 & 0.031 & 1.062 & 0.392 & 0 & 133.387 & 134.184 & 5.103 & 7.192 & 119.436 & 147.459 & 313.923 & 0.06 & 9.582 & 47.937 & 2.898 & 0 & M17 \\ 
SDSS J152922.49+362142.2 & 0.038 & 0.099 & 1.737 & 0.165 & 0.093 & 0 & 87.701 & 0.1 & -5.682 & 0 & -16.526 & 188.467 & 291.389 & 0.062 & 12.603 & 96.205 & 2.334 & 0 & G15, M17 \\ 
SDSS J153437.61+251311.4 & 0.043 & 0.034 & 0.988 & 0.405 & 0.702 & 0.249 & 123.848 & 126.986 & -0.524 & 0.478 & -83.024 & 44.279 & 68.887 & 0.2 & 9.594 & 38.04 & 4.8 & 0 & M17 \\ 
SDSS J153452.95+290919.8 & 0.049 & 0.201 & 0.006 & 0.281 & 1.39 & 0.088 & 139.345 & 144.185 & 13.816 & 1.474 & 142.085 & 556.791 & 611.721 & 0.034 & 10.919 & 53.238 & 1.454 & 0 & M17 \\ 
SDSS J155902.70+230830.4 & 0.043 & 0.193 & 0.03 & 5.304 & 6.144 & 0 & 129.962 & 0.1 & 11.028 & 0 & 41.072 & 434.017 & 438.944 & 0.03 & 13.141 & 33.587 & 2.755 & 0 & M17 \\ 
SDSS J160246.39+524358.3 & 0.577 & 0.106 & 71.53 & 0.036 & 0.162 & 0.062 & 120.743 & 83.752 & -36.014 & 2.066 & -202.735 & 448.433 & 624.031 & 0.009 & 3.361 & 982.676 & 0.01 & 0 & G15, M17 \\ 
SDSS J160332.08+171155.3 & 0.278 & 0.034 & 0.05 & 0.708 & 0.714 & 0.479 & 93.436 & 91.249 & 1.083 & 0.698 & -4.526 & 46.516 & 70.037 & 0.069 & 3.39 & 450.742 & 0.044 & 0 & G15, M17 \\ 
SDSS J160338.06+155402.5 & 0.1 & 0.11 & 0.07 & 0.271 & 0.231 & 0 & 93.104 & 94.01 & 5.676 & 4.5 & -30.413 & 348.911 & 447.286 & 0.121 & 31.673 & 98.432 & 0.392 & 0 & G15, M17 \\ 
SDSS J160952.60+133148.0 & 0.034 & 0.036 & 0.037 & 6.863 & 0.553 & 0.104 & 129.624 & 129.235 & 7.63 & 1.719 & 40.389 & 267.235 & 287.481 & 0.144 & 22.235 & 21.888 & 16.406 & 0 & M17 \\ 
SDSS J161217.62+282546.4 & 0.078 & 0.053 & 0.051 & 0.475 & 0.616 & 0 & 92.911 & 0.1 & 1.996 & 0 & -5.693 & 76.496 & 112.209 & 0.045 & 3.548 & 133.051 & 0.528 & 0 & G15, M17 \\ 
SDSS J161740.53+350015.1 & 0.141 & 0.03 & 0 & 6.729 & 1.375 & $1.29 \times 10^5$ & 116.327 & 116.807 & 3.328 & 0.631 & -200.05 & 111.317 & 122.348 & 0.02 & 1.399 & 119.448 & 0.233 & 0 & M17 \\ 
SDSS J163804.02+264329.1 & 0.041 & 0.065 & 0.047 & 0.294 & 0.39 & 0.034 & 130.403 & 129.427 & 1.878 & 3.805 & 54.594 & 71.887 & 246.621 & 0.098 & 11.721 & 48.122 & 3.625 & 0 & M17 \\ 
SDSS J163844.80+275439.1 & 0.032 & 0.104 & 0.43 & 0.502 & 0.087 & 0 & 99.221 & 0.1 & -0.446 & 0 & -427.919 & 136.493 & 234.159 & 0.143 & 22.247 & 18.932 & 20.832 & 0 & M17 \\ 
SDSS J163956.07+112757.4 & 0.159 & 0.079 & 0.251 & 0.954 & 1.089 & 0 & 126.38 & 0.1 & 1.141 & 0 & -27.121 & 43.654 & 63.401 & 0.226 & 10.198 & 59.713 & 1.156 & 0 & M17 \\ 
SDSS J170815.25+211117.7 & 0.034 & 0.224 & 0.156 & 0.376 & 0.206 & 0 & 128.791 & 0.1 & 4.197 & 0 & 28.331 & 183.398 & 268.024 & 0.14 & 26.512 & 31.035 & 7.751 & 0 & M17 \\ 
SDSS J091927.61+014603.0 & 0.183 & 1.273 & 0.026 & 0.034 & 0.021 & 0.015 & 277.171 & 268.992 & 9.236 & 0.946 & -2.449 & 129.994 & 198.64 & 0.013 & 1.785 & 382.282 & 0.32 & 0 & D17b \\ 
SDSS J152134.17+550857.2  & 0.195 & 1.07 & 0.041 & 0.696 & 0.076 & 0.004 & 255.158 & 262.361 & -0.984 & 3.167 & -2.507 & 8.312 & 44.195 & 0.023 & 0.437 & 326.081 & 0.701 & 0 & D20 \\ 
PKS 0201+113 & 0.422 & 3.388 & 0.048 & 1.613 & 0.534 & 0 & 33.108 & 0.1 & 1.329 & 0 & -2.837 & 20.938 & 31.04 & 0.043 & 0.933 & 164.795 & 0.495 & 1 & K07 \\ 
0235+164 & 2.12 & 0.524 & $7.74 \times 10^9$ & 0.362 & 0.016 & 0 & -77.619 & 0.1 & -146.505 & 0 & -46.778 & 22.424 & 40.968 & 0.156 & 4.29 & 347.165 & 2.188 & 1 & K03 \\ 
0237-233 & 6.7 & 1.672 & 232.535 & 7.734 & 0.257 & 0 & 35.367 & 0.1 & -4.702 & 0 & -4.351 & 1.461 & 3.016 & 0.056 & 0.106 & 318.013 & 0.361 & 1 & K09b \\ 
TXS 0311+430 & 5.96 & 2.289 & 0.129 & 0.131 & 0.029 & 0 & 211.759 & 0.1 & -13.446 & 0 & 62.13 & 56.514 & 94.536 & 0.01 & 0.661 & 1757.133 & 0.013 & 1 & K13 \\ 
PKS 0438-436 & 7.3 & 2.347 & 0.321 & 0.895 & 0.103 & 0 & 20.37 & 0.1 & -6.624 & 0 & 21.208 & 30.921 & 54.916 & 0.006 & 0.202 & 1216.297 & 0.007 & 1 & K06 \\ 
PKS 0458-020 (high z) & 3 & 2.039 & 0.063 & 0.247 & 0.266 & 0.163 & 58.944 & 61.272 & 7.695 & 1.032 & -2.423 & 27.017 & 34.561 & 0.12 & 2.825 & 721.455 & 0.104 & 1 & K03 \\ 
0458-020 (low z) & 2.2 & 1.561 & 0.029 & 0.401 & 0.318 & 0.021 & 114.504 & 116.365 & 0.243 & 4.053 & 0.718 & 5.258 & 9.926 & 0.024 & 0.157 & 958.966 & 0.114 & 1 & K09b \\ 
0738+313 (high z) & 2 & 0.221 & 0.122 & 0.085 & 0.16 & 0 & 499.6 & 0.1 & 5.061 & 0 & 11.691 & 4.973 & 7.714 & 0.078 & 0.416 & 2904.017 & 0.039 & 1 & K01b \\ 
0738+313 (low z) & 2 & 0.091 & 0.471 & 0.158 & 0.181 & 0 & 66.055 & 0.1 & -0.004 & 0 & 9.321 & 3.953 & 6.087 & 0.118 & 0.501 & 6993.24 & 0.037 & 1 & K03 \\ 
0801+303 & 2.07 & 1.191 & 0.004 & 4.654 & 0.077 & 0 & 36.332 & 0.1 & 10.238 & 0 & 31.156 & 74.613 & 102.91 & 0.004 & 0.298 & 1312.054 & 0.012 & 1 & K09b \\ 
0827+243 & 0.9 & 0.525 & 0.006 & 0.205 & 0.34 & 0.034 & 51.898 & 42.009 & 2.67 & 1.012 & 10.971 & 35.987 & 55.4 & 0.006 & 0.227 & 1108.11 & 0.015 & 1 & K01a \\ 
0952+179 & 1.4 & 0.238 & 0.01 & 0.218 & 0.25 & 0 & 71.532 & 0.1 & 8.484 & 0 & 2.828 & 8.678 & 11.278 & 0.01 & 0.089 & 3730.845 & 0.017 & 1 & K01a \\ 
PKS 1127-145 & 5.285 & 0.313 & 0.096 & 0.081 & 0.334 & 0 & 62.896 & 0.1 & 8.947 & 0 & 4.995 & 40.959 & 59.843 & 0.089 & 3.756 & 4929.531 & 0.752 & 1 & K03 \\ 
1157+014 & 0.89 & 1.944 & 0.046 & 5.291 & 0.367 & 0 & 32.167 & 0.1 & 2.194 & 0 & 6.012 & 16.858 & 23.404 & 0.045 & 0.774 & 252.556 & 0.768 & 1 & K09a \\ 
1229-021 & 1.65 & 0.395 & $0.61 \times 10^9$ & 0.555 & 0.072 & 0 & 8.22 & 0.1 & -30.538 & 0 & -0.783 & 17.5 & 28.021 & 0.061 & 1.195 & 1279.878 & 0.049 & 1 & K09a \\ 
PKS 1243-072 & 0.48 & 0.437 & 89.74 & 0.242 & 0.051 & 0 & 34.521 & 0.1 & -20.428 & 0 & -0.161 & 10.59 & 17.082 & 0.07 & 0.829 & 588.919 & 1.259 & 1 & K02 \\ 
MC3 1331+305 & 19 & 0.692 & $2.546 \times 10^9$ & 0.018 & 0.096 & 0 & 384.765 & 0.1 & -122.927 & 0 & 9.026 & 8.52 & 13.282 & 0.144 & 1.339 & 3025.573 & 0.055 & 1 & K03 \\ 
1429+400 & 0.21 & 0.604 & 0.143 & 0.195 & 0.107 & 0.01 & 119.537 & 122.253 & 0.421 & 3.594 & -383.773 & 22.123 & 26.408 & 0.194 & 2.98 & 171.592 & 4.955 & 1 & E12 \\ 
$\ast$1430-178 & 1.05 & 1.327 & 0.002 & 0.038 & 0.998 & 0 & 187.251 & 0.1 & 42.632 & 0 & -61.192 & 37.672 & 44.893 & 0.002 & 0.088 & 42082.385 & 0.006 & 1 & K09b \\ 
1621+074 & 0.142 & 1.337 & 0.277 & 0.182 & 4.696 & 0 & 28.482 & 0.1 & -1.575 & 0 & 7.259 & 15.742 & 30.654 & 0.051 & 0.975 & 385.076 & 0.328 & 1 & G09 \\ 
PKS 1629+120 & 2.35 & 0.532 & 1.803 & 0.15 & 0.027 & 0.053 & 38.64 & 72.005 & -25.94 & 1.288 & -7.08 & 10.858 & 23.314 & 0.035 & 0.496 & 4496.995 & 0.025 & 1 & K03 \\ 
1755+578 & 0.377 & 1.97 & 0.027 & 3.775 & 4.258 & 0 & 47.077 & 0.1 & 0.836 & 0 & 160.309 & 50.967 & 61.558 & 0.027 & 1.402 & 144.893 & 0.322 & 1 & K14 \\ 
$\ast$1830-211 & 10.5 & 0.886 & 0.003 & 0.517 & 0.298 & 1.789 & 21.133 & 25.419 & 5.518 & 1.148 & -72.411 & 215.611 & 322.826 & 0.048 & 11.262 & 414.327 & 0.088 & 1 & K03 \\ 
1850+402 & 0.65 & 1.989 & 1.022 & 0.401 & 0.032 & 0 & 154.098 & 0.1 & -13.887 & 0 & 127.241 & 30.278 & 56.227 & 0.082 & 2.991 & 299.418 & 1.504 & 1 & K14 \\ 
2003-025 & 3.7 & 1.411 & 0.003 & 1.369 & 0.567 & 0.242 & 63.695 & 61.489 & 5.355 & 0.631 & 54.334 & 43.152 & 49.039 & 0.005 & 0.21 & 2378.203 & 0.006 & 1 & K09b \\ 
2039+187 & 1.92 & 2.192 & $0.322 \times 10^9$ & 0.416 & 0.018 & 0 & -17.33 & 0.1 & -125.944 & 0 & -65.098 & 13.137 & 24.14 & 0.025 & 0.411 & 696.253 & 0.187 & 1 & K13 \\ 
2337-011 & 0.063 & 1.361 & 0.355 & 1.159 & 1.671 & 0 & 40.322 & 0.1 & 1.446 & 0 & 33.644 & 5.643 & 7.357 & 0.353 & 1.997 & 85.556 & 9.334 & 1 & K09b \\ 
2351+456 & 1.99 & 0.78 & 0.459 & 0.143 & 0.073 & 0 & 205.838 & 0.1 & 6.171 & 0 & 6.604 & 56.366 & 87.41 & 0.303 & 18.281 & 315.117 & 0.299 & 1 & D04 \\ 
2355-106 & 0.42 & 1.173 & 0.01 & 0.511 & 4.109 & 0 & 106.424 & 107.516 & 0.774 & 8.519 & 44.744 & 4.663 & 7.263 & 0.045 & 0.245 & 505.466 & 0.431 & 1 & K09b \\ 
SDSS J084957.97+510829.0 & 0.233 & 0.312 & 0.018 & 0.832 & 0.7 & 0.03 & 237.065 & 237.848 & 4.637 & 3.254 & -5.935 & 6.757 & 26.376 & 0.082 & 0.946 & 302.085 & 9.021 & 1 & D17a \\ 
SDSS J092136.22+621552.5 & 1.332 & 1.104 & 0.012 & 5.378 & 4.944 & 0.503 & 257.12 & 254.782 & 3.872 & 0.993 & 0.655 & 7.847 & 14.289 & 0.048 & 0.399 & 365.25 & 0.607 & 1 & D17b \\ 
SDSS J124157.54+633241.6 & 0.68 & 0.143 & 0.002 & 1.008 & 0.555 & 0.261 & 59.255 & 59.843 & 3.762 & 2.439 & -11.084 & 9.769 & 39.261 & 0.014 & 0.25 & 1068.227 & 0.225 & 1 & D17a \\ 
SDSS J124355.78+404358.4 & 0.196 & 0.017 & 0.104 & 0.277 & 4.637 & 0.111 & 46.168 & 44.065 & 0.574 & 4.434 & 9.378 & 2.483 & 4.654 & 0.439 & 2.678 & 250.379 & 0.542 & 1 & D17a \\ 
SDSS J125531.75+181750.9 & 0.856 & 0.758 & 0 & 0.323 & 6.854 & 2535.58 & 966.391 & 975.233 & 14.453 & 1.37 & -1.315 & 25.089 & 39.57 & 0.078 & 2.109 & 321.044 & 0.526 & 1 & D17b \\ 
SDSS J132720.97+432627.9 & 0.647 & 0.954 & 0.038 & 3.763 & 4.552 & 0 & 1118.64 & 0.1 & 0.405 & 0 & 338.648 & 1.838 & 2.515 & 0.037 & 0.069 & 336.622 & 0.546 & 1 & D17b \\ 
SDSS J134224.31+511012.4 & 0.157 & 1.488 & 0.066 & 0.11 & 0.387 & 0 & 234.553 & 0.1 & 12.428 & 0 & 3.201 & 53.767 & 68.766 & 0.066 & 3.543 & 143.385 & 2.677 & 1 & D17b \\ 
SDSS J142846.41+210336.6 & 0.137 & 0.394 & $0.597 \times 10^6$ & 0.341 & 0.745 & 0.056 & 236.719 & 233.156 & -5.341 & 6.767 & 3.957 & 10.107 & 16.845 & 0.25 & 2.3 & 258.821 & 0.929 & 1 & D17c \\ 
SDSS J143806.79+175805.4 & 0.053 & 0.147 & 0.314 & 0.208 & 0.387 & 0 & 249.837 & 0.1 & 3.985 & 0 & -3.108 & 17.199 & 25.277 & 0.277 & 4.951 & 70.167 & 10.427 & 1 & D17a \\ 
SDSS J144304.53+021419.3 & 0.163 & 0.372 & $5.226 \times 10^9$ & 0.09 & 0.556 & 0 & 274.913 & 0.1 & -24.985 & 0 & 0.228 & 8.428 & 13.278 & 0.334 & 3.088 & 438.749 & 0.423 & 1 & D17a \\ 
SDSS J163956.35+112758.7 & 0.155 & 0.079 & 0.481 & 0.21 & 0.136 & 0.02 & 238.244 & 253.009 & 13.134 & 0.985 & -1.4 & 23.008 & 30.306 & 0.661 & 15.045 & 305.535 & 13.051 & 1 & D17a \\ 
SDSS J221930.79+022945.4 & 0.2 & 0.981 & 0.289 & 0.222 & 0.222 & 0.089 & 573.187 & 572.297 & -1.562 & 3.562 & 4.283 & 11.094 & 36.077 & 0.251 & 3.946 & 104.312 & 4.987 & 1 & D17b \\ 
1331+170 & 0.62 & 1.776 & $0.053 \times 10^9$ & 0.011 & 0.053 & 0 & 307.951 & 0.1 & -192.536 & 0 & 2.753 & 24.369 & 37.928 & 0.035 & 0.92 & 917.834 & 0.466 & 1 & K09a \\ 
SDSS J004125.98-014324.6 & 0.215 & 0.018 & 0.066 & 0.519 & 0.44 & 0.036 & 228.776 & 229.988 & 2.034 & 1.008 & -1.854 & 8.359 & 12.395 & 0.058 & 0.501 & 258.145 & 1.029 & 1 & D17a \\ 
SDSS J155121.13+071357.7 & 0.058 & 0.329 & 0.268 & 9.502 & 0.536 & 0 & 230.31 & 0.1 & 0.271 & 0 & 0.763 & 2.768 & 5.027 & 0.162 & 0.512 & 92.546 & 3.822 & 1 & D17c \\  
\hline
\end{longtable}
\end{landscape}

\begin{longtable}{c c c c c c }
\caption{The spectral features of 30 new \hi\ 21-cm absorption spectra of the FLASH pilot surveys. The column $z_\text{HI}$ denotes \hi\ absorber redshift, $\tau_{\text{int}}$ denotes integrated optical depth of the spectral line and `Linewidth' denotes the FWHM linewidth from a single Gaussian fit. These are taken from Table 3 of \citet{Yoon2025}. The column `ML predicted class' denotes the absorber type predicted by the random forest model trained using $w_{20}$ and $\tau_{\text{int}}$ on our data sample. The column `Agreement flag' denotes the absorber type ML classification agreement with \citet{Yoon2025} per the logistic regression model of \citet{Curran2021} trained using Gaussian function fitted spectral parameters.}
\label{tab:b2} \\
\hline
Spectra name & $z_\text{HI}$ & $\tau_{\text{int}}$ & Linewidth & ML predicted class & Agreement flag \\
\hline
MRC 0023-482 & 0.6745 & $4.63\substack{+0.47\\-0.46}$ & $70.0\substack{+7.8\\-7.1}$ & 0 & No \\ 
NVSS J014141-231511 & 0.6707 & $17.42\substack{+1.91\\-1.92}$ & $142.2\substack{+18.7\\-16.1}$ & 0 & Yes \\
NVSS J015516-251423 & 0.7251 & $94.53\substack{+4.31\\-4.08}$ & $51.6\substack{+1.7\\-1.6}$ & 1 & No \\
PKS 0253-259 & 0.6564 & $1.12\substack{+0.13\\-0.12}$ & $19.7\substack{+2.6\\-2.3}$ & 1 & Yes\\ 
SUMSS J045501-423858 & 0.6525 & $7.38\substack{+0.70\\-0.67}$ & $145.7\substack{+17.5\\-16.4}$ & 0 & Yes\\ 
NVSS J090331+010846 & 0.5218 & $106.74\substack{+6.93\\-6.68}$ & $61.1\substack{+2.4\\-2.4}$ & 0 & Yes\\ 
NVSS J090425+012015 & 0.8004 & $37.42\substack{+2.79\\-2.74}$ & $32.1\substack{+3.5\\-3.0}$ & 1 & Yes \\ 
NVSS J091256+030021 & 0.8592 & $64.93\substack{+7.09\\-6.64}$ & $51.7\substack{+5.7\\-5.4}$ & 1 & Yes \\
PKS 0917+18 & 0.9044 & $1.37\substack{+0.17\\-0.16}$ & $171.5\substack{+30.4\\-25.4}$ & 0 & Yes \\ 
NVSS J092012+161238 & 0.4362 & $5.28\substack{+0.55\\-0.53}$ & $28.0\substack{+3.0\\-2.6}$ & 1 & Yes \\ 
NVSS J113622+004850 & 0.5632 & $6.57\substack{+0.61\\-0.61}$ & $46.9\substack{+5.3\\-4.5}$ & 1 & Yes \\ 
PKS 2007-245 & 0.6778 & $1.06\substack{+0.07\\-0.06}$ & $19.0\substack{+1.8\\-1.7}$ & 1 & Yes \\ 
NVSS J215924-241752 & 0.8679 & $2.69\substack{+0.35\\-0.32}$ & $95.0\substack{+19.1\\-14.9}$ & 0 & No \\
NVSS J223605-251919 & 0.4974 & $11.29\substack{+0.45\\-0.47}$ & $27.2\substack{+1.3\\-1.3}$ & 1 & Yes \\ 
NVSS J223620-222430 & 0.7846 & $15.32\substack{+1.26\\-1.18}$ & $97.7\substack{+9.6\\-8.3}$ & 0 & Yes \\ 
MRC 2234-254 & 0.4641 & $3.06\substack{+0.39\\-0.35}$ & $42.9\substack{+8.1\\-7.6}$ & 1 & Yes \\ 
PKS 2311-477 & 0.5811 & $9.65\substack{+0.22\\-0.21}$ & $114.6\substack{+2.8\\-2.7}$ & 0 & Yes \\ 
SUMSS J233432-585646 & 0.5769 & $37.38\substack{+3.60\\-3.77}$ & $102.1\substack{+11.1\\-10.0}$ & 0 & Yes \\
PKS 0011-023 & 0.6785 & $7.90\substack{+0.20\\-0.20}$ & $36.6\substack{+1.2\\-1.2}$ & 1 & Yes \\ 
NVSS J002331+010114 & 0.5159 & $32.30\substack{+3.20\\-3.06}$ & $74.8\substack{+9.6\\-8.4}$ & 0 & Yes \\ 
PKS 0405-280 & 0.7280 & $3.34\substack{+0.22\\-0.22}$ & $94.6\substack{+9.4\\-9.1}$ & 0 & Yes \\ 
NVSS J051806-245502 & 0.5538 & $17.85\substack{+0.68\\-0.70}$ & $114.2\substack{+3.9\\-4.1}$ & 0 & Yes \\ 
MRC 0531-237 & 0.8508 & $75.82\substack{+0.35\\-0.35}$ & $199.1\substack{+0.8\\-0.7}$ & 0 & Yes \\ 
NVSS J094650-202044 & 0.9134 & $10.61\substack{+0.54\\-0.54}$ & $78.2\substack{+4.2\\-4.0}$ & 0 & No \\ 
NVSS J100238-195917 & 0.4815 & $17.16\substack{+2.31\\-2.24}$ & $34.2\substack{+5.4\\-4.3}$ & 1 & Yes \\ 
NVSS J150506+022927 & 0.8085 & $11.93\substack{+1.54\\-1.50}$ & $106.4\substack{+18.8\\-16.0}$ & 0 & Yes \\ 
NVSS J170135-294918 & 0.6299 & $4.43\substack{+0.30\\-0.30}$ & $29.8\substack{+2.1\\-2.1}$ & 1 & Yes \\ 
NVSS J205147+021740 & 0.8884 & $29.91\substack{+2.35\\-2.35}$ & $81.0\substack{+7.3\\-6.9}$ & 0 & No \\ 
NVSS J223317-015739 & 0.6734 & $5.91\substack{+0.62\\-0.63}$ & $79.0\substack{+9.4\\-7.8}$ & 0 & No \\ 
NVSS J233702-015209 & 0.7645 & $9.26\substack{+1.23\\-1.21}$ & $127.8\substack{+23.5\\-19.2}$ & 0 & Yes \\
\hline
\end{longtable}

\bsp	
\label{lastpage}

\end{document}